\documentclass[12pt,a4paper]{report}

\usepackage[utf8]{inputenc}
\usepackage[T1]{fontenc}
\usepackage{textcomp}
\usepackage{lmodern}
\usepackage{microtype}
\usepackage{amsmath,amssymb,bm}
\usepackage{graphicx}
\usepackage{booktabs}
\usepackage{array}
\usepackage{tabularx}
\usepackage{longtable}
\usepackage{multirow}
\usepackage{enumitem}
\usepackage{caption}
\usepackage{subcaption}
\usepackage{float}
\usepackage{setspace}
\usepackage{geometry}
\usepackage{xcolor}
\usepackage{tikz}
\usepackage{titlesec}
\usepackage{tocloft}
\usepackage{fancyhdr}
\usepackage{hyperref}
\usepackage[open,openlevel=1]{bookmark}

\usetikzlibrary{calc}
\graphicspath{{figure/}}
\DeclareGraphicsExtensions{.pdf,.png,.jpg,.jpeg}

\definecolor{ITUNavy}{HTML}{172A53}
\definecolor{ITULight}{HTML}{EEF3FA}
\definecolor{ITUGray}{HTML}{5D6470}

\DeclareUnicodeCharacter{0130}{\.I}
\DeclareUnicodeCharacter{0131}{\i}
\DeclareUnicodeCharacter{011E}{\u{G}}
\DeclareUnicodeCharacter{011F}{\u{g}}
\DeclareUnicodeCharacter{015E}{\c{S}}
\DeclareUnicodeCharacter{015F}{\c{s}}
\DeclareUnicodeCharacter{2013}{--}
\DeclareUnicodeCharacter{2014}{---}
\DeclareUnicodeCharacter{2018}{`}
\DeclareUnicodeCharacter{2019}{'}
\DeclareUnicodeCharacter{201C}{``}
\DeclareUnicodeCharacter{201D}{''}

\hypersetup{
  hidelinks,
  plainpages=false,
  hypertexnames=false,
  pdftitle={Communication Channel Modeling of Unmanned Aerial Vehicles},
  pdfauthor={Necati Kagan Erkek, Emre Balci, Berkin Halay},
  pdfsubject={Bachelor Thesis},
  pdfkeywords={UAV, channel modeling, channel sounding, path loss, CIR, PDP}
}

\titleformat{\chapter}[display]
  {\normalfont\bfseries\color{ITUNavy}}
  {\raggedleft\Large\chaptertitlename\ \thechapter}
  {0.8ex}
  {\Huge\raggedright}
\titlespacing*{\chapter}{0pt}{-8pt}{28pt}

\titleformat{\section}
  {\normalfont\Large\bfseries\color{ITUNavy}}
  {\thesection}{1em}{}
\titleformat{\subsection}
  {\normalfont\large\bfseries\color{ITUNavy}}
  {\thesubsection}{1em}{}

\fancypagestyle{thesisstyle}{%
  \fancyhf{}
  \fancyhead[L]{\includegraphics[height=0.85cm]{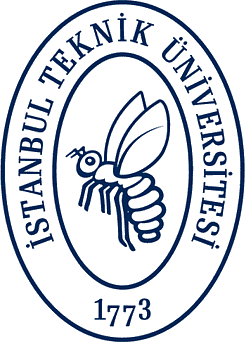}}
  \fancyhead[C]{\footnotesize\nouppercase{\leftmark}}
  \fancyhead[R]{\footnotesize\thepage}

}
\fancypagestyle{plain}{%
  \fancyhf{}
  \fancyhead[L]{\includegraphics[height=0.85cm]{figure/logo_header_itu.png}}
  \fancyhead[C]{\footnotesize\nouppercase{\leftmark}}
  \fancyhead[R]{\footnotesize\thepage}

}
\newcommand{\ThesisTitle}{Communication Channel Modeling of Unmanned Aerial Vehicles}
\newcommand{\ThesisSubtitle}{Measurement-Based Characterization of G2G, A2G, and A2A Links}
\newcommand{\ThesisType}{Bachelor Thesis}
\newcommand{\UniversityName}{Istanbul Technical University}
\newcommand{\FacultyName}{Faculty of Electrical and Electronics Engineering}
\newcommand{\ProgrammeName}{Electronics and Communication Engineering Programme}
\newcommand{\SupervisorName}{Prof. Dr. Hakan Ali Çırpan}
\newcommand{\ThesisDate}{June 2022}

\newcommand{\AuthorBlock}{%
\begin{tabular}{rl}
Necati Kağan Erkek & 040170082\\
Emre Balcı & 040170113\\
Berkin Halay & 040180732
\end{tabular}}

\newcommand{\frontchapter}[1]{%
  \cleardoublepage
  \phantomsection
  \chapter*{#1}
  \addcontentsline{toc}{chapter}{#1}
  \markboth{#1}{#1}
}

\newcommand{\makecustomcover}{%
\begin{titlepage}
\thispagestyle{empty}
\begin{tikzpicture}[remember picture,overlay]

  \fill[ITUNavy] (current page.south west) rectangle ([xshift=1.35cm]current page.north west);
  \fill[ITUNavy] ([yshift=-7.2cm]current page.north west) rectangle ([yshift=-7.35cm]current page.north east);
  \node[opacity=0.025,anchor=center] at ([xshift=4.8cm,yshift=-6.8cm]current page.center)
    {\includegraphics[width=0.78\paperwidth]{figure/logo_header_itu.png}};
  \node[anchor=north west] at ([xshift=7.25cm,yshift=-0.40cm]current page.north west)
    {\includegraphics[width=0.38\paperwidth]{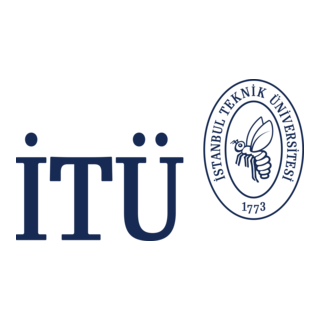}};
;
\end{tikzpicture}

\vspace*{4.55cm}
\begin{center}
  {\color{ITUNavy}\fontsize{25}{25}\selectfont\bfseries \UniversityName}\par
  \vspace{0.5cm}
  {\large \FacultyName}\par
  {\large \ProgrammeName}\par
  \vspace{1.25cm}
  {\color{ITUNavy}\fontsize{25}{25}\selectfont\bfseries {\ThesisTitle}\par}
  \vspace{0.35cm}
  {\large\itshape \ThesisSubtitle}\par
  \vspace{1.05cm}
  {\Large\bfseries \ThesisType}\par
  \vspace{0.75cm}
  {\large\bfseries Authors}\par
  \vspace{0.25cm}
  {\large \AuthorBlock}\par
  \vspace{0.60cm}
  {\large\bfseries Supervisor}\par
  \vspace{0.25cm}
  {\large \SupervisorName}\par
  \vfill
  {\large\bfseries \ThesisDate}\par
\end{center}
\end{titlepage}
}

\begin{document}

\makecustomcover

\pagenumbering{roman}
\setcounter{page}{1}

\frontchapter{FOREWORD}
\hspace{0.95cm} We would like to express our sincere gratitude to our supervisor, \SupervisorName, for his guidance, encouragement, and valuable technical feedback throughout this thesis. His support helped us define the scope of the project, improve the measurement methodology, and interpret the results more carefully.

We also thank the TÜBİTAK BİLGEM Hisar Laboratory personnel for their assistance during the experimental stages of the work. Their experience and support were particularly valuable during the preparation of the measurement setup, the verification tests, and the outdoor campaigns. We are also grateful to our friends and families for their patience, motivation, and continuous support during the completion of this bachelor thesis.

\vspace{1.5cm}
\noindent\ThesisDate

\vspace{1.0cm}
\noindent Necati Kağan Erkek\\
Emre Balcı\\
Berkin Halay

\frontchapter{ABSTRACT}

\hspace{0.95cm} Unmanned aerial vehicles (UAVs) are increasingly used in civil, industrial, and defense-oriented applications. As their operational roles expand, reliable wireless communication becomes a fundamental requirement for safe control, data exchange, and mission performance. This need has made UAV communication channel modeling an important research topic, particularly for links in which the transmitter, the receiver, or both platforms are mobile.

This thesis investigates UAV communication channels through a measurement-based approach. The study focuses not only on large-scale path-loss behavior, but also on small-scale channel characteristics such as the channel impulse response (CIR) and the power delay profile (PDP). These parameters are important because received power alone cannot fully describe the delay-domain structure of the channel or the effect of multipath propagation.

The work begins with a review of UAV channel modeling studies and channel sounding techniques. Based on this review, a software-defined radio (SDR)-based measurement system is used to collect in-phase and quadrature (IQ) data. The transmitted sounding sequence and receiver processing are selected so that path loss, CIR, and PDP can be extracted from the measured data. Preliminary verification tests are performed before the UAV-based measurements in order to evaluate the reliability of the setup.

Three measurement scenarios are considered: ground-to-ground (G2G), air-to-ground (A2G), and air-to-air (A2A). The G2G campaign is used as an initial validation stage. The A2G campaign extends the setup by mounting the transmitter on a UAV, while the A2A campaign represents the most complex case because both communication terminals are aerial platforms. For each scenario, the collected data are processed and interpreted in terms of the main channel statistics.

The results show that path loss provides useful information about average attenuation, while the PDP and CIR provide additional insight into the multipath structure of the propagation environment. Therefore, a more realistic UAV channel model should include both large-scale and small-scale statistics. The thesis concludes with recommendations for future measurements, including higher sounding bandwidth, improved synchronization, different environments, and a more detailed investigation of Doppler behavior.

\frontchapter{ÖZET}
\hspace{0.95cm} İnsansız hava araçları (İHA), sivil, endüstriyel ve savunma odaklı birçok uygulamada giderek daha yaygın biçimde kullanılmaktadır. Bu kullanım alanları genişledikçe, güvenli kontrol, veri aktarımı ve görev başarımı için güvenilir kablosuz haberleşme temel bir gereksinim haline gelmektedir. Bu nedenle, özellikle verici, alıcı veya her iki platformun hareketli olduğu durumlarda İHA haberleşme kanallarının modellenmesi önemli bir araştırma konusu olmuştur.

Bu tezde, İHA haberleşme kanalları ölçüm tabanlı bir yaklaşımla incelenmiştir. Çalışmada yalnızca büyük ölçekli yol kaybı davranışı değil, aynı zamanda kanal darbe cevabı (CIR) ve güç gecikme profili (PDP) gibi küçük ölçekli kanal özellikleri de ele alınmıştır. Bu parametreler önemlidir; çünkü yalnızca alınan güç değeri, kanalın gecikme alanındaki yapısını ve çok yollu yayılım etkilerini tam olarak açıklamak için yeterli değildir.

Çalışma, İHA kanal modelleme araştırmalarının ve kanal kestirim tekniklerinin incelenmesiyle başlamaktadır. Bu incelemeye dayanarak, yazılım tanımlı radyo (SDR) tabanlı bir ölçüm sistemi kullanılmış ve eş fazlı/dik fazlı (IQ) veriler toplanmıştır. Gönderilen kanal kestirim dizisi ve alıcı işleme adımları, ölçülen verilerden yol kaybı, CIR ve PDP parametrelerinin çıkarılmasına uygun olacak şekilde belirlenmiştir. İHA tabanlı ölçümlerden önce ölçüm düzeneğinin güvenilirliğini değerlendirmek amacıyla ön doğrulama testleri gerçekleştirilmiştir.

Tez kapsamında üç ölçüm senaryosu incelenmiştir: yerden yere (G2G), havadan yere (A2G) ve havadan havaya (A2A). G2G ölçümleri başlangıç doğrulama aşaması olarak kullanılmıştır. A2G senaryosunda verici bir İHA üzerine yerleştirilmiş, A2A senaryosunda ise her iki haberleşme terminali de hava platformu olarak kullanılmıştır. Her senaryo için toplanan veriler işlenmiş ve temel kanal istatistikleri açısından yorumlanmıştır.

Sonuçlar, yol kaybının ortalama zayıflama hakkında yararlı bilgi sağladığını; ancak PDP ve CIR parametrelerinin yayılım ortamındaki çok yollu yapıyı anlamak için ek ve önemli bilgiler sunduğunu göstermektedir. Bu nedenle, daha gerçekçi bir İHA kanal modeli hem büyük ölçekli hem de küçük ölçekli istatistikleri birlikte içermelidir. Tez, daha yüksek kanal kestirim bant genişliği, gelişmiş senkronizasyon, farklı çevre koşulları ve Doppler davranışının daha ayrıntılı incelenmesi gibi gelecek çalışma önerileriyle tamamlanmaktadır.

\frontchapter{ABBREVIATIONS}
\begin{longtable}{@{}p{3.0cm}p{11.0cm}@{}}
\textbf{A2A} & Air-to-Air\\
\textbf{A2G} & Air-to-Ground\\
\textbf{CIR} & Channel Impulse Response\\
\textbf{CSS} & Chirp Spread Spectrum\\
\textbf{DSSS} & Direct Sequence Spread Spectrum\\
\textbf{FSPL} & Free-Space Path Loss\\
\textbf{G2G} & Ground-to-Ground\\
\textbf{GM} & Graph Model\\
\textbf{GPS} & Global Positioning System\\
\textbf{ISI} & Inter-Symbol Interference\\
\textbf{LFSR} & Linear Feedback Shift Register\\
\textbf{LNA} & Low-Noise Amplifier\\
\textbf{LoRa} & Long Range\\
\textbf{LOS} & Line-of-Sight\\
\textbf{LTE} & Long-Term Evolution\\
\textbf{MIMO} & Multiple-Input Multiple-Output\\
\textbf{MPC} & Multipath Component\\
\textbf{NLOS} & Non-Line-of-Sight\\
\textbf{OFDM} & Orthogonal Frequency-Division Multiplexing\\
\textbf{OLOS} & Obstructed Line-of-Sight\\
\textbf{PA} & Power Amplifier\\
\textbf{PAPR} & Peak-to-Average Power Ratio\\
\textbf{PDP} & Power Delay Profile\\
\textbf{PL} & Path Loss\\
\textbf{PLE} & Path-Loss Exponent\\
\textbf{PN} & Pseudo-Noise\\
\textbf{RF} & Radio Frequency\\
\textbf{RMS} & Root Mean Square\\
\textbf{Rx} & Receiver\\
\textbf{SDR} & Software-Defined Radio\\
\textbf{SIMO} & Single-Input Multiple-Output\\
\textbf{SISO} & Single-Input Single-Output\\
\textbf{Tx} & Transmitter\\
\textbf{UAV} & Unmanned Aerial Vehicle\\
\textbf{UDP} & User Datagram Protocol\\
\textbf{VNA} & Vector Network Analyzer\\
\textbf{VSWR} & Voltage Standing Wave Ratio\\
\textbf{Wi-Fi} & Wireless Fidelity\\
\end{longtable}

\frontchapter{SYMBOLS}
\begin{longtable}{@{}p{3.0cm}p{11.0cm}@{}}
\textbf{$c$} & Speed of light\\
\textbf{$f$} & Frequency\\
\textbf{$t$} & Time\\
\textbf{$R_b$} & Bit rate\\
\textbf{$V_f$} & Velocity factor\\
\textbf{$\eta$} & Path-loss exponent\\
\textbf{$\epsilon_r$} & Relative dielectric constant\\
\textbf{$h(t)$} & Channel impulse response\\
\textbf{$P(\tau)$} & Power delay profile as a function of delay\\
\textbf{$\tau$} & Propagation delay or mean excess delay, depending on context\\
\textbf{$\tau_0$} & First-arrival delay\\
\textbf{$\tau_x$} & Maximum excess delay\\
\textbf{$\sigma_\tau$} & RMS delay spread\\
\textbf{$d$} & Transmitter--receiver separation distance\\
\textbf{$P_r$} & Received power\\
\textbf{$P_t$} & Transmit power\\
\end{longtable}

\cleardoublepage
\phantomsection
\tableofcontents
\markboth{CONTENTS}{CONTENTS}

\cleardoublepage
\phantomsection
\addcontentsline{toc}{chapter}{LIST OF FIGURES}
\listoffigures
\markboth{LIST OF FIGURES}{LIST OF FIGURES}

\cleardoublepage
\phantomsection
\addcontentsline{toc}{chapter}{LIST OF TABLES}
\listoftables
\markboth{LIST OF TABLES}{LIST OF TABLES}

\cleardoublepage
\pagenumbering{arabic}
\setcounter{page}{1}

%%%%%%%%%%%%%%%%%%%%%%%%%%%%%%%%%%%%%%%%%%%%%%%%%%%%%%%%%%%%%%%%%
\chapter{PROJECT DETAILS}\label{Ch1}
%%%%%%%%%%%%%%%%%%%%%%%%%%%%%%%%%%%%%%%%%%%%%%%%%%%%%%%%%%%%%%%%%
\section{Project Statement}
\subsection{Purpose of the Thesis}\label{purposeofthesis}

\hspace{0.95cm} The main purpose of this thesis is to develop a more realistic measurement-based channel characterization for unmanned aerial vehicle (UAV) communication links. Many earlier studies on UAV channels emphasize large-scale path-loss behavior, whereas the present study also considers small-scale channel characteristics such as the channel impulse response (CIR) and the power delay profile (PDP). By evaluating these parameters together, the propagation environment can be described in a more complete way, especially for links in which reflections, scattering, and mobility affect the received signal.

This project was selected because UAV communication channel modeling is still an active and developing research area. As UAVs are increasingly used in monitoring, delivery, emergency response, mapping, and military applications, reliable wireless links become essential for safe and effective operation. A channel model based only on received power is not sufficient to describe how multipath components arrive at the receiver or how the signal spreads in time. Therefore, this thesis aims to contribute a practical measurement methodology and a clearer interpretation of UAV channel statistics. The required theoretical background was obtained from communication systems courses, and the experimental work was supported by the technical experience and facilities available at TUBITAK.

\subsection{Project Work Plan}

\hspace{0.95cm} The work was carried out in several stages. First, the system elements used in UAV communication channels were identified through a literature review. This review covered channel types, measurement scenarios, channel sounding methods, and the parameters commonly used in channel characterization. Next, suitable channel sounding techniques were investigated for measuring CIR, PDP, and path loss (PL). The mathematical definitions of these parameters were examined so that the measurements could be processed consistently.

After the theoretical investigation, the required hardware and software components were selected for the measurement setup. The transmitter and receiver structures were designed, the software-defined radio (SDR) units were configured, and the signal sequences used for channel sounding were tested. Before the outdoor campaigns, preliminary tests were performed to verify that the system could detect multipath components and provide meaningful delay-domain information.

The final stage consisted of outdoor measurement campaigns under three scenarios: ground-to-ground (G2G), air-to-ground (A2G), and air-to-air (A2A). The measurement setup was modified for each scenario because the mechanical structure, link distance, antenna placement, and synchronization requirements changed as the system moved from a fixed terrestrial link to UAV-based links. The collected in-phase and quadrature (IQ) data were then processed to obtain the corresponding channel parameters, and the results were interpreted separately for each scenario.

\section{Literature Review}

\subsection{Wireless Communication Channel Modeling Scenarios}

\hspace{0.95cm} Communication channel models for unmanned aerial vehicles can mainly be grouped into air-to-ground (A2G) and air-to-air (A2A) channel models. A2G channels describe the link between a UAV and a ground station, while A2A channels describe the link between two aerial platforms. Several A2G studies have been reported in the literature [1--6]. These studies can be classified in different ways. One classification is based on whether the model is obtained from simulations only or from both simulations and real-time measurements. Another classification is based on whether the model is deterministic or stochastic.

Detailed measurements and simulation studies are necessary because UAV links differ from conventional terrestrial links. UAV channels often include a strong line-of-sight (LOS) component, but the received signal can still be affected by ground reflections, buildings, terrain, antenna orientation, and UAV motion. A2G studies generally include both simulation-based analyses and real measurement campaigns. In contrast, the number of comprehensive A2A measurement campaigns is more limited, mainly because flying two UAVs simultaneously introduces additional safety, synchronization, and regulatory constraints. Some A2A studies are presented in [7--10]. These works usually consider low-altitude UAVs, and their channel characterization is performed at the physical layer using parameters extracted from transmitted and received data packets.

\subsection{Measurement Methods}

\hspace{0.95cm} In [11], a large-scale fading model between UAVs and cellular base stations was studied at the 800 MHz band. Other studies have focused on path-loss prediction, including approaches in which machine learning algorithms are used with ray-tracing data for A2A scenarios. In practical measurement campaigns, the selection of equipment depends strongly on the budget, the desired channel parameters, the available bandwidth, and the complexity of the measurement environment. UAV measurements also require additional safety procedures and flight permissions compared with conventional ground-based measurements.

Some channel sounding methods employ a vector network analyzer (VNA). Although VNAs provide accurate frequency-domain measurements, they generally require a wired connection between the transmitter and receiver or a heavy measurement setup. This makes them less suitable for UAV applications, where payload weight and battery capacity are limited. Software-defined radio offers a more flexible alternative because the transmitted waveform, sampling rate, carrier frequency, and receiver processing can be adjusted in software. For this reason, SDR-based channel sounding is appropriate for experimental UAV links, especially when the measurement system must be portable and adaptable.

Considering the physical restrictions of UAV platforms and the need for practical A2A measurements, this thesis uses an SDR-based channel sounding approach. A sounder operating around 2.57 GHz is used in the measurement campaigns, and the received IQ data are processed to extract PDP and path-loss behavior. Although UAV communication includes many propagation mechanisms, LOS-dominant large-scale behavior remains important, particularly for A2A scenarios. At the same time, small-scale parameters such as delay spread and multipath structure must also be examined to understand the channel more accurately.

\subsection{Applied Modules in Recent Studies}

\hspace{0.95cm} Some earlier UAV channel studies rely only on path loss for channel modeling. This approach is useful for estimating received power as a function of distance, but it does not fully describe the delay-domain structure of the channel. A more complete channel model should also include the power delay profile and the channel impulse response, because these parameters reveal how multiple copies of the signal arrive at different delays.

\

\

The LOS component is usually dominant in UAV channels because aerial links often have fewer obstructions than terrestrial links. However, terrain, buildings, vegetation, ground reflections, and antenna orientation can still produce non-line-of-sight (NLOS) or obstructed-line-of-sight (OLOS) components. Therefore, path-loss exponents obtained under LOS conditions may not represent every UAV operating environment. In the literature, path-loss exponents are commonly estimated by fitting measured power values to a log-distance model. This fitting procedure provides a compact representation of the average attenuation trend, but it must be supported with delay-domain information when multipath behavior is important.

Different communication modules have also been used in UAV channel measurements. For example, Wi-Fi modules operating at 2.4 GHz and LoRa modules operating at 868 MHz have been compared in terms of path-loss behavior between transmitting and receiving UAVs. The measurements can vary with ground reflection conditions, UAV speed, link distance, and antenna gain. In general, Wi-Fi can provide higher data rates at the same transmit power, whereas LoRa can offer better link budget and longer communication range because of its lower data-rate operation and stronger spreading gain. These trade-offs show that the selected wireless module must be matched to the purpose of the UAV application and the required channel measurement parameters.

%%%%%%%%%%%%%%%%%%%%%%%%%%%%%%%%%%%%%%%%%%%%%%%%%%%%%%%%%%%%%%%%%
\chapter{CHANNEL STATISTICS}
%%%%%%%%%%%%%%%%%%%%%%%%%%%%%%%%%%%%%%%%%%%%%%%%%%%%%%%%%%%%%%%%%
\section{Communication Channel Statistics of UAVs}

\hspace{0.95cm} Modeling a UAV communication channel requires both large-scale and small-scale statistical analysis. Large-scale parameters describe the average behavior of the received signal over distance, while small-scale parameters describe the rapid variations caused by multipath propagation, delay dispersion, and mobility. In mobile radio communication, even small changes in the transmitter position, receiver position, or surrounding environment can change the phase and amplitude of the received multipath components. For this reason, path loss, CIR, PDP, coherence bandwidth, and Doppler spread are important parameters for designing and evaluating UAV measurement setups.

\subsection{Path Loss}

\hspace{0.95cm} Path loss ($PL$) is the attenuation experienced by a radio-frequency signal as it propagates from the transmitting antenna to the receiving antenna. It represents the reduction in signal power caused by geometric spreading, reflection, diffraction, scattering, absorption, antenna characteristics, and the distance between the transmitter and receiver. As the propagation distance increases, the received signal power generally decreases, and the path loss increases.

For UAV links, path loss is strongly affected by the probability of line-of-sight (LOS) propagation. In many aerial links, the LOS component is dominant; however, reflections from the ground, nearby buildings, or other objects may still contribute to the received signal. A commonly used LOS-based path-loss expression can be:

\begin{equation}
\Lambda(d)=\underbrace{20 \times \log \frac{4 \pi d_{0}}{\lambda}}_{P L_{0}}+10 \eta \log \left(d / d_{0}\right)+\mu_{L O S}
\end{equation}
where $d$ is the transmitter-receiver distance, $d_0$ is the reference distance, $\lambda$ is the wavelength, $\eta$ is the path-loss exponent, and $\mu_{LOS}$ represents the additional LOS-related loss term. In the measurement chapters of this thesis, the path-loss exponent is estimated from measured received-power values by fitting the measurement data to a log-distance model.

\subsection{Channel Impulse Response}

\hspace{0.95cm} The channel impulse response (CIR) is a fundamental tool for wideband channel characterization. It describes how a transmitted signal is transformed by the propagation channel in the delay domain. In a multipath environment, the transmitted signal reaches the receiver through several paths with different delays, amplitudes, and phases. Therefore, the CIR provides detailed information about the time of arrival and strength of each resolvable multipath component.

For UAV communication, the CIR is especially important because the channel may vary with time due to UAV movement, antenna orientation, and changing propagation conditions. The channel is therefore treated as a linear time-varying system. If $e(t)$ denotes the transmitted excitation signal and $s(t)$ denotes the received signal, the input-output relation can be expressed as:

\begin{equation}
s(t)=\int_{-\infty}^{\infty} e(t-\tau) h(\tau, t) \mathrm{d} \tau
\end{equation}

where $h(\tau,t)$ is the time-varying channel impulse response. This expression shows that the received signal is obtained from the convolution of the transmitted signal with the impulse response of the channel. A multipath channel can be represented as:

\begin{equation}
h(\tau, t)=\sum_{k=1}^{N(t)} \alpha_{k}(t) \delta\left(\tau-\tau_{k}(t)\right)
\end{equation}

where $\alpha_k(t)$ is the complex amplitude of the $k$th multipath component at time $t$, $\tau_k(t)$ is the corresponding delay, and $N(t)$ is the number of resolvable paths. In [12], the CIR of a C-band A2G channel was measured by separating the direct path from scattered and reflected components. The direct component was observed to dominate the response, with a signal level more than 20 dB stronger than many of the non-direct components.

In [13], a UAV channel measurement campaign estimated the CIR from multipath components using a high-resolution algorithm based on the space-alternating generalized expectation-maximization (SAGE) principle. The measurement results were then compared with a simulation-based graph model. In that approach, the channel impulse response was calculated as:

\begin{equation}
h(t, \tau)=\frac{\int r(t) s^{*}(t-\tau) \mathrm{d} t}{\int|s(t)|^{2} \mathrm{~d} t}
\end{equation}
where $r(t)$ is the received signal and $s(t)$ is the transmitted pseudo-noise sequence. The SAGE algorithm is used to estimate delays, Doppler frequencies, and complex amplitudes of multipath components. After applying this algorithm, the CIR can be:

\begin{equation}
h(t, \tau)=\sum_{\ell=1}^{L(t)} \alpha_{\ell}(t) \delta\left(\tau-\tau_{\ell}(t)\right) \exp \left\{j 2 \pi \int_{0}^{t} \nu_{\ell}(u)\,\mathrm{d}u\right\}+n(t, \tau)
\end{equation}
where $\tau_{\ell}(t)$ is the delay of the $\ell$th path, $\nu_{\ell}(t)$ is its Doppler frequency, $\alpha_{\ell}(t)$ is its complex amplitude, $L(t)$ is the number of paths, and $n(t,\tau)$ denotes additive white Gaussian noise.

\subsection{Power Delay Profile}

\hspace{0.95cm} The power delay profile (PDP) describes how the received signal power is distributed over propagation delay. It is directly related to the channel impulse response and is commonly obtained by taking the squared magnitude of the CIR. In a UAV channel, the first strong component is often associated with the LOS path, while later components may be caused by reflections from the ground, buildings, trees, or other obstacles. Figure~2.1 illustrates the general multipath propagation mechanism.
\begin{figure}[htbp]
\centering\includegraphics[width=0.92\textwidth,height=0.70\textheight,keepaspectratio]{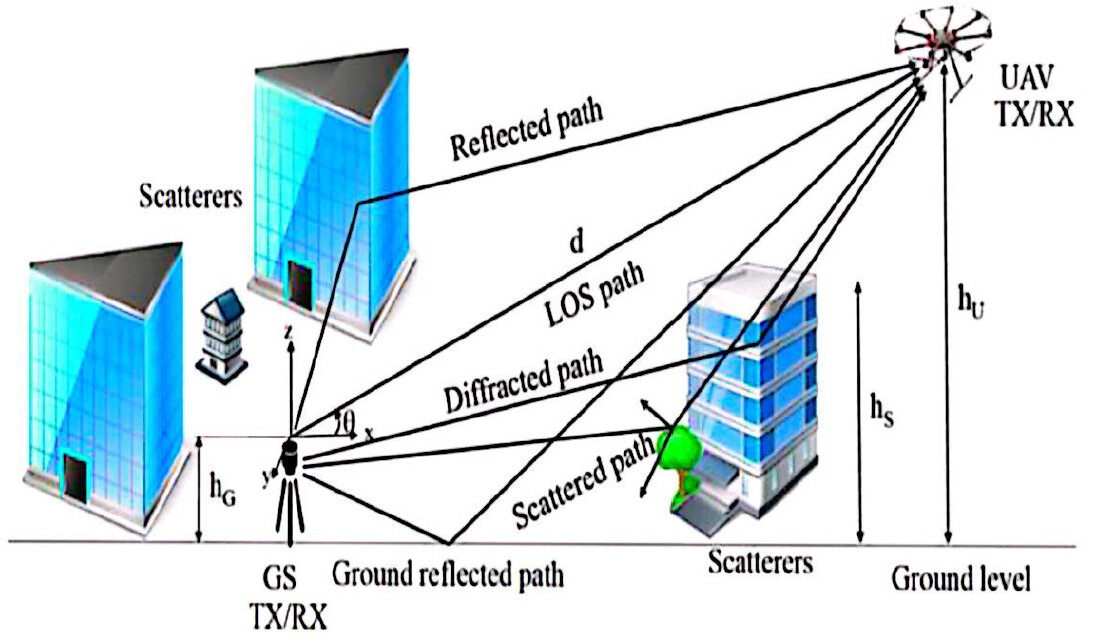}
\caption{Propagation fading mechanism in a multipath environment}
\end{figure}
In a multipath channel, the PDP gives the received signal power as a function of propagation delay. If the probing pulse used to sound the channel is much shorter than the impulse response of the channel, the PDP can be approximated by

\begin{equation}
\phi(\tau)=|h(\tau)|^{2}=\sum_{i=0}^{L-1}\left|a_{i}\right|^{2} \delta\left(\tau-\tau_{i}\right)
\end{equation}
where $a_i$ is the amplitude of the $i$th path and $\tau_i$ is its delay. The PDP makes it possible to identify the dominant arrival, weaker delayed components, and the delay range over which significant multipath energy is present. Several time-dispersion parameters are extracted from the PDP. The most common ones are mean excess delay, root-mean-square (RMS) delay spread, and maximum excess delay. The mean excess delay, denoted by $\bar{\tau}$, is the first moment of the PDP and can be calculated as:

\begin{equation}
\bar{\tau}=\frac{\sum_{k} p\left(\tau_{k}\right)\tau_{k}}{\sum_{k} p\left(\tau_{k}\right)}
\end{equation}
For a local area, the PDP can also be related to the band-limited impulse response by
\begin{equation}
p(\tau) \approx k\left|h_{b}(t ; \tau)\right|^{2}
\end{equation}
where $k$ is a proportionality constant and $h_b(t;\tau)$ is the band-limited channel response. The second important time-dispersion parameter is RMS delay spread. Different multipath components travel different path lengths and therefore arrive at the receiver at different times. As a result, a transmitted pulse spreads in time when it reaches the receiver. RMS delay spread quantifies this temporal spreading and is defined as the square root of the second central moment of the PDP:

\begin{equation}
\begin{aligned}
\sigma_{\tau} &=\sqrt{\overline{\tau^{2}}-(\bar{\tau})^{2}} 
\end{aligned}
\end{equation}

\begin{equation}
\begin{aligned}
\overline{\tau^{2}} &=\frac{\sum_{k} a_{k}^{2} \tau_{k}^{2}}{\sum_{k} a_{k}^{2}}=\frac{\sum_{k} p\left(\tau_{k}\right)\left(\tau_{k}^{2}\right)}{\sum_{k} p\left(\tau_{k}\right)}
\end{aligned}
\end{equation}

A larger RMS delay spread indicates stronger delay dispersion and a greater probability of intersymbol interference in high-data-rate communication systems. Therefore, the delay spread is an important parameter when evaluating whether a UAV communication link can support a given bandwidth and data rate. The maximum excess delay describes the time interval between the first significant arriving component and the last multipath component that remains above a selected threshold. If the threshold is chosen as $X$ dB below the strongest component, the maximum excess delay can be expressed as:
\begin{equation}
M \cdot E \cdot D \cdot(X d B)=\tau_{x}-\tau_{0}
\end{equation}
where $\tau_0$ is the delay of the first significant arriving component and $\tau_x$ is the maximum delay at which the multipath component is still within $X$ dB of the strongest arriving path. The strongest component is not always the first component, especially in environments where the LOS path is obstructed and a reflected path becomes dominant. Thus, maximum excess delay is useful for describing the temporal extent of multipath energy.

\

\

\

When the received signal is processed, the delay and power of each resolvable component can be estimated, and a PDP is obtained. Figure~2.2 illustrates the basic structure of a power delay profile. The time axis is divided into delay bins, and the resolution of these bins depends mainly on the sounding bandwidth. A wider bandwidth provides finer delay resolution, making it easier to separate closely spaced multipath components. However, increasing the bandwidth is not always possible because of hardware limitations, regulatory restrictions, and signal-to-noise ratio constraints.

\begin{figure}[htbp]
\centering\includegraphics[width=0.92\textwidth,height=0.70\textheight,keepaspectratio]{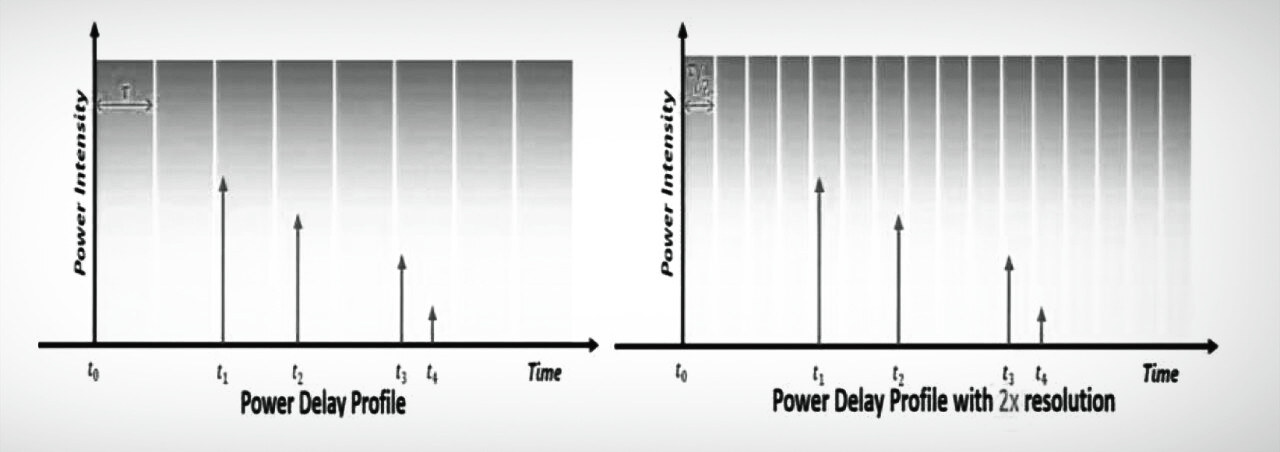}
\caption{Power delay profile representation}
\end{figure}

Figure~2.3 shows an example PDP obtained from measured data. The horizontal axis represents excess delay in nanoseconds, and the vertical axis represents normalized received power in dB. From this type of graph, the mean excess delay, RMS delay spread, and maximum excess delay can be obtained after selecting a suitable noise threshold. The maximum excess delay is determined by the time difference between the first and last components that exceed the threshold level [14].

\begin{figure}[htbp]
\centering\includegraphics[width=0.92\textwidth,keepaspectratio]{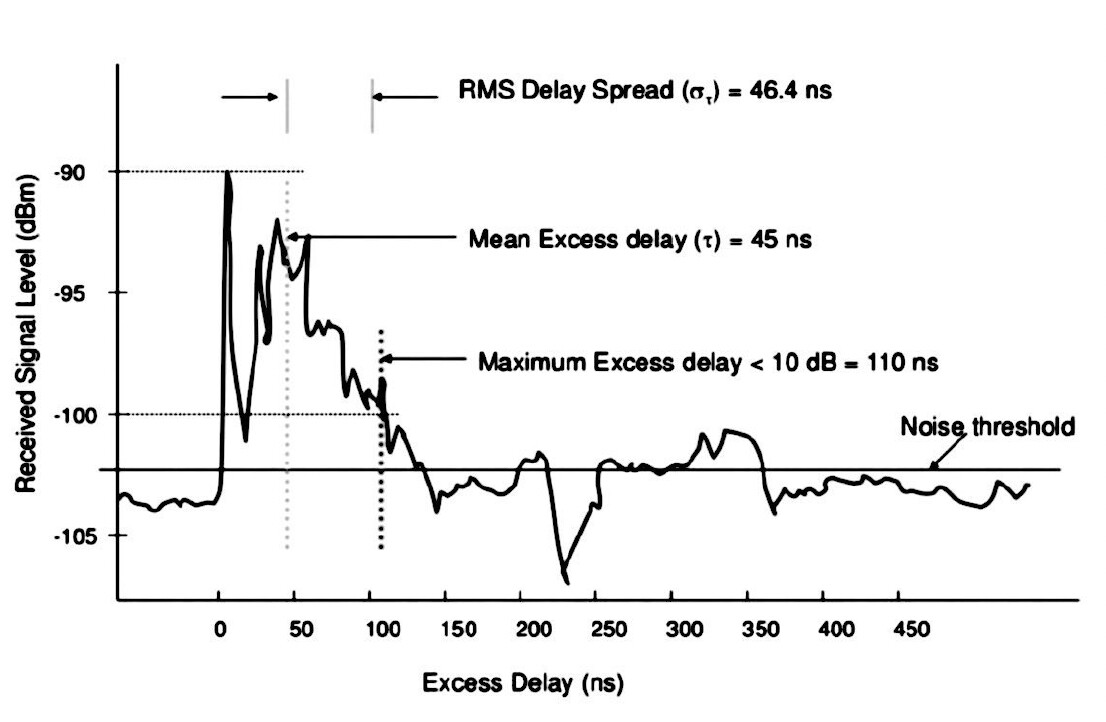}
\caption{Example measured power delay profile}
\end{figure}

\subsection{Coherence Bandwidth and Doppler Spread}

\hspace{0.95cm} Coherence bandwidth is a frequency-domain parameter related to RMS delay spread. It indicates the frequency range over which the channel response can be considered approximately flat. If the signal bandwidth is smaller than the coherence bandwidth, all frequency components of the signal experience similar fading. If the signal bandwidth is larger than the coherence bandwidth, the channel becomes frequency-selective and different parts of the signal spectrum experience different attenuation and phase shifts. Common approximate relation between coherence bandwidth and RMS delay spread is:

\begin{equation}
B_c \approx \frac{1}{5\sigma_{\tau}}
\end{equation}
where $B_c$ is the coherence bandwidth and $\sigma_{\tau}$ is the RMS delay spread. This relation shows that channels with larger delay spread have smaller coherence bandwidth.

Doppler spread is associated with relative motion between the transmitter, receiver, and scatterers. In UAV communication, motion can be significant because the UAV may change position and orientation during the measurement. The maximum Doppler shift can be approximated by

\begin{equation}
f_D=\frac{v}{\lambda}
\end{equation}
where $v$ is the relative velocity and $\lambda$ is the wavelength. Doppler spread determines how rapidly the channel changes with time. Therefore, together with PDP and path loss, it provides an important basis for evaluating the reliability and time selectivity of UAV communication channels.

%%%%%%%%%%%%%%%%%%%%%%%%%%%%%%%%%%%%%%%%%%%%%%%%%%%%%%%%%%%%%%%%%
\chapter{CHANNEL SOUNDING}
%%%%%%%%%%%%%%%%%%%%%%%%%%%%%%%%%%%%%%%%%%%%%%%%%%%%%%%%%%%%%%%%%
\section{Sounding Methods}

\hspace{0.95cm} Channel sounding is the process of transmitting a known signal through a communication channel and analyzing the received signal to estimate channel characteristics. The selected sounding waveform must have suitable correlation, bandwidth, power, and implementation properties. Figure~3.1 shows several sounding signals commonly used in wireless channel research: chirp signals, RF Gaussian pulses, pseudo-random number (PRN) sequences, and orthogonal frequency division multiplexing (OFDM) signals.

\begin{figure}[htbp]
\centering\includegraphics[width=0.92\textwidth,keepaspectratio]{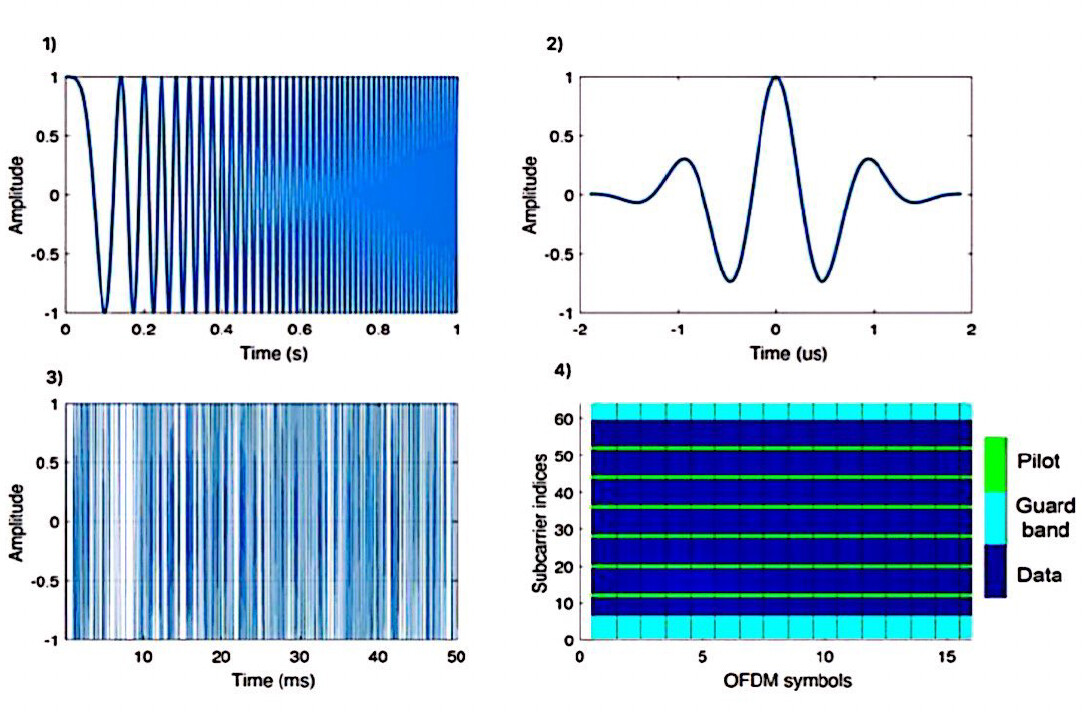}
\caption{Common signal types used for channel sounding}
\end{figure}

\subsection{Direct Sequence Spread Spectrum}

\hspace{0.95cm} Direct sequence spread spectrum (DSSS) is a widely used modulation and transmission method in wireless communications. In spread-spectrum systems, a narrowband information signal is intentionally spread over a much wider bandwidth. DSSS performs this spreading by combining the original data with a high-rate pseudo-noise (PN) sequence. At the transmitter, the data signal enters a spreading modulator, where the PN sequence is multiplied with the original data. The resulting signal occupies a broader bandwidth than the original information signal.

The elements of the PN sequence are often called chips. Their rate is much higher than the information bit rate, and this difference causes the transmitted signal to resemble a low-power noise-like waveform over a wide frequency band. At the receiver, the same PN sequence is used for despreading. If the transmitter and receiver codes are synchronized, the desired signal is compressed back into its original bandwidth, while many forms of interference remain spread and are therefore reduced.

\begin{figure}[htbp]
\centering\includegraphics[width=0.92\textwidth,height=0.70\textheight,keepaspectratio]{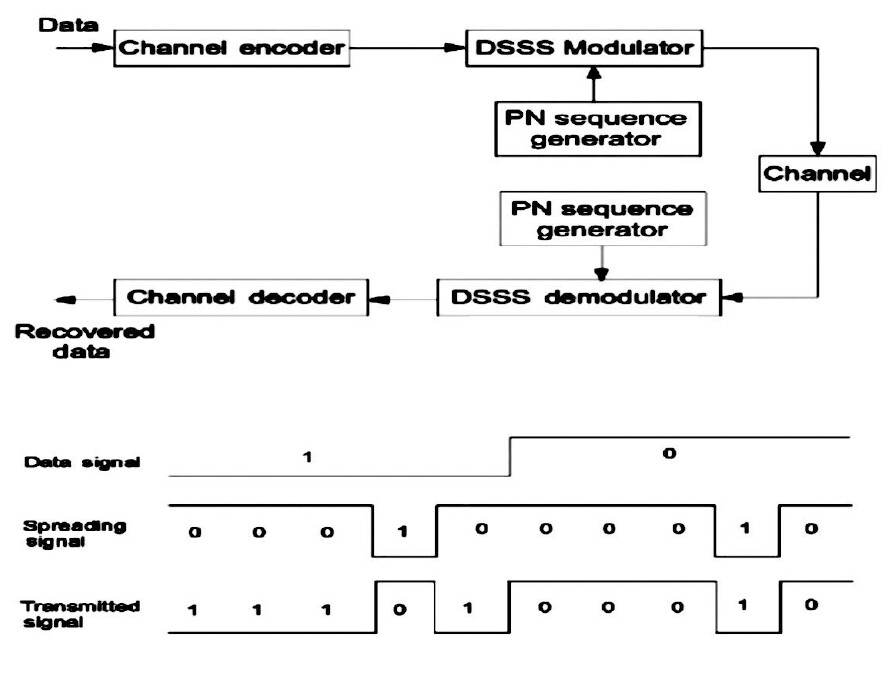}
\caption{Diagram of direct sequence spread spectrum}
\end{figure}

The process illustrated in Figure~3.2 provides two major benefits. First, each data bit is represented by several chips, which gives the receiver processing gain. Second, the signal energy is distributed over a wider frequency band, reducing the power spectral density. Narrowband communication signals are generally easier to jam or interfere with because their energy is concentrated in a small bandwidth. After spreading, the total signal power may remain the same, but the power is distributed across a larger spectrum. This makes DSSS signals more robust against narrowband interference and more difficult to detect.

DSSS has therefore been used in applications where interference resistance, low probability of detection, and secure communication are important. It is also useful in channel sounding because PN sequences can provide favorable autocorrelation properties. A sharp autocorrelation peak allows the receiver to estimate the channel delay response by correlating the received waveform with the known transmitted code.

One measurement campaign in [15] used a dual-band DSSS system. The system included one transmitter and two receivers in each band and operated in both the L-band (0.9 GHz--1.2 GHz) and the C-band (5.30 GHz--5.91 GHz). The center frequencies were 968 MHz in the L-band and 5060 MHz in the C-band, which yielded delay resolutions of 200 ns and 20 ns, respectively. A PN sequence with a length of 1023 produced a maximum delay span of 204.6 microseconds in the L-band and 20.46 microseconds in the C-band. The receiver outputs were interpreted as power delay profiles (PDPs) [16].

\subsection{Continuous Wave}

\hspace{0.95cm} A continuous wave (CW) signal is an electromagnetic wave with constant amplitude and frequency, usually represented by a sinusoid. CW transmission was one of the earliest radio transmission methods. In early radio telegraphy, the carrier was switched on and off to transmit information, for example by using Morse code. Such signals were historically called undamped waves to distinguish them from damped waves produced by spark-gap transmitters.

In the context of channel measurements, CW signals can be useful for measuring received power, path loss, and Doppler behavior at a single frequency. However, a pure CW signal does not provide delay resolution by itself because it does not occupy a wide bandwidth. Therefore, CW methods are more suitable for narrowband characterization, whereas wideband sounding signals such as PN sequences, chirps, or OFDM are preferred when CIR and PDP information is required.

\subsection{Orthogonal Frequency Division Multiplexing}

\hspace{0.95cm} Orthogonal frequency division multiplexing (OFDM) is a digital transmission technique in which information is carried by many closely spaced orthogonal subcarriers. Each subcarrier has a narrow bandwidth, and the subcarriers overlap in the frequency domain while remaining mathematically orthogonal. As shown in Figure~3.3, OFDM divides a wideband signal into several parallel narrowband channels. Because data symbols are transmitted in parallel over many subcarriers, OFDM can reduce the effect of frequency-selective fading and can efficiently use the available bandwidth. A cyclic prefix can also be inserted to reduce intersymbol interference caused by multipath delay spread. These advantages make OFDM attractive for many modern wireless systems.

\begin{figure}[htbp]
\centering\includegraphics[width=0.75\textwidth,height=0.65\textheight,keepaspectratio]{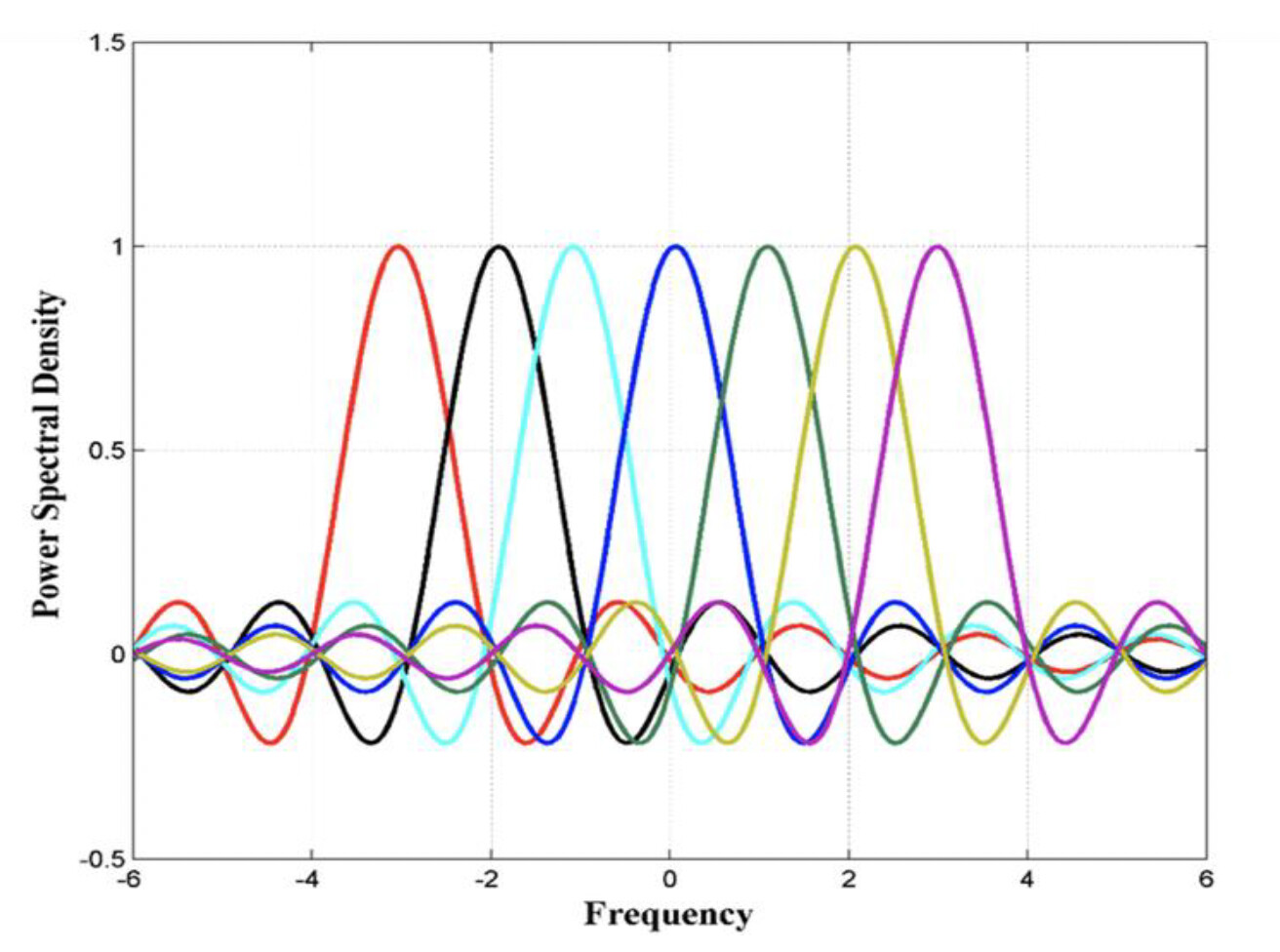}
\caption{OFDM signal structure}
\end{figure}

Despite these advantages, OFDM also has limitations for UAV channel sounding. It is sensitive to Doppler shift and carrier-frequency offset, which may be significant in fast-moving aerial platforms. It also requires accurate synchronization between the transmitter and receiver. In addition, OFDM signals can have a high peak-to-average power ratio (PAPR), which requires a linear RF power amplifier and may reduce power efficiency. For fast-moving UAV scenarios, these drawbacks must be carefully considered before selecting OFDM as the sounding waveform.

In [18], a 4-by-4 MIMO-OFDM airborne system was compared with a SISO system in an A2G environment in terms of data throughput and channel measurements. The measurement results showed that significant MIMO gains in throughput and range can be achieved compared with SISO systems. OFDM was selected partly because short-term waveform deviations between adjacent symbols could be monitored under high-SNR conditions. Flight trials were also conducted in [19] to investigate the PDP, Doppler profile, and time-of-arrival estimation for the L-band A2G radio channel.

A multitone signal can also be used for channel sounding. It has similarities to OFDM, but it may use constant subcarrier amplitudes, pseudorandom phases, and no cyclic prefix. In [20], UAV A2G performance measurements were conducted using different antenna orientations. The results showed that DSSS signals based on pseudo-random sequences can be used to estimate the channel impulse response. Chirp sounding was also found useful because it provides high frequency resolution and can sweep across a large frequency range.

OFDM has attracted attention as a channel sounding method because of its relatively flat spectrum. In A2G measurement campaigns, carrier frequencies have ranged from hundreds of megahertz to several gigahertz, while sounding bandwidths have varied from narrowband values of tens of megahertz to ultra-wideband values of several gigahertz. Wider bandwidth produces finer time resolution, which improves the ability to separate multipath components. In another measurement campaign, the path loss of IEEE 802.11a-based UAV links was analyzed to observe the effect of antenna orientation on received signal strength.

IEEE 802.11a is often preferred over 802.11b/g or IEEE 802.15.4 in some UAV experiments because it can offer lower interference probability and higher achievable data rates at 5 GHz. In one access-point implementation, a Netgear WNDR3700 version 2 device with an Atheros AR7161 680 MHz CPU and 64 MB RAM was used. It included two Atheros AR9280-based wireless cards, allowing simultaneous use of the 2.4 GHz and 5 GHz bands [21].

In a further IEEE 802.11a study, three-dimensional mobility was investigated for A2G communication networks. Signal-strength samples obtained at 5 GHz were used to model path loss and small-scale fading. The measurement results indicated that data rates up to 12 Mbps were possible at distances on the order of 300 m. The quadrotor communicated through an IEEE 802.11a wireless LAN with an access point on the ground. The Linux wireless subsystem was used to record received signal strength, transmission rate, and the number of retransmissions [22].

\subsection{Chirp Spread Spectrum}

\hspace{0.95cm} A chirp signal is a waveform whose frequency increases or decreases over time. Chirp signals are commonly used in radar, sonar, laser systems, and chirp spread spectrum (CSS) communication systems. In [23], the effects of multipath on a communication channel were shown to depend on signal bandwidth. Signals whose bandwidth is larger than the channel coherence bandwidth experience frequency-selective fading, whereas narrowband signals may suffer from Rayleigh fading with deep fades.

The UMIST chirp sounder was used in [23] to obtain measurements with bandwidths of 10 MHz, 36 MHz, and 72 MHz. A chirp sounder transmits a linearly frequency-swept signal over a duration $T$ and bandwidth $B$. After compression, the chirp response has a sinc-like form, and the processing gain is commonly related to $10\log_{10}(BT)$. Chirp compression can be performed by a matched filter or by a heterodyne detector. In the matched-filter method, the filter response is matched to the transmitted waveform, which compresses the received chirp in the time domain.

\

\

The output of a chirp sounder consists of pulses corresponding to distinct multipath echoes. IQ demodulation can be used to obtain both the PDP and Doppler information. In the heterodyne technique, the received signal is multiplied by a locally generated copy of the transmitted signal in the frequency domain [23]. The average delay and delay spread can be calculated as

\begin{equation}
\begin{aligned}
A D &=\frac{\sum_{i=1}^{N} \tau_{i} P_{h}\left(\tau_{i}\right)}{\sum_{i=1}^{N} P_{h}\left(\tau_{i}\right)}-\tau_{A} 
\end{aligned}
\end{equation}

\begin{equation}
\begin{aligned}
D S &=\sqrt{\frac{\sum_{i=1}^{N}\left(\tau_{i}-A D\right)^{2} P_{h}\left(\tau_{i}\right)}{\sum_{i=1}^{N} P_{h}\left(\tau_{i}\right)}}
\end{aligned}
\end{equation}

Industrial, scientific, and medical (ISM) bands can be used for radio communication without an individual license in many countries, provided that the system satisfies the relevant power and spectrum regulations. In many ISM-band applications, the allowed transmit power is limited. When long communication distance is more important than high data rate, low-power wide-area network (LPWAN) technologies can be considered. LPWAN systems generally provide lower data rates than broadband wireless systems, but they can support longer communication ranges [24].

CSS uses wideband linearly frequency-modulated chirp pulses to encode information, commonly in sub-GHz bands such as 868 MHz. The spreading factor is selected according to the desired trade-off among data rate, bandwidth, and signal-to-noise ratio (SNR). A higher spreading factor increases robustness against noise but decreases the bit rate. CSS symbols can be described by parameters such as spreading factor $S$, input bits, minimum frequency $f_{min}$, maximum frequency $f_{max}$, and starting frequency $f_0$. If $t_s$ is the symbol duration, then:
\begin{equation}
\begin{array}{l}
\mathrm{t}_{\mathrm{s}}=\mathrm{S} / \mathrm{B}=\mathrm{S} /\left(\mathrm{f}_{\max }-\mathrm{f}_{\min }\right)
\end{array}
\end{equation}
According to this expression, the bit rate $R_b(S)$ is:
\begin{equation}
\begin{aligned}
\mathrm{R}_{\mathrm{b}}(\mathrm{S})=\log _{2} \mathrm{~S} / \mathrm{t}_{\mathrm{s}}
\end{aligned}
\end{equation}
Since CSS symbols are approximately orthogonal, a correlation-based decoder can be used to recover the transmitted symbols. Each received symbol is compared with candidate CSS symbols, and the decoder selects the symbol with the maximum correlation [24].

\section{Channel Sounding Applications}

\hspace{0.95cm} The UAV channel measurements in this thesis require the estimation of CIR, PDP, and path loss. Therefore, the sounding signal must have strong autocorrelation properties so that the receiver can clearly identify the dominant peak and delayed multipath components. A waveform with a sharp autocorrelation peak improves delay estimation, while low sidelobes reduce ambiguity between the LOS component and weaker reflections.

Several sounding techniques were reviewed in the previous section. Based on this comparison, two candidate sequences were selected for experimental evaluation: the maximum length sequence (M-sequence) and the Zadoff--Chu sequence. Both sequences are known for useful correlation properties. In the experiments, ADALM-Pluto SDR units were used as transmitter and receiver. The sequences were generated in Python, loaded into the SDRs, transmitted through the channel, and then correlated at the receiver to evaluate their suitability for measurement campaigns.

\subsection{Maximum Length Sequence}

\hspace{0.95cm} A maximum length sequence, or M-sequence, is a type of pseudo-noise code. A pseudo-noise sequence is a deterministic binary sequence that has noise-like properties and favorable autocorrelation behavior. Other well-known pseudo-noise codes include Gold codes, Kasami codes, and Barker codes. M-sequences are generated by linear feedback shift registers (LFSRs), and when the feedback taps are selected correctly, the sequence achieves the maximum possible period for the register length.

M-sequences have three important properties: balance, run, and correlation. The balance property means that the numbers of ones and zeros are nearly equal during one period. The run property describes the distribution of consecutive ones and zeros. The correlation property means that the autocorrelation function has one dominant peak and low values elsewhere. This makes M-sequences suitable for channel sounding, because the receiver can locate the main arrival time and identify delayed multipath components.

Figures~3.4 and 3.5 show the frequency-domain and time-domain forms of the received M-sequence signal, respectively. Figure~3.6 shows the autocorrelation result. The length of the M-sequence used in this demonstration is 127. The single dominant autocorrelation peak indicates that the transmitter and receiver signals are well aligned and that the sequence is suitable for estimating the delay response of the channel.

\begin{figure}[htbp]
\centering\includegraphics[width=0.92\textwidth,height=0.70\textheight,keepaspectratio]{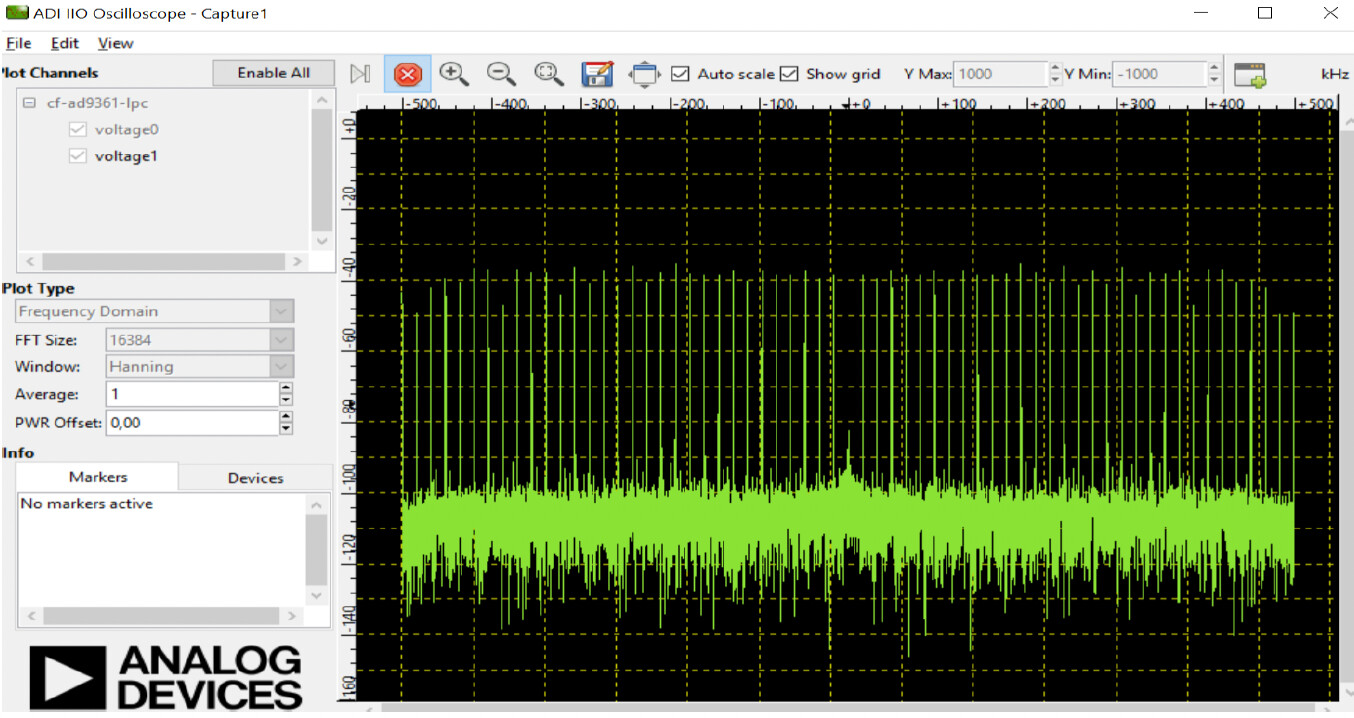}
\caption{Frequency-domain representation of the M-sequence signal}
\end{figure}

\begin{figure}[htbp]
\centering\includegraphics[width=0.92\textwidth,height=0.70\textheight,keepaspectratio]{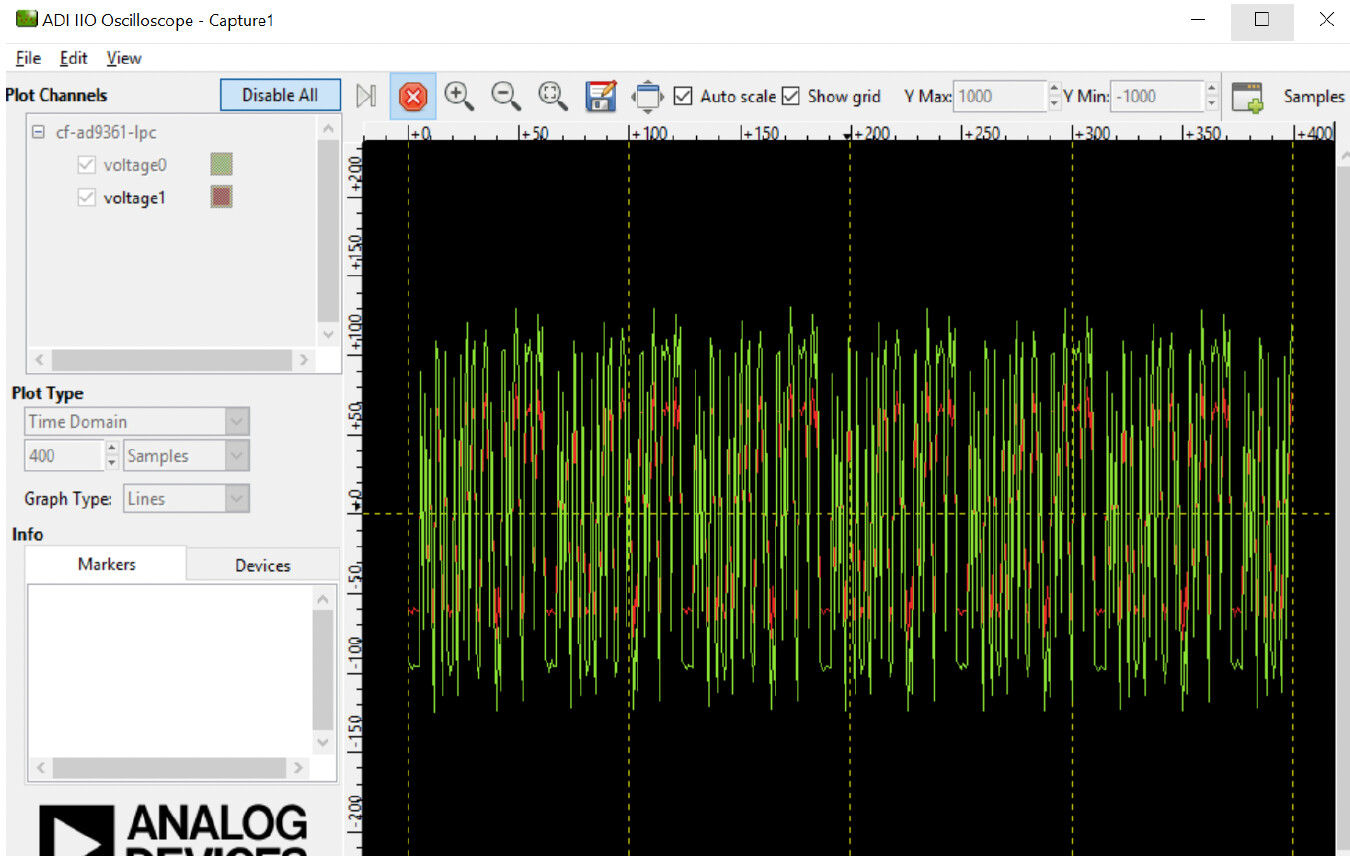}
\caption{Time-domain representation of the M-sequence signal}
\end{figure}

\begin{figure}[htbp]
\centering\includegraphics[width=0.92\textwidth,height=0.70\textheight,keepaspectratio]{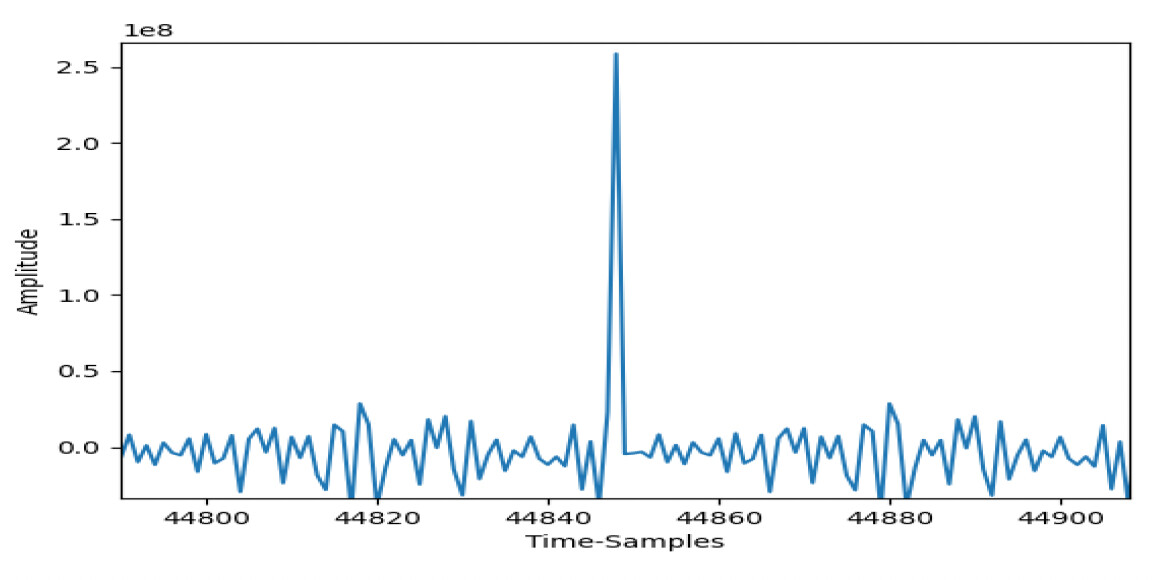}
\caption{Autocorrelation of the M-sequence signal}
\end{figure}

\subsection{Zadoff--Chu Sequence}

\hspace{0.95cm} A Zadoff--Chu sequence is a complex-valued mathematical sequence commonly used in synchronization and preamble generation. Its main advantage is that it has a constant amplitude in both the time and frequency domains under ideal conditions. Cyclically shifted versions of a Zadoff--Chu root sequence can also have very low cross-correlation with one another, which makes them attractive for multiuser and synchronization applications. 

\begin{figure}[htbp]
\centering\includegraphics[width=0.92\textwidth,keepaspectratio]{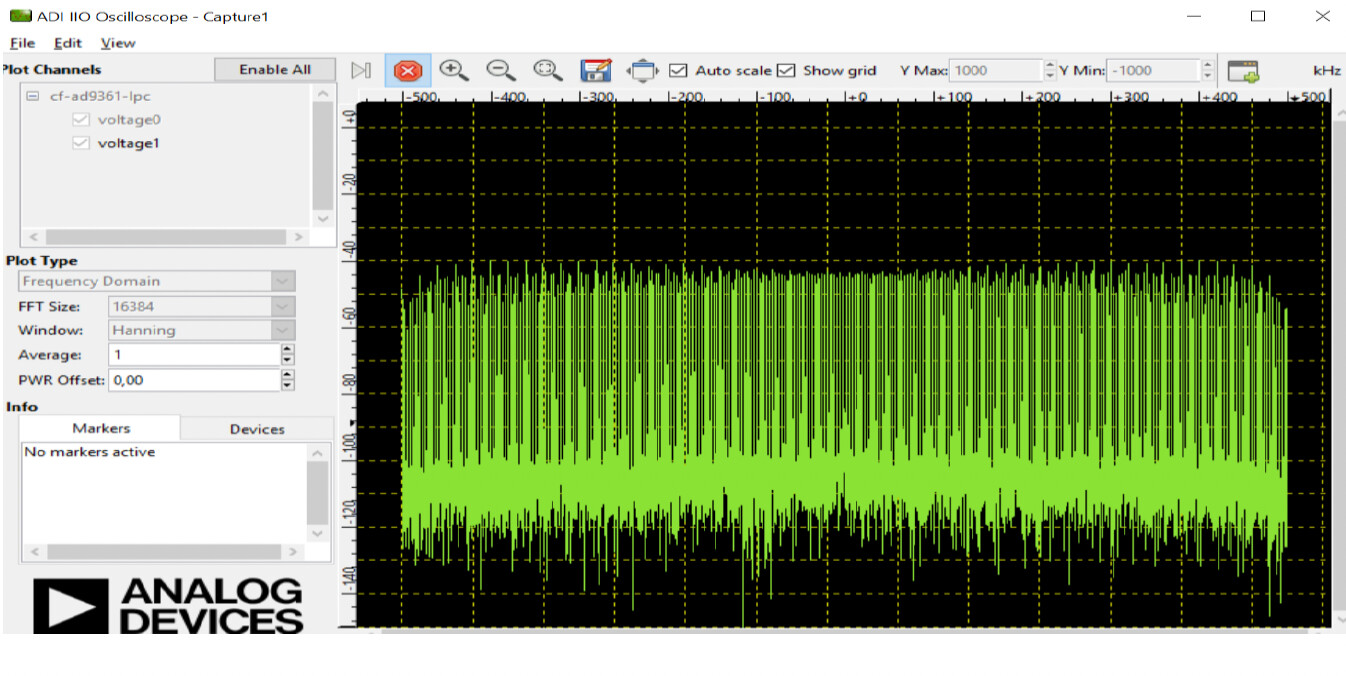}
\caption{Frequency-domain representation of the Zadoff--Chu sequence signal}
\end{figure}

\begin{figure}[t!]
\centering\includegraphics[width=0.92\textwidth,keepaspectratio]{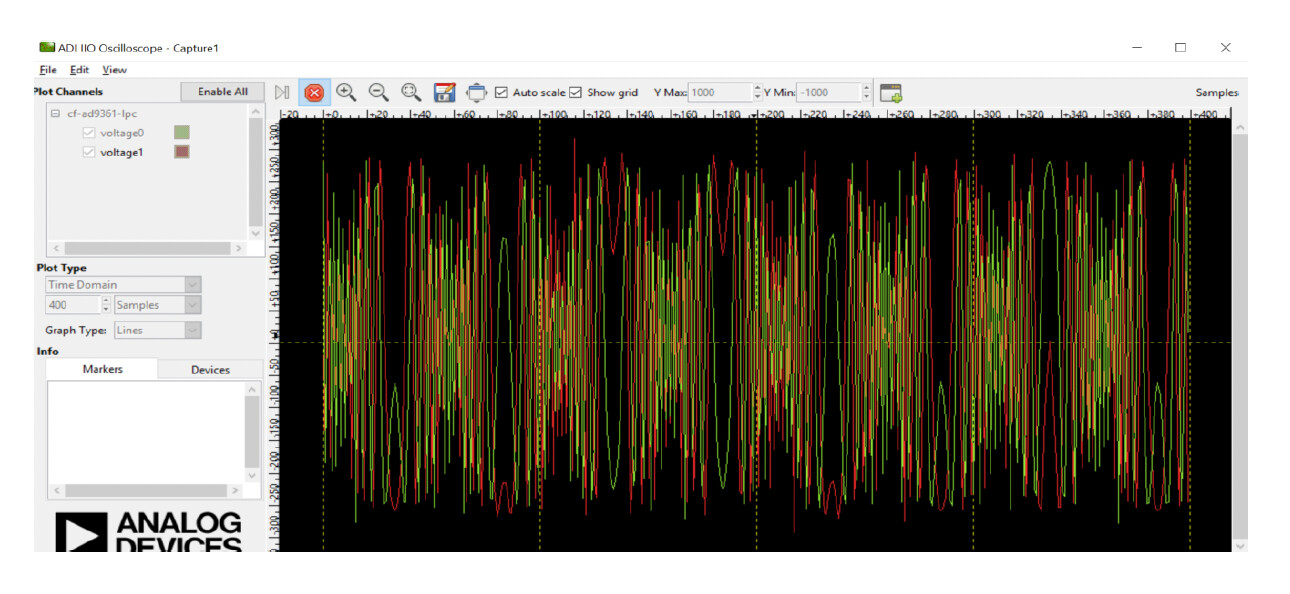}
\caption{Time-domain representation of the Zadoff--Chu sequence signal}
\end{figure}

\

\

The complex value at the $n$th position of a root Zadoff--Chu sequence parameterized by $u$ can be expressed as:
\begin{equation}
x_{u}(n)=\exp \left(-j \frac{\pi u n\left(n+c_{f}+2 q\right)}{N_{Z C}}\right)
\end{equation}
where $N_{ZC}$ is the sequence length, $u$ is the root index, $q$ is an integer shift parameter, and $c_f$ depends on the selected sequence convention.

\begin{figure}[h!]
\centering\includegraphics[width=0.8\textwidth,keepaspectratio]{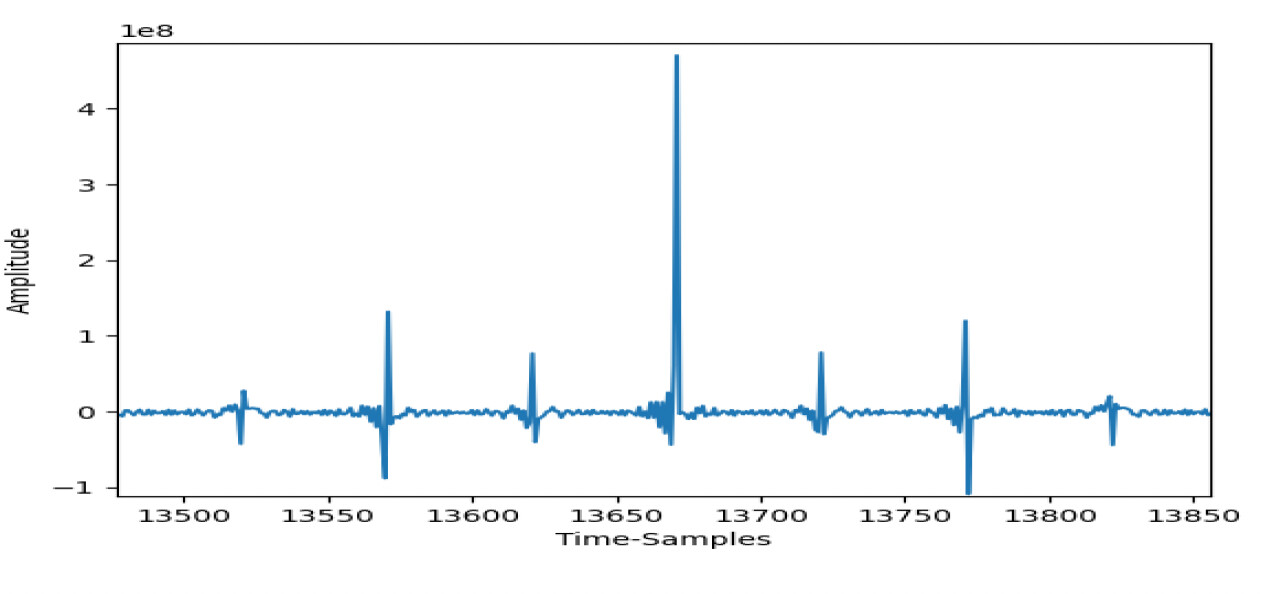}
\caption{Autocorrelation of the Zadoff--Chu sequence signal}
\end{figure}

Zadoff--Chu sequences are used in LTE systems for uplink control channels, uplink traffic channels, and sounding reference signals. Their low cross-correlation helps reduce inter-cell interference and improves synchronization performance. However, frequency offset can degrade the correlation result and may create additional peaks, especially at high Doppler frequencies. For this reason, the number of preamble sequences generated from a root sequence may need to be limited in practical systems.

Figures~3.7 and 3.8 show the frequency-domain and time-domain forms of the received Zadoff--Chu signal, respectively. Figure~3.9 shows the autocorrelation result for a Zadoff--Chu sequence with a length of 353. The autocorrelation includes one dominant peak, but smaller additional peaks are also visible. These additional peaks may create ambiguity when the channel includes weak multipath components.

Both the M-sequence and the Zadoff--Chu sequence are strong candidates for UAV channel sounding. However, for the measurement campaign in this thesis, the M-sequence was selected because it produced a clearer autocorrelation response under the tested conditions and was less sensitive to the frequency-offset effects observed in the Zadoff--Chu experiment. The final measurements were therefore conducted at 2.57 GHz using the M-sequence as the channel sounding signal.

%%%%%%%%%%%%%%%%%%%%%%%%%%%%%%%%%%%%%%%%%%%%%%%%%%%%%%%%%%%%%%%%%
\chapter{Ground to Ground UAV Wireless Communication Channel Modeling}\label{Ch4}
%%%%%%%%%%%%%%%%%%%%%%%%%%%%%%%%%%%%%%%%%%%%%%%%%%%%%%%%%%%%%%%%%

\hspace{0.95cm} This chapter presents the first outdoor measurement campaign of the thesis. After the sounding-signal comparison described in Chapter~3, the maximum length sequence was selected for channel sounding because it provided a clear autocorrelation peak and suitable delay-domain behavior. The ground-to-ground (G2G) measurements were conducted before the UAV-based campaigns in order to validate the measurement setup under controlled outdoor conditions. This stage was essential because it allowed the transmitter, receiver, antennas, amplifier, and processing software to be tested before the system was mounted on a UAV.

The G2G campaign had two main purposes. The first purpose was to verify that the setup could detect a known reflected path by using a physical reflector. The second purpose was to collect received data at several transmitter locations and then estimate the PDP and path-loss behavior of the outdoor channel. The following sections describe the measurement setup, the measurement campaign, and the obtained results.

\section{Measurement Setup}

\hspace{0.95cm} In order to model the UAV communication channel, a practical transmitter-receiver structure was prepared. The main system components were the software-defined radio units, antennas, power amplifier, computer, flash disk, and power bank. These components were selected according to the operating frequency, sampling-rate requirement, portability, and ability to collect IQ data for post-processing.

\subsection{Software-Defined Radio}

\hspace{0.95cm} Software-defined radio (SDR) was used to generate, transmit, receive, and record the sounding signal. At the beginning of the study, ADALM-Pluto was selected as the SDR for both the transmitter and the receiver. However, the first measurement results showed that the available resolution was not sufficient to observe the multipath effect clearly in the channel.

\

\

\begin{figure}[htbp]
\centering\includegraphics[width=0.82\textwidth,height=0.60\textheight,keepaspectratio]{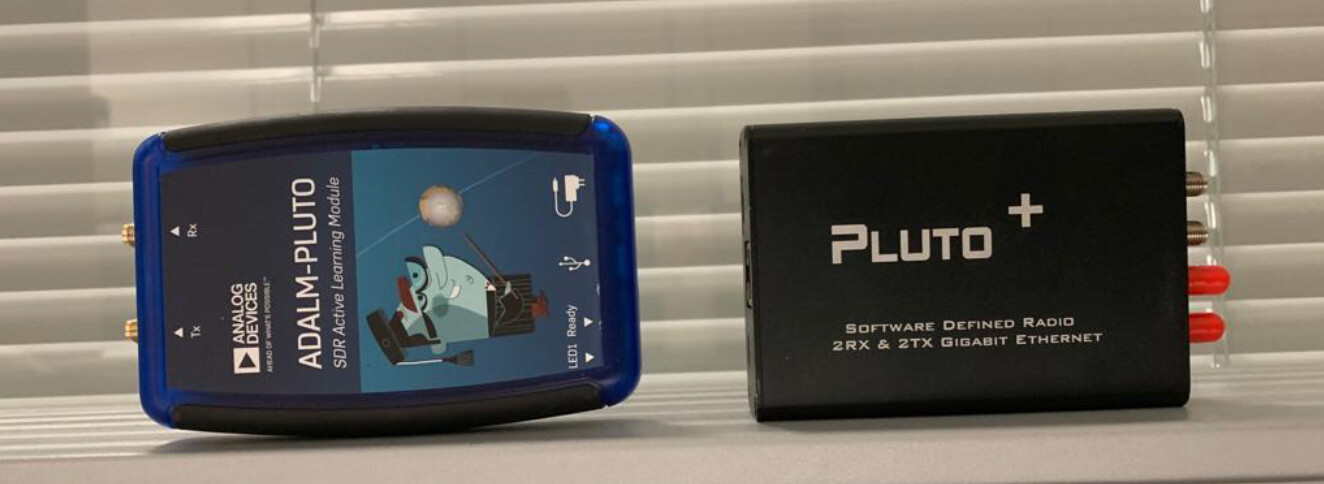}
\caption{ADALM-Pluto and Pluto Plus SDRs}
\end{figure}

For this reason, Pluto Plus was adopted as the main SDR platform. The sampling rate was increased by modifying the Pluto Plus firmware settings through the PuTTY control interface. The ADC configuration in Pluto Plus was changed from AD9364 to AD9361, allowing the SDR to operate at higher sample rates and frequencies. After this modification, the obtained distance resolution was approximately 4.8 m, which was sufficient to observe resolvable multipath components in the measurement environment.

\subsection{Antenna and Power Amplifier}

\hspace{0.95cm} Antenna selection was important because both the transmitter and receiver had to operate around 2.57 GHz. The ANT-LTE-WS-SMA antenna manufactured by Linx Technologies was selected because it can operate between 2.5 GHz and 2.7 GHz. Although this antenna was suitable for the operating frequency, the received signal level was not always high enough to overcome the channel noise at larger distances. Therefore, a ZX60-V63+ power amplifier manufactured by Mini-Circuits was added to the setup. This component can also be used as a low-noise amplifier in suitable configurations. The power amplifier significantly improved the received signal level and reduced the negative effect of noise on the channel measurements. As shown in Figure~4.2, the amplifier was connected between the antenna and the Pluto Plus SDR on the transmitter side and provided approximately 16 dB gain.

\begin{figure}[htbp]
\centering\includegraphics[width=0.92\textwidth,height=0.70\textheight,keepaspectratio]{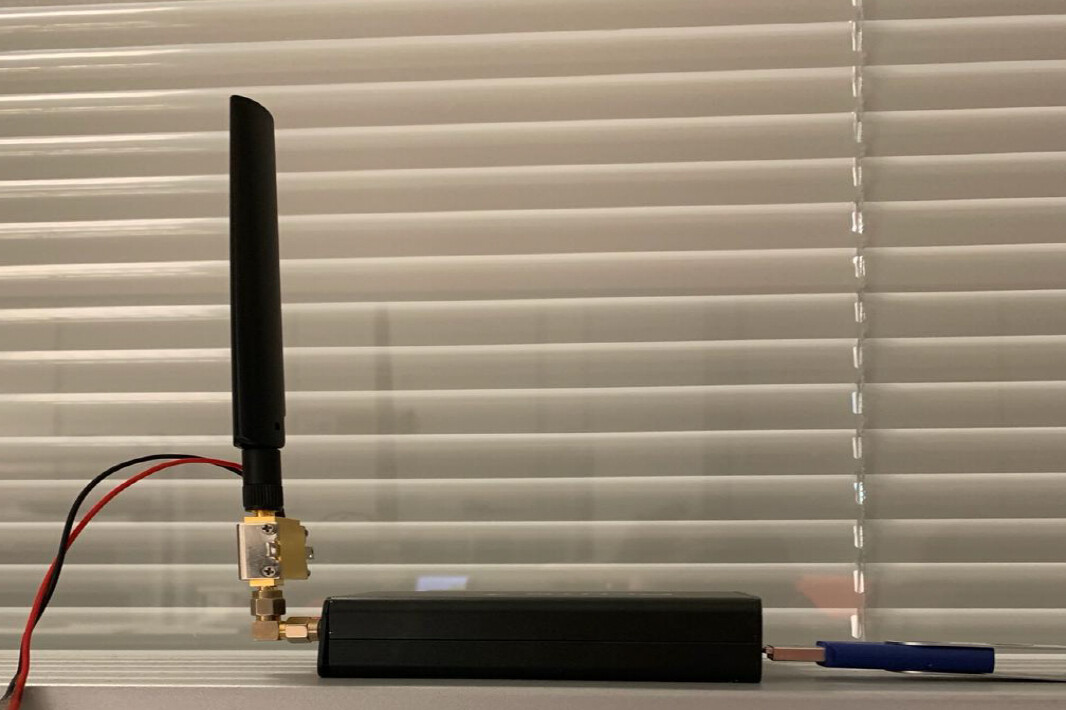}
\caption{ANT-LTE-WS antenna and ZX60-V63+ power amplifier}
\end{figure}

\begin{figure}[htbp]
\centering\includegraphics[width=0.92\textwidth,height=0.70\textheight,keepaspectratio]{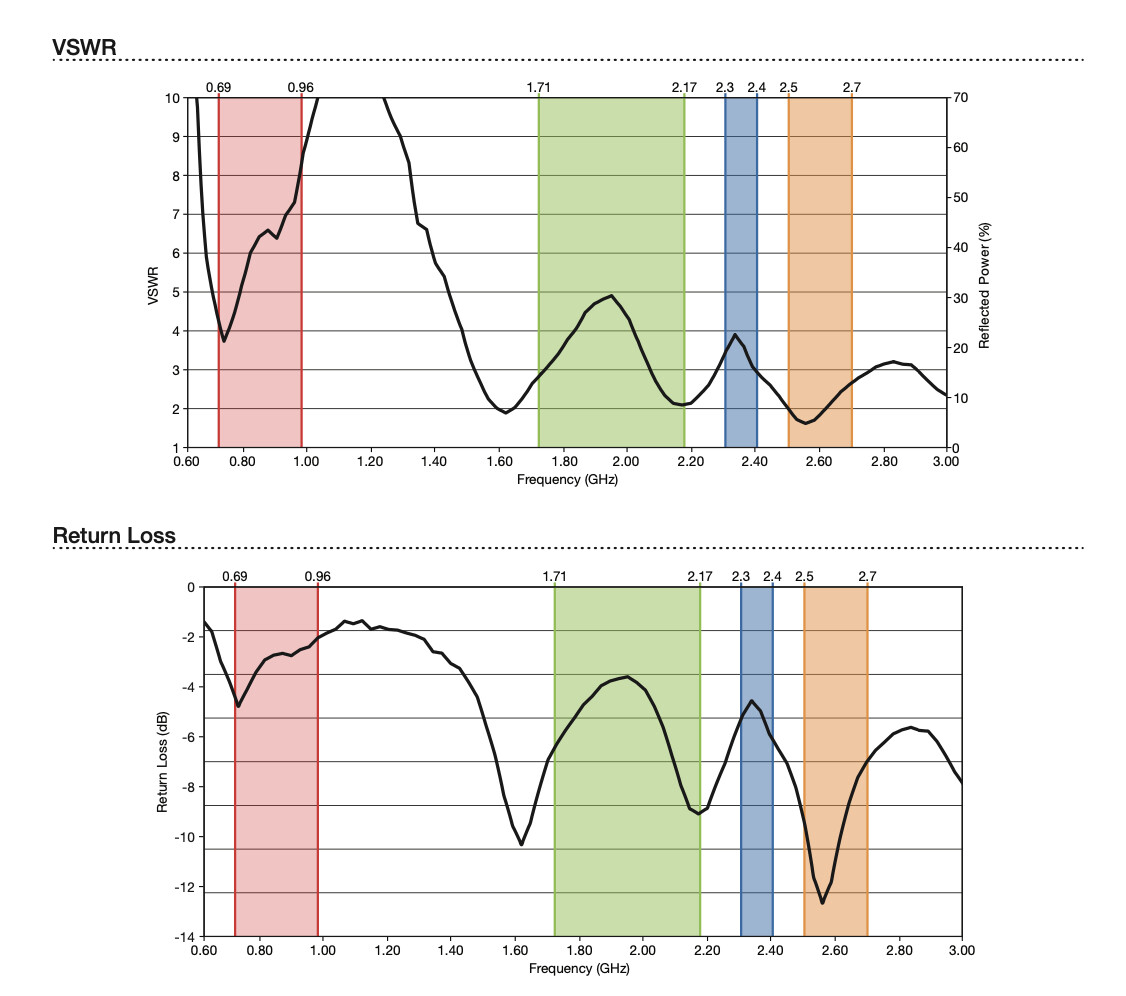}
\caption{VSWR and return loss of the antenna}
\end{figure}

\begin{figure}[htbp]
\centering\includegraphics[width=0.92\textwidth,height=0.70\textheight,keepaspectratio]{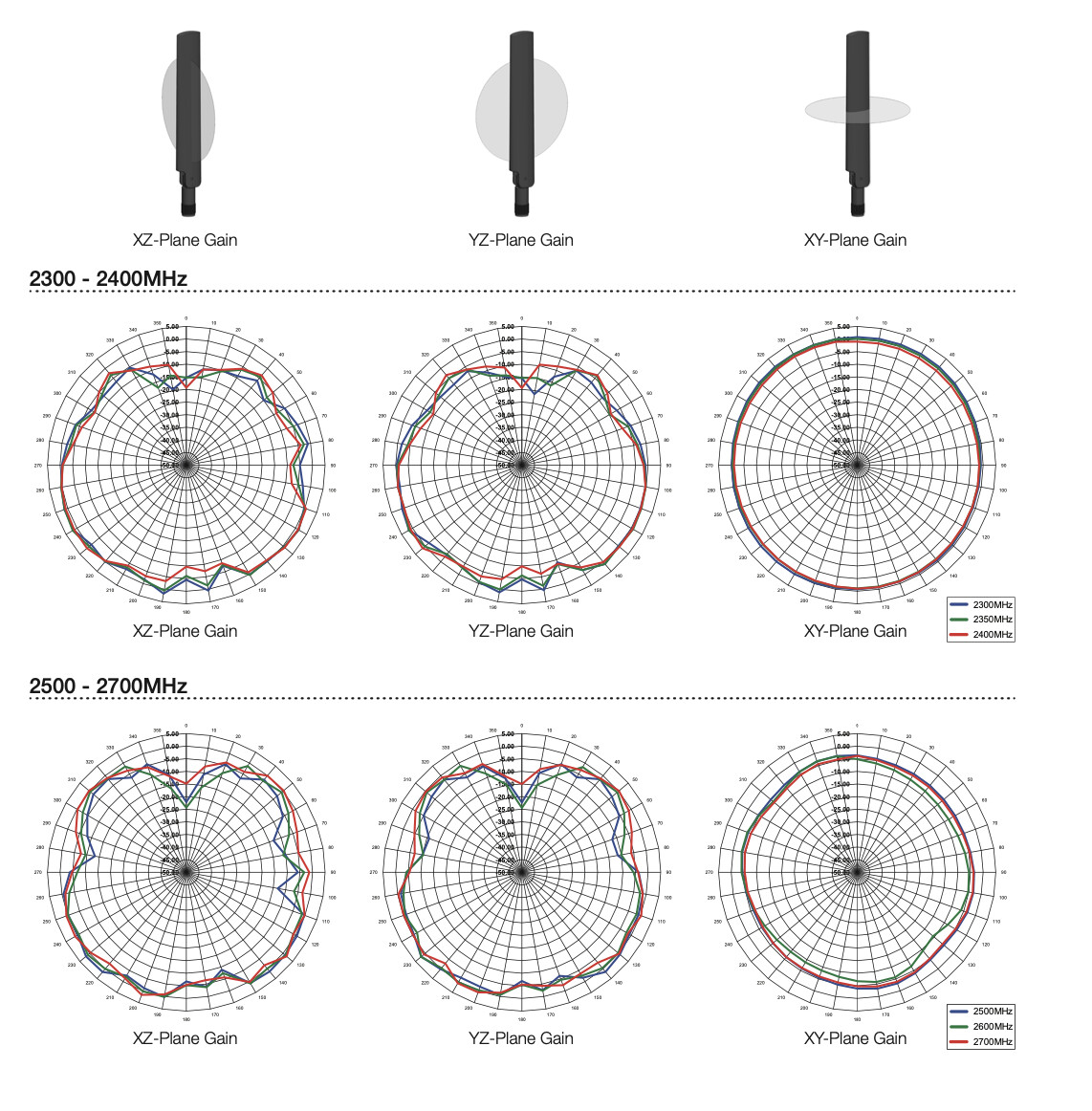}
\caption{Gain plots of the antenna}
\end{figure}

\subsection{Computer, Flash Disk, and Power Bank}

\hspace{0.95cm} A computer was used on the receiver side to record the received IQ data. On the transmitter side, a flash disk was used to store and transmit the prepared sounding sequence. A power bank supplied the power amplifier with 5 V. This simple structure made the G2G setup portable and allowed the system to be moved easily between different measurement points.

\section{Measurement Campaign}

\hspace{0.95cm} The G2G measurement campaign was performed at the TUBITAK Gebze campus near the library area. The measurement environment is shown in Figure~4.5. The latitude and longitude values of the transmitter and receiver locations were obtained using Google Earth. During the campaign, the transmitter and receiver antennas were placed at a height of 1.5 m above the ground.

\begin{figure}[htbp]
\centering\includegraphics[width=0.92\textwidth,height=0.70\textheight,keepaspectratio]{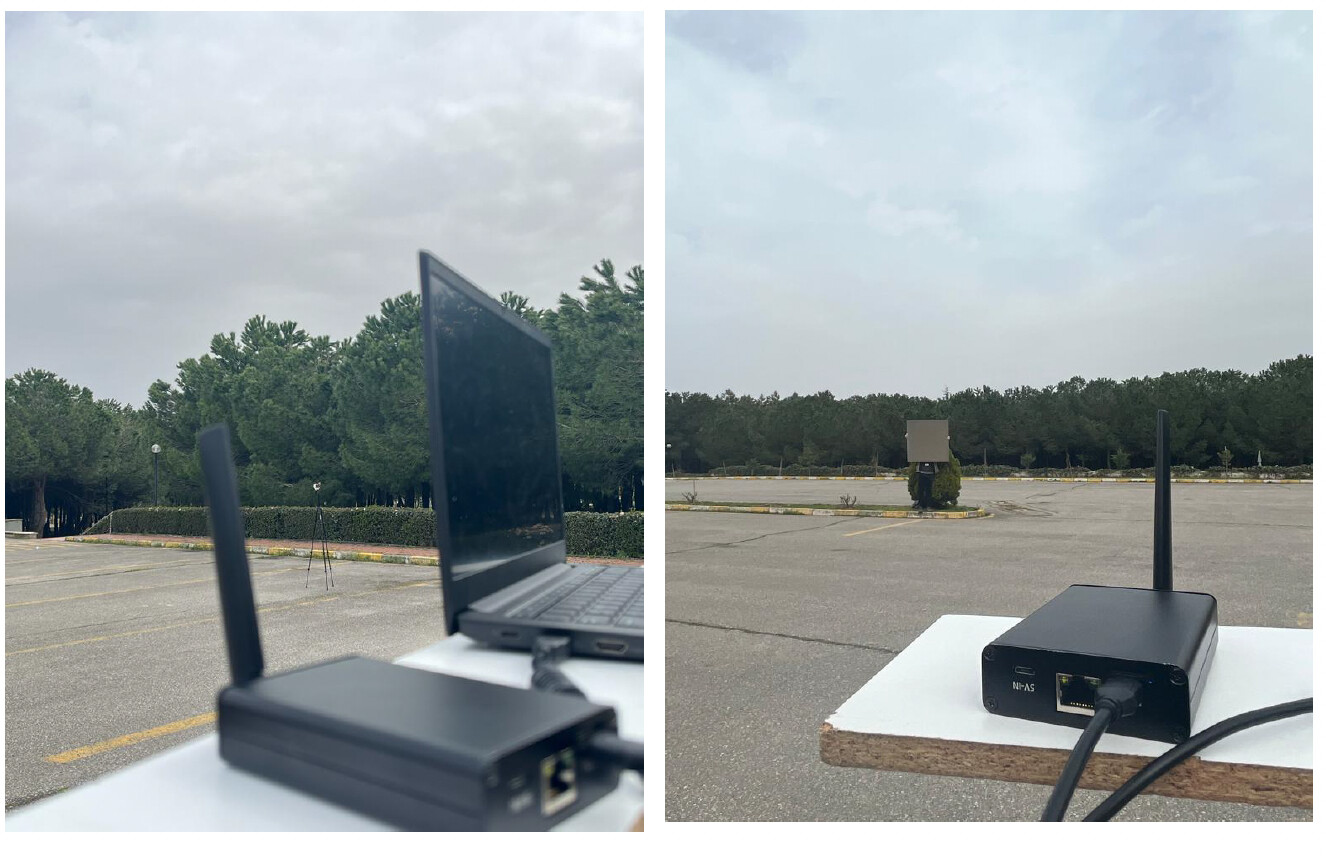}
\caption{G2G measurement environment}
\end{figure}

Before collecting data at multiple locations, a reflector was placed in the measurement area to verify that the setup could detect an intentionally created multipath component. The reflector position is shown in Figure~4.6. When the M-sequence was transmitted, the reflected path appeared as an additional peak in the correlation response, confirming that the system could resolve the delay difference between the direct and reflected components.

\begin{figure}[htbp]
\centering\includegraphics[width=0.92\textwidth,height=0.70\textheight,keepaspectratio]{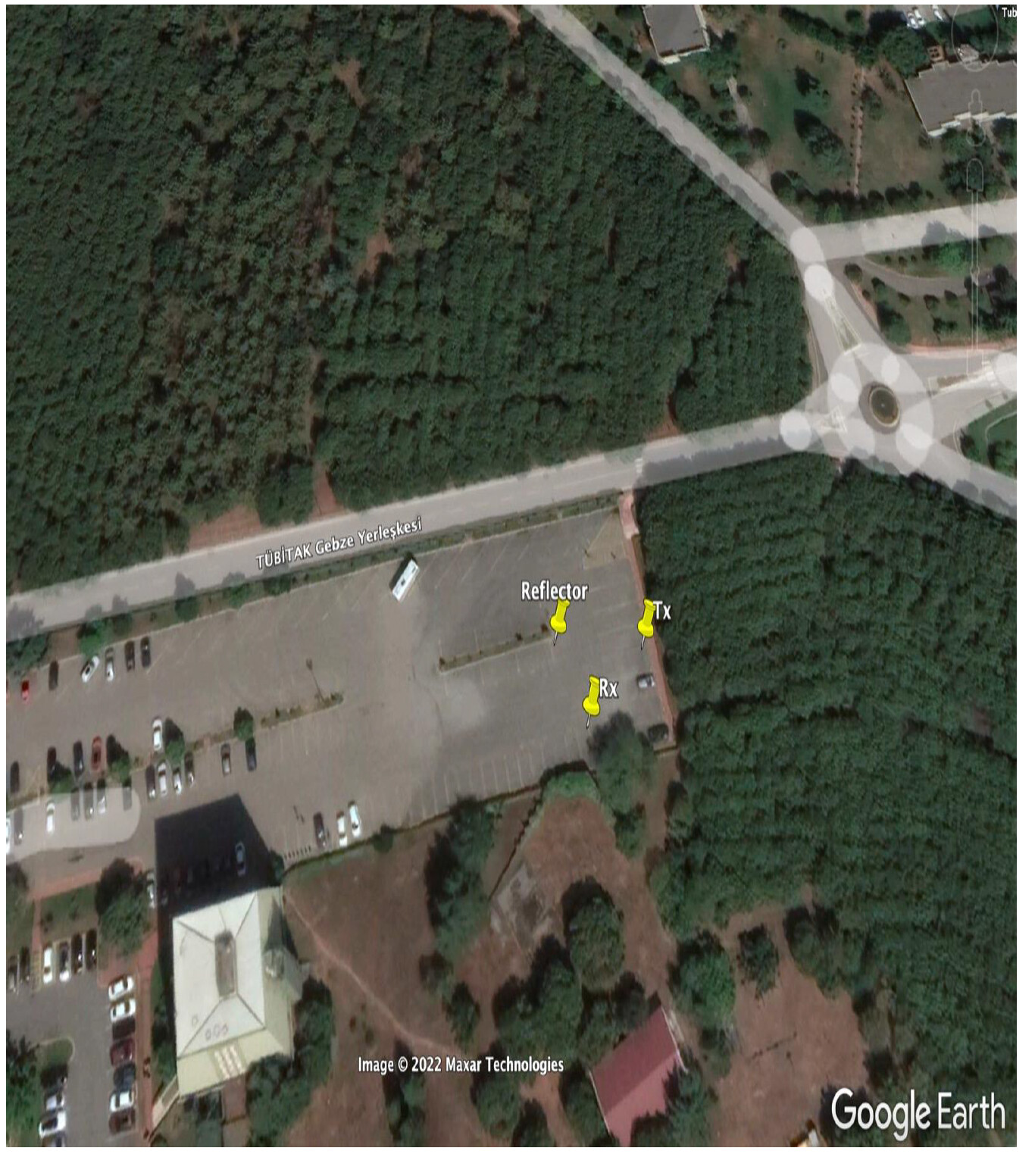}
\caption{Measurement environment with the reflector}
\end{figure}

After verifying the measurement setup with the reflector, the G2G measurements were repeated without the reflector at eight transmitter locations. The receiver remained fixed, while the transmitter was moved from TX8 to TX1, as shown in Figure~4.7. The coordinates and receiver-transmitter distances are given in Table~4.1. These measurements were then used to calculate the PDP and path-loss characteristics of the G2G channel.

\begin{figure}[htbp]
\centering\includegraphics[width=0.92\textwidth,height=0.70\textheight,keepaspectratio]{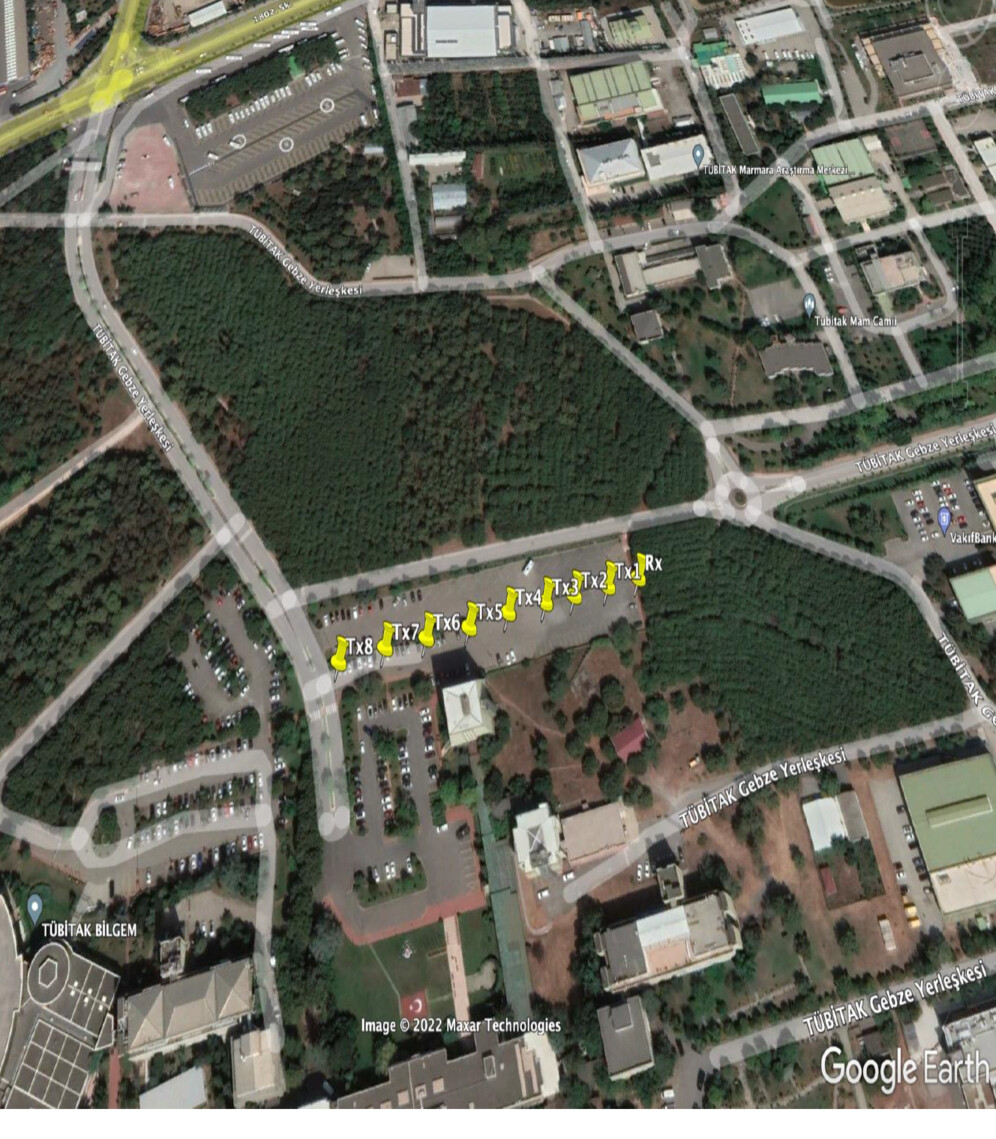}
\caption{Measurement points at eight different G2G locations}
\end{figure}

\begin{table}[htbp]
\caption{G2G Measurement Locations}
$$
\begin{array}{|c|c|c|c|}
\hline \text { Devices } & \text { Latitude } & \text { Longitude } & \text { Receiver-Transmitter Distance } \\
\hline \text { Receiver } & 40.78708 \mathrm{~N}  & 29.45049 \mathrm{E} & - \\
\hline \text { TX8 } & 40.78687 \mathrm{~N} & 29.44878 \mathrm{E} & \mathbf{146.4 m} \\
\hline \text { TX7 } & 40.78694 \mathrm{~N} & 29.44909 \mathrm{E} & \mathbf{119.3 m} \\
\hline \text { TX6 } & 40.78697 \mathrm{~N} & 29.44930 \mathrm{E} & \mathbf{101.3 m} \\
\hline \text { TX5 } & 40.78699 \mathrm{~N} & 29.44958 \mathrm{E} & \mathbf{77.6 m} \\
\hline \text { TX4 } & 40.78702 \mathrm{~N} & 29.44980 \mathrm{E} & \mathbf{58.8 m} \\
\hline \text { TX3 } & 40.78706 \mathrm{~N} & 29.45000 \mathrm{E} & \mathbf{41.5 m} \\
\hline \text { TX2 } & 40.78708 \mathrm{~N} & 29.45017 \mathrm{E} & \mathbf{27 m} \\
\hline \text { TX1 } & 40.78711 \mathrm{~N} & 29.45033 \mathrm{E} & \mathbf{13.7 m}\\
\hline
\end{array}
$$
\end{table}

\section{Measurement Results}

\hspace{0.95cm} Two G2G measurement cases were analyzed. The first case used the reflector to validate the ability of the system to detect a known multipath component. The second case used eight transmitter locations without the reflector to characterize the outdoor G2G channel. The results are interpreted in terms of the PDP and path loss.

\subsection{Power Delay Profiles of the G2G Measurements}

\hspace{0.95cm} In the reflector test, the multipath component created by the reflector appeared as a second peak in the correlation response. Figure~4.8 shows the correlation between the transmitted and received signals. The existence of the second peak confirms that the channel sounder can distinguish the LOS component from a delayed reflected component.

\begin{figure}[htbp]
\centering\includegraphics[width=0.92\textwidth,height=0.70\textheight,keepaspectratio]{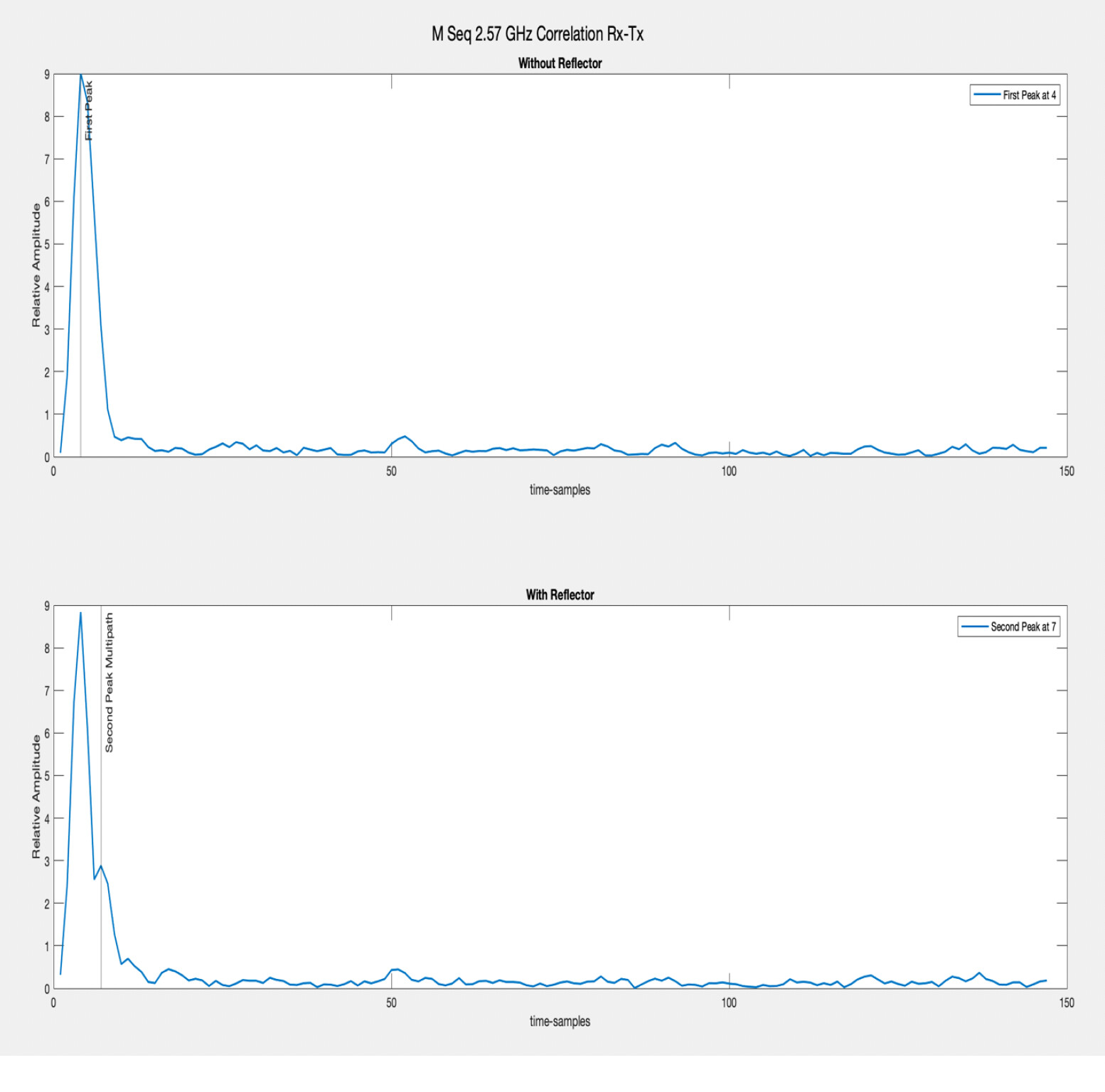}
\caption{Correlation of the system with a reflector}
\end{figure}

Figure~4.9 shows the corresponding PDP for the reflector test. The first peak appears at approximately 50 ns, while the second peak appears at approximately 99 ns. Therefore, the delay difference between the direct component and the reflected component is about 49 ns. Using the speed of light as approximately $3\times10^8~\mathrm{m/s}$, this delay corresponds to an additional propagation distance of approximately 14.7 m.

\

\

In the physical setup shown in Figure~4.6, the distance between the reflector and receiver was about 14 m, and the distance between the reflector and transmitter was about 16 m. Therefore, the reflected path was approximately 30 m, while the direct path was approximately 15 m. The resulting path-length difference is about 15 m, which is consistent with the 14.7 m value calculated from the measured delay. This agreement validates the measurement setup and the post-processing method.

\begin{figure}[htbp]
\centering\includegraphics[width=0.92\textwidth,height=0.70\textheight,keepaspectratio]{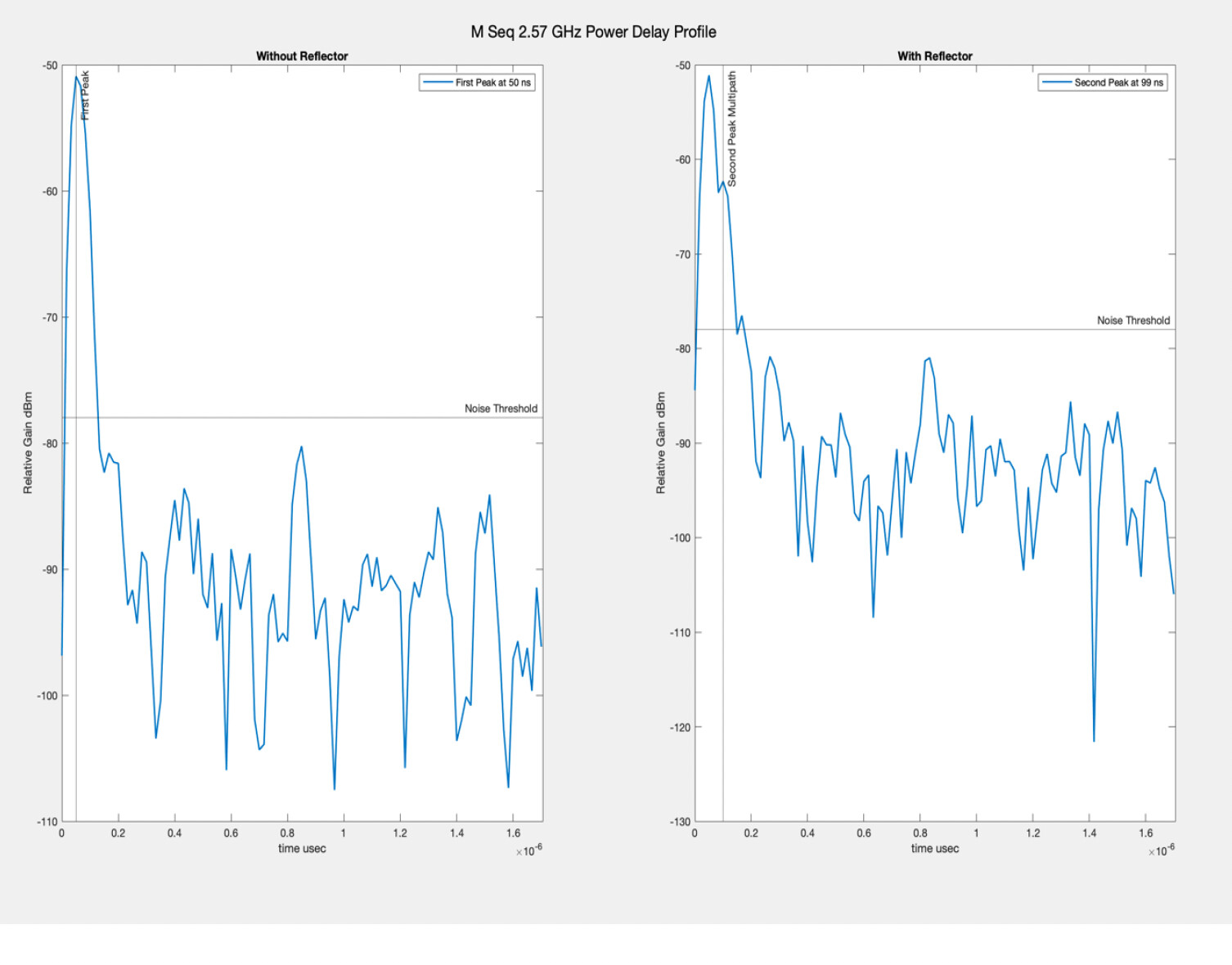}
\caption{Power delay profile of the system with a reflector}
\end{figure}

\begin{figure}[htbp]
\centering\includegraphics[width=0.92\textwidth,height=0.70\textheight,keepaspectratio]{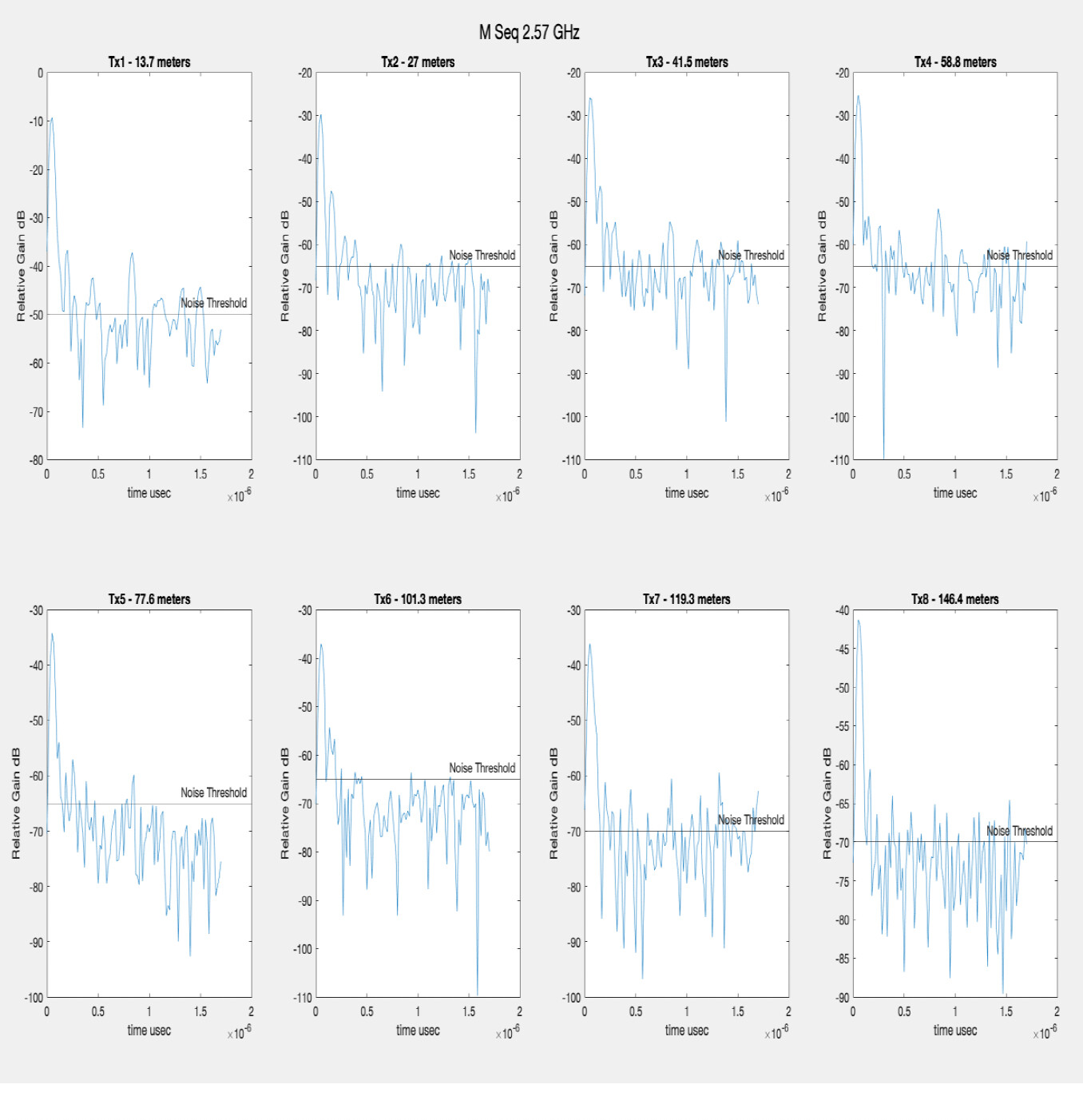}
\caption{Power delay profiles for eight G2G locations}
\end{figure}

Figure~4.10 shows the PDPs obtained from the eight transmitter locations. As expected, the relative gain decreases as the receiver-transmitter distance increases. TX1, which is 13.7 m from the receiver, produces the strongest peak. Other transmitter locations include several delayed components above the noise threshold, indicating the presence of multipath reflections in the outdoor environment.

\begin{figure}[htbp]
\centering\includegraphics[width=0.92\textwidth,height=0.70\textheight,keepaspectratio]{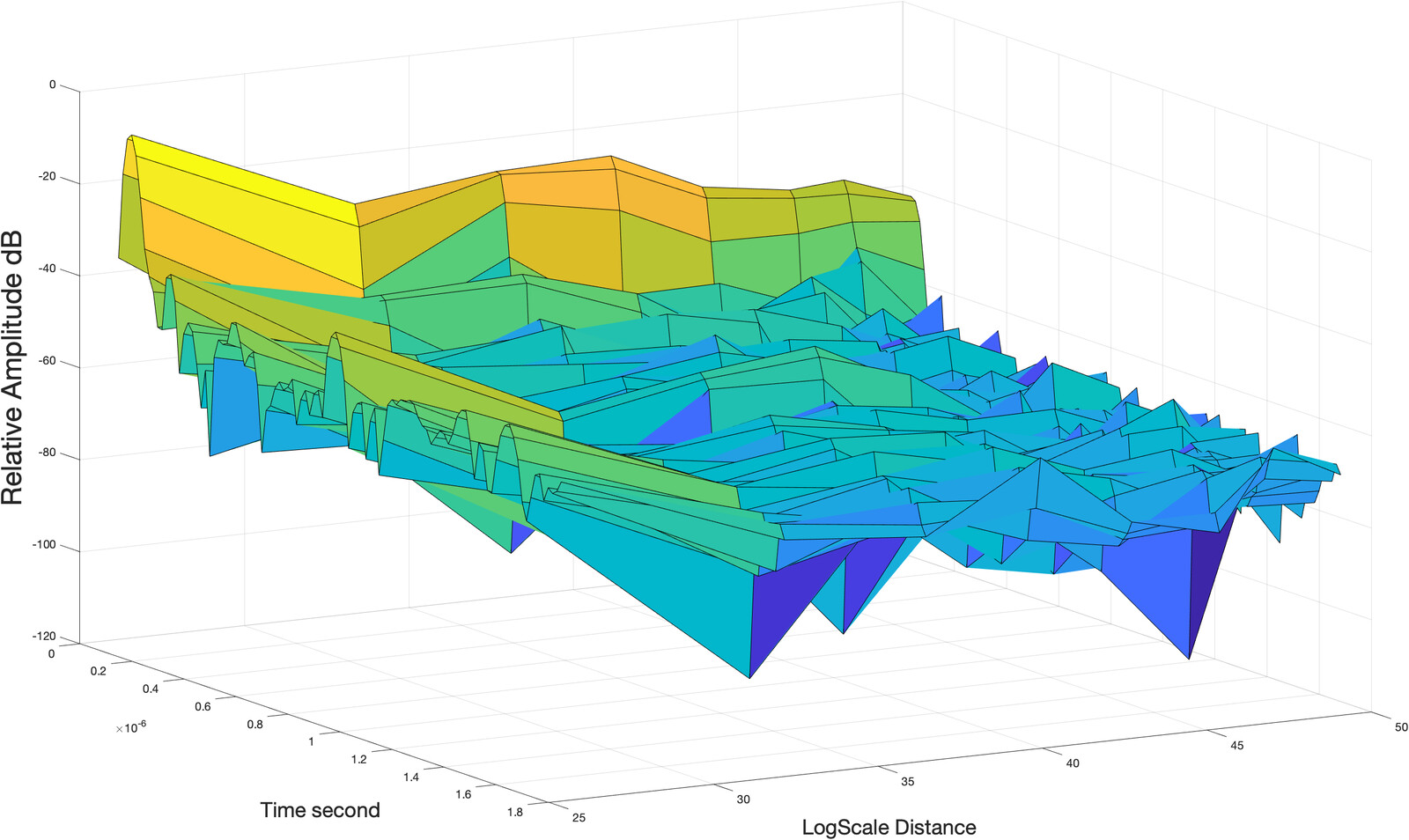}
\caption{Three-dimensional representation of the G2G power delay profiles}
\end{figure}

Figure~4.11 presents the PDPs in a three-dimensional form. The x-axis represents time in microseconds, the y-axis represents the receiver-transmitter distance on a logarithmic scale, and the z-axis represents relative power in dB. This representation makes it easier to compare the delay responses of the eight locations. Each response starts with a dominant peak and continues with delayed multipath components whose amplitudes depend on the local propagation environment.

\subsection{Path Loss of the G2G Measurements}

\hspace{0.95cm} For the G2G path-loss analysis, TX1 was selected as the reference point. The received powers at the other transmitter locations were compared with this reference, and the relative power values were plotted against distance on a logarithmic scale. The path-loss exponent $\eta$ was calculated using a least-squares fitting method in MATLAB. The receiver gain was adjusted to 30 dB during the measurements. The estimated path-loss exponent is $\eta=-2.063$, and the fitted path-loss expression is

$$
P_{L}(d B)=-2.063 \times d-5.53
$$

\begin{figure}[htbp]
\centering\includegraphics[width=0.92\textwidth,height=0.70\textheight,keepaspectratio]{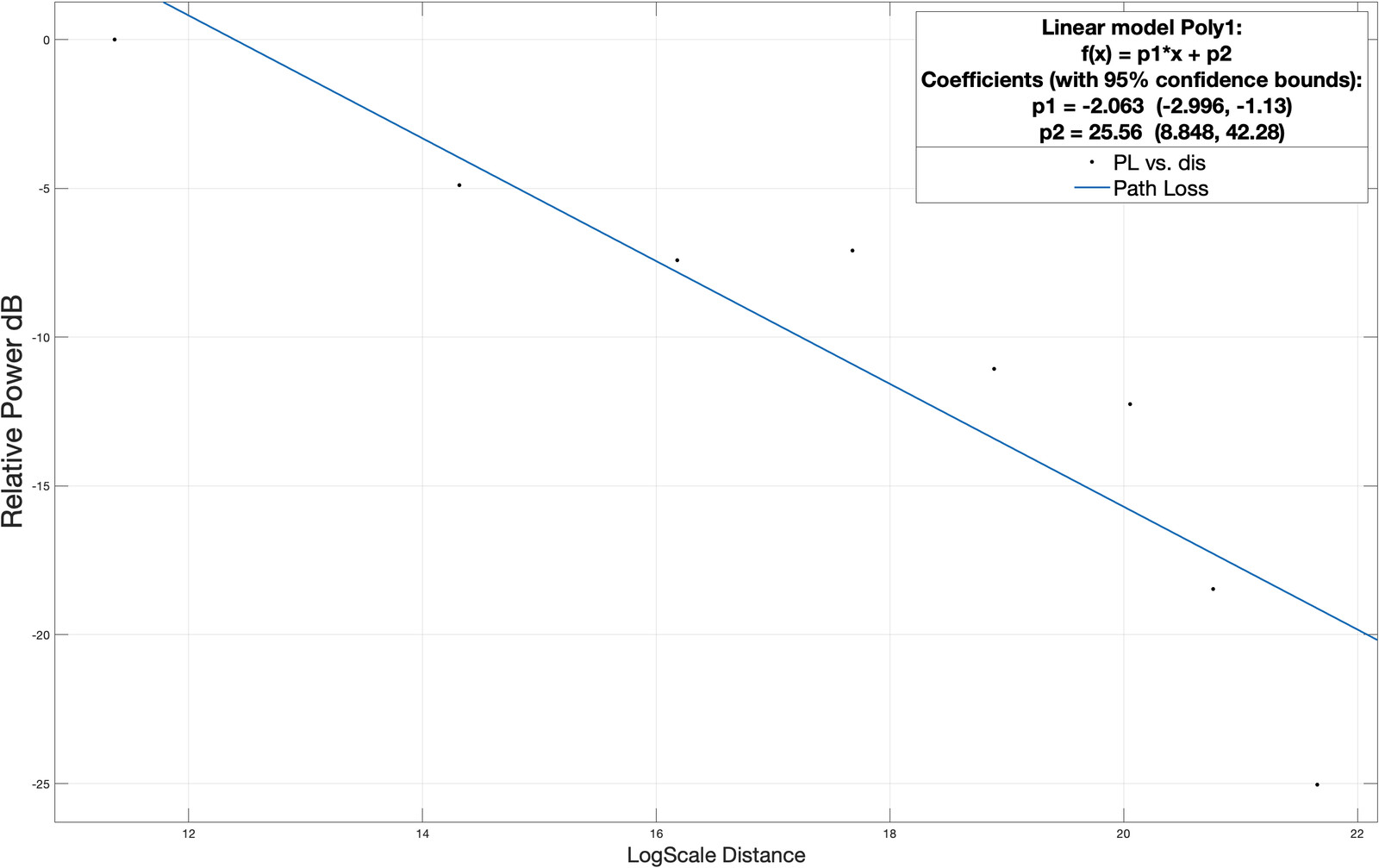}
\caption{Measured G2G path loss}
\end{figure}

The fitted curve in Figure~4.12 summarizes the average attenuation trend of the G2G channel. Although path loss gives useful information about large-scale signal reduction, the PDP results show that the channel also contains several delayed components. Therefore, both large-scale and small-scale channel statistics are required for a complete interpretation of the measurement campaign.

%%%%%%%%%%%%%%%%%%%%%%%%%%%%%%%%%%%%%%%%%%%%%%%%%%%%%%%%%%%%%%%%%
\chapter{Air to Ground UAV Wireless Communication Channel Modeling}\label{Ch5}
%%%%%%%%%%%%%%%%%%%%%%%%%%%%%%%%%%%%%%%%%%%%%%%%%%%%%%%%%%%%%%%%%

\hspace{0.95cm} This chapter presents the air-to-ground (A2G) UAV communication channel measurements. The A2G campaign was performed after the ground-to-ground outdoor measurements described in Chapter~4. The same general channel-sounding approach was used, and the maximum length sequence was selected as the sounding waveform because of its strong autocorrelation property and clear delay-domain response. Compared with the G2G campaign, the A2G campaign required a more portable and mechanically stable transmitter structure because the transmitter had to be mounted on a UAV.

The A2G measurement setup remained similar to the G2G setup in terms of the SDR, antenna, power amplifier, and signal-processing method. However, a dedicated measurement box was designed to carry the transmitter-side equipment safely on the drone. Before the outdoor A2G campaign, a multipath verification test was also performed in an anechoic chamber. This test confirmed that the measurement system could detect a known delay difference accurately before it was used in experiment.

\section{Measurement Setup}

\hspace{0.95cm} Designing the measurement setup was one of the most important parts of the A2G campaign. The transmitter had to be compact, stable, and light enough to be placed on the UAV. For this reason, a measurement box was designed to hold the SDR, storage unit, power components, and RF connections. The following subsections explain the multipath test and the measurement box used in the outdoor campaign.

\subsection{Multipath Testing}

\hspace{0.95cm} Multipath testing was performed to verify the sensitivity and delay resolution of the designed system. In the G2G campaign, a reflector was used to check whether the measurement setup could detect a reflected path. For the A2G campaign, a more controlled test was performed in an anechoic chamber at YITAL, TUBITAK. The anechoic chamber provided a low-reflection environment, which made it possible to create and measure a known artificial delay.

The theoretical distance resolution of the system was approximately 4.9 m. This value was obtained by assuming the speed of light as $3\times10^8~\mathrm{m/s}$ and using a sampling rate of 60 MS/s. With this sampling rate, the time interval between two consecutive samples is approximately 16.66 ns. Therefore, each sample corresponds to approximately 4.9 m of propagation distance.

As shown in Figure~5.1, the multipath test setup included two 50-foot cables, producing a total cable length of 100 feet. The transmitted signal was divided into two paths by a power splitter. One path was transmitted directly, while the other passed through the cable, creating a known delay. The main components used in the test were CBL-50FT-SMSM+ cables and a ZC16PD-S+ power splitter. The ANT-LTE-WS-SMA antennas and Pluto SDRs were the same components used in the G2G measurement setup.

\begin{figure}[htbp]
\centering\includegraphics[width=0.92\textwidth,height=0.70\textheight,keepaspectratio]{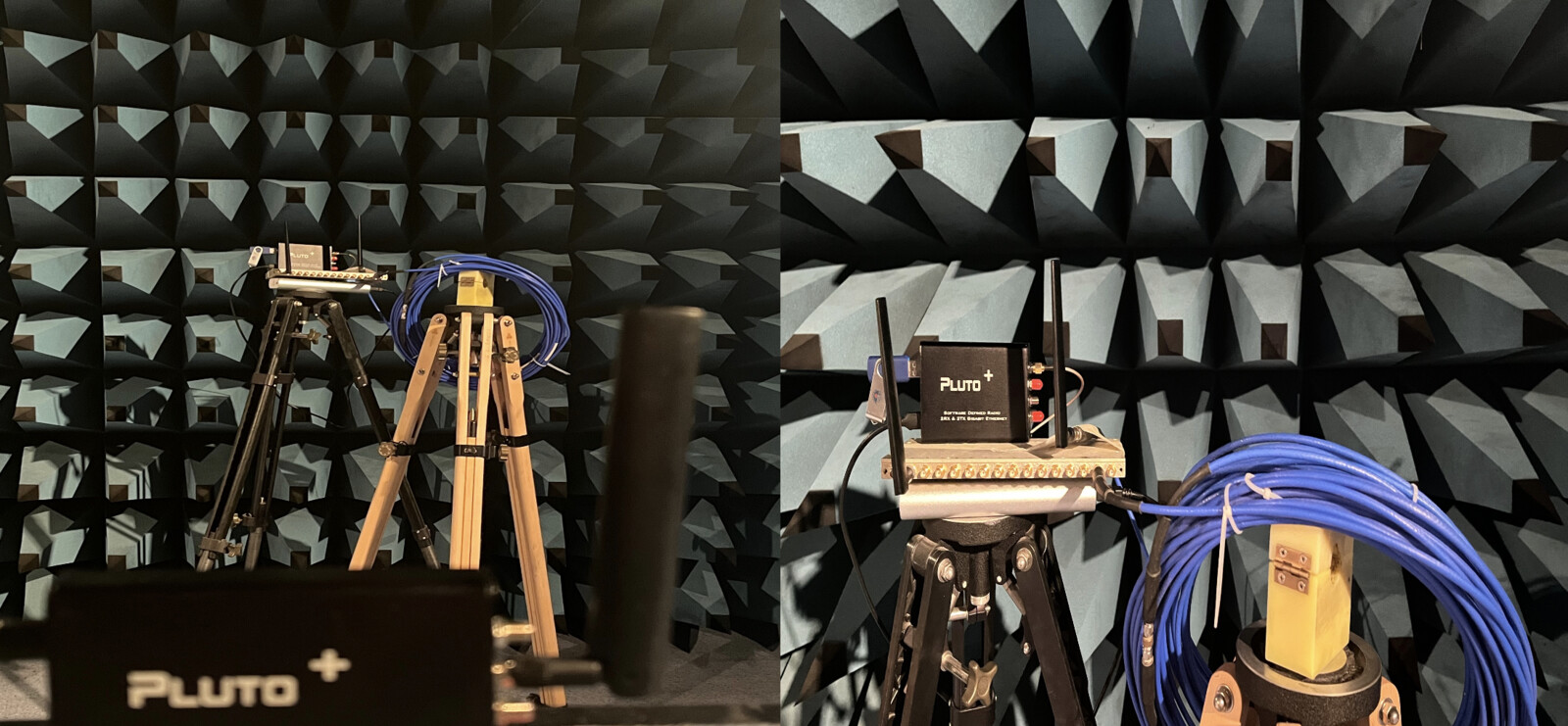}
\caption{Anechoic chamber setup for multipath testing}
\end{figure}
The dielectric constant of the CBL-50FT-SMSM+ cable, which uses solid PTFE, is approximately 2.2. The velocity factor of the cable can therefore be calculated as:
\begin{equation}
V_{f}=\frac{1}{\sqrt{\varepsilon_r}}=\frac{1}{\sqrt{2.2}}=0.677
\end{equation}
Using the total cable length, the expected delay can be calculated as:

\begin{equation}
D_{t}=\frac{L}{V_f c}=\frac{100 \times 0.3048}{0.677 \times 3\times10^8}=150.07~\mathrm{ns}
\end{equation}

\begin{figure}[htbp]
\centering\includegraphics[width=0.92\textwidth,height=0.70\textheight,keepaspectratio]{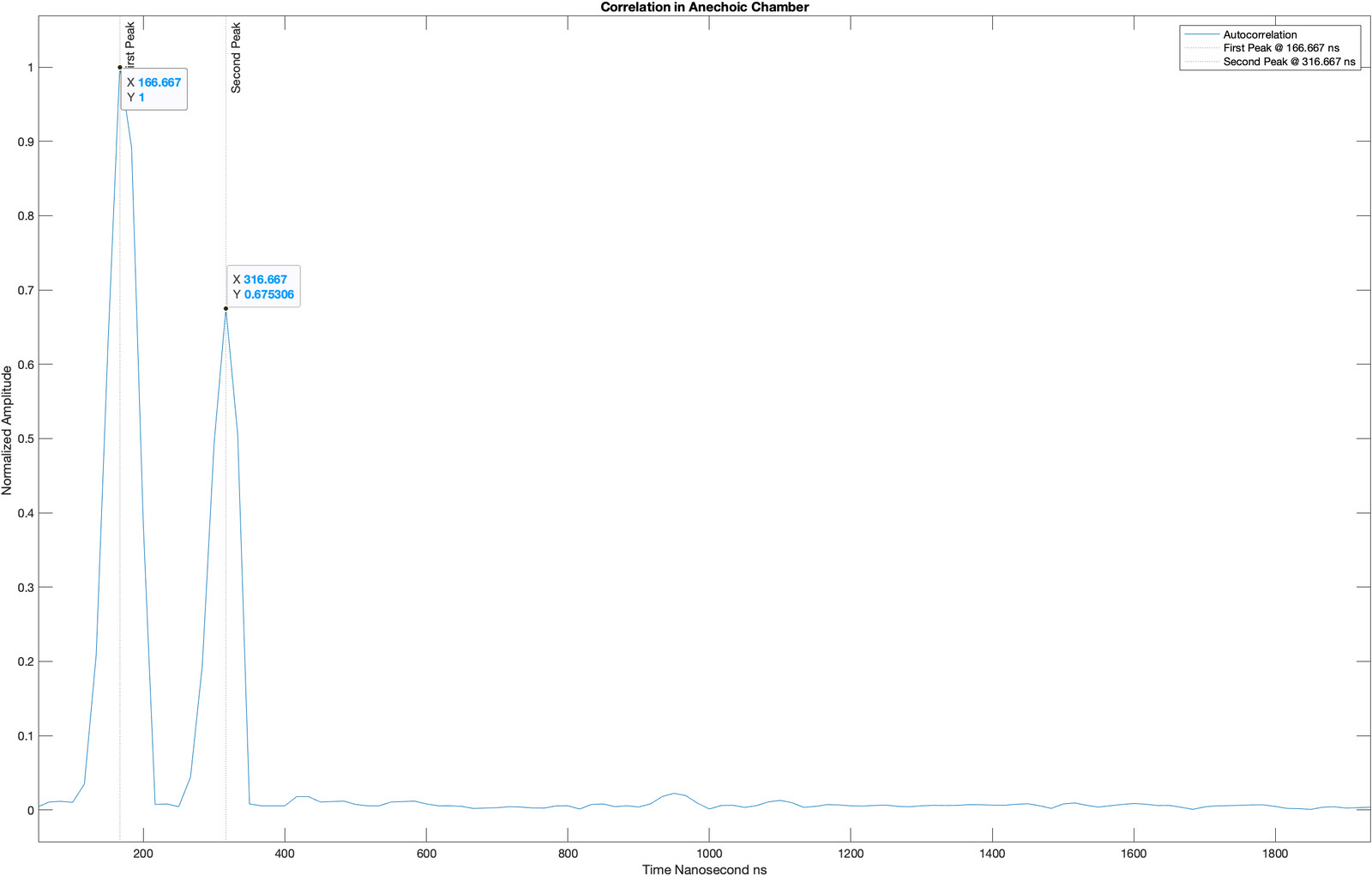}
\caption{Correlation results of the multipath test}
\end{figure}

As shown in Figure~5.2, two signal peaks were observed. The first peak corresponds to the direct output of the splitter, and the second peak corresponds to the delayed cable path. The first peak appeared around the 11th sample, and the second peak appeared around the 20th sample. The measured delay can be estimated as

\begin{equation}
 D_t = \text{samples} \times \text{sampling period} = 9 \times 16.66 \approx 150\,\mathrm{ns}
\end{equation}

The measured delay is very close to the theoretical delay. Therefore, the anechoic chamber test confirmed that the system could detect the expected multipath delay correctly. This result increased confidence in the outdoor A2G measurement campaign.

\subsection{Measurement Box}

\hspace{0.95cm} In the G2G outdoor measurements, a measurement box was not required because the transmitter and receiver could be placed on tripods. In the A2G measurements, however, the transmitter had to be carried by the UAV. Therefore, a compact measurement box was designed to stabilize and protect the transmitter-side components during flight.

\begin{figure}[htbp]
\centering\includegraphics[width=0.92\textwidth,height=0.70\textheight,keepaspectratio]{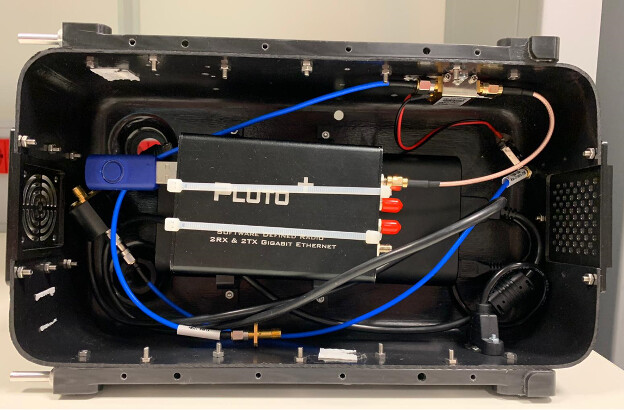}
\caption{Measurement box}
\end{figure}

\begin{figure}[htbp]
\centering\includegraphics[width=0.92\textwidth,height=0.70\textheight,keepaspectratio]{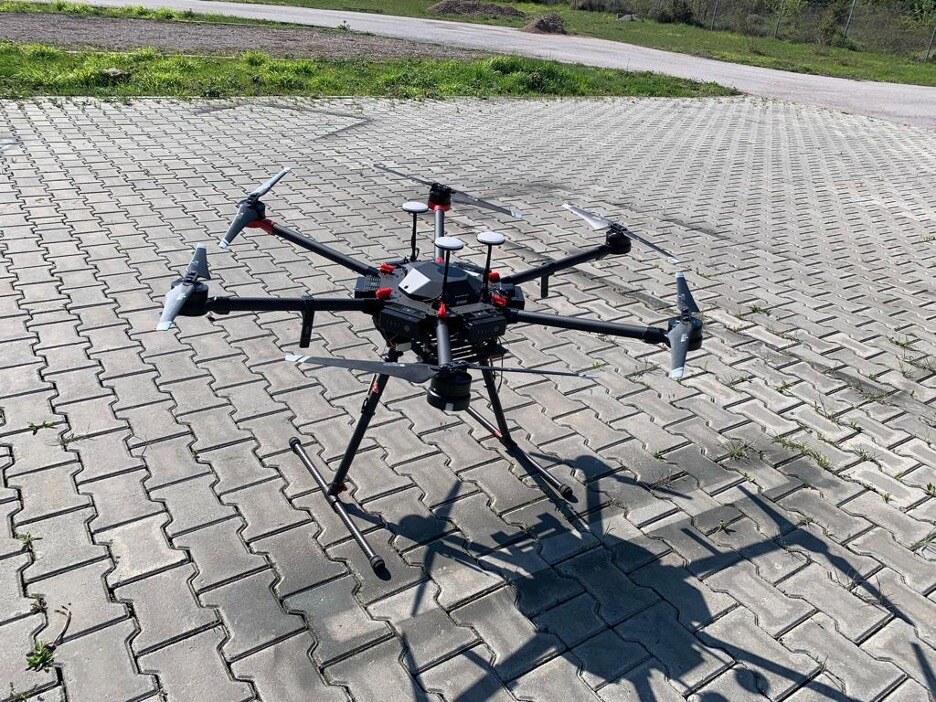}
\caption{Drone carrying the measurement box}
\end{figure}

During the A2G campaign, the receiver remained on the ground, while the transmitter was placed on the drone. The measurement box shown in Figure~5.3 was mounted on the UAV, as shown in Figure~5.4. The box was designed to keep the SDR, RF cables, antenna connections, storage unit, and power components secure during the flight. The main components of the measurement box are described below.

\subsubsection{Software-Defined Radio}

\hspace{0.95cm} ADALM-Pluto was initially selected as the SDR platform for both the transmitter and receiver. However, the first tests showed that its available resolution was not sufficient for observing multipath components clearly. As in the G2G campaign, Pluto Plus was adopted because its sampling rate could be increased through firmware modification. The ADC setting was changed from AD9364 to AD9361 using the PuTTY control interface, allowing the SDR to operate at higher sample rates and frequencies. With this configuration, the obtained distance resolution was approximately 4.9 m, which was suitable for the A2G measurements.

\subsubsection{Antenna, Power Amplifier, and Coaxial Cables}

\hspace{0.95cm} The ANT-LTE-WS-SMA antenna manufactured by Linx Technologies was used on both the transmitter and receiver sides because it supports the 2.5 GHz--2.7 GHz operating range. Since the UAV moved away from the receiver during the measurement campaign, the received power decreased with distance. To improve the link budget, a ZX60-V63+ power amplifier manufactured by Mini-Circuits was added to the transmitter-side setup. The amplifier increased the signal level by approximately 16 dB and reduced the effect of noise on the received data.

Coaxial cables were used to connect the antenna outside the measurement box to the SDR inside the box. Because the available cables were not long enough to allow every component to remain inside the box, the power amplifier was mounted on the side of the box, as shown in Figure~5.3. This arrangement provided a practical balance between mechanical stability and RF connectivity.

\subsubsection{Computer, Flash Disk, and Power Bank}

\hspace{0.95cm} A computer was used on the receiver side to record the received IQ data. On the transmitter side, a flash disk stored the sounding signal to be transmitted through the channel. A power bank supplied the power amplifier with 5 V. This arrangement allowed the transmitter system to operate independently while mounted on the UAV.

\subsection{Measurement Campaign}

\hspace{0.95cm} The A2G measurement campaign was conducted at 2.57 GHz for 16 transmitter locations. The receiver was fixed on the ground, and the transmitter was carried by the UAV at an altitude of approximately 50 m. The UAV moved horizontally in 25 m steps, and measurements were collected until the horizontal separation between the transmitter and receiver reached 400 m. Figure~5.5 shows A2G measurement environment.

\begin{figure}[htbp]
\centering\includegraphics[width=0.92\textwidth,height=0.70\textheight,keepaspectratio]{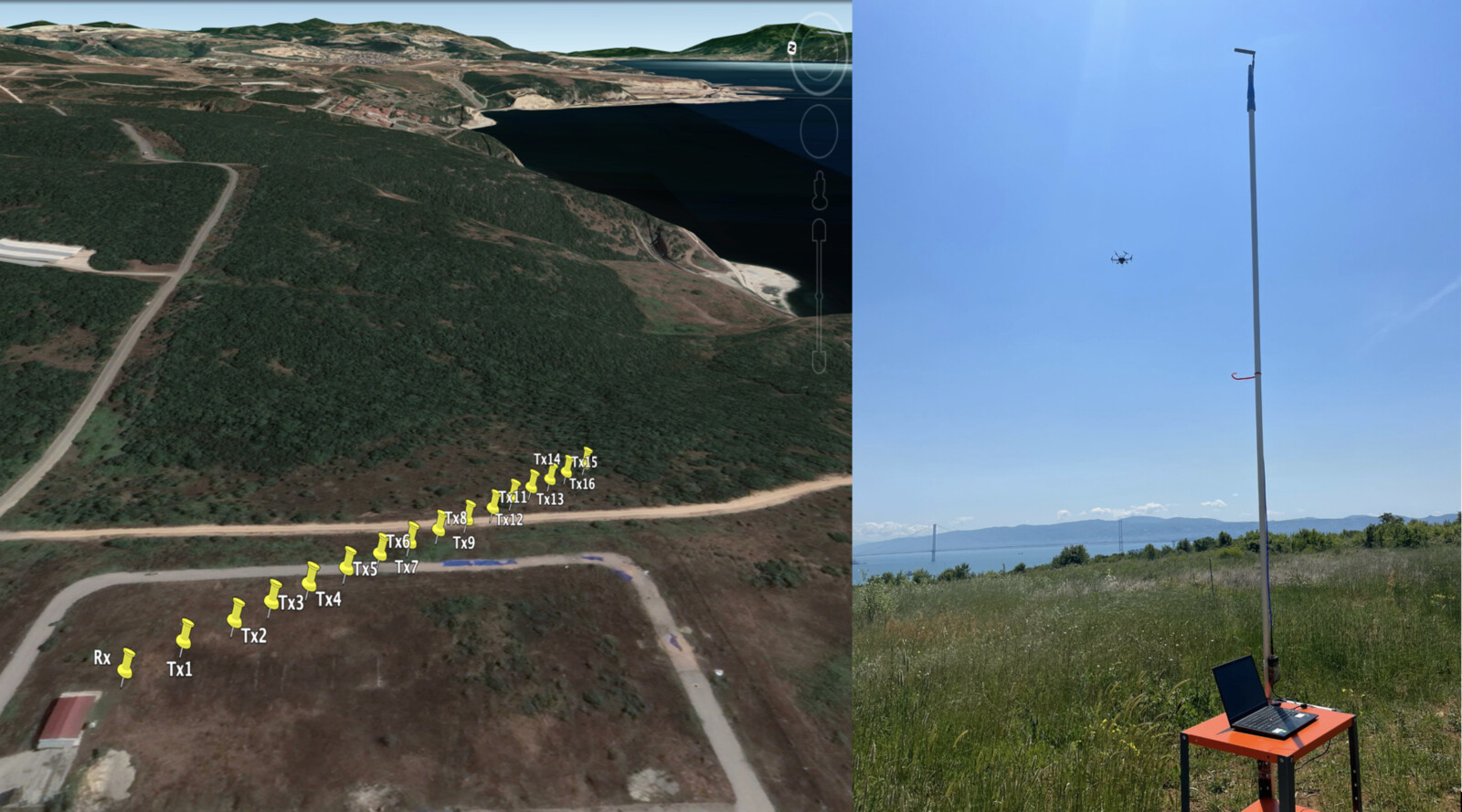}
\caption{A2G measurement environment}
\end{figure}

\begin{table}[htbp]
\caption{A2G Measurement Locations}
$$
\begin{array}{|c|c|c|c|}
\hline \text { Devices } & \text { Latitude } & \text { Longitude } & \text { Receiver-Transmitter Distance } \\
\hline \text { Receiver } & 40.78708 \mathrm{~N}  & 29.45049 \mathrm{E} & - \\
\hline \text { TX1 } & 40.78687 \mathrm{~N} & 29.44878 \mathrm{E} & \mathbf{25 m} \\
\hline \text { TX2 } & 40.78694 \mathrm{~N} & 29.44909 \mathrm{E} & \mathbf{50 m} \\
\hline \text { TX3 } & 40.78697 \mathrm{~N} & 29.44930 \mathrm{E} & \mathbf{75 m} \\
\hline \text { TX4 } & 40.78699 \mathrm{~N} & 29.44958 \mathrm{E} & \mathbf{100 m} \\
\hline \text { TX5 } & 40.78702 \mathrm{~N} & 29.44980 \mathrm{E} & \mathbf{125 m} \\
\hline \text { TX6 } & 40.78706 \mathrm{~N} & 29.45000 \mathrm{E} & \mathbf{150 m} \\
\hline \text { TX7 } & 40.78708 \mathrm{~N} & 29.45017 \mathrm{E} & \mathbf{175 m} \\
\hline \text { TX8 } & 40.78711 \mathrm{~N} & 29.45033 \mathrm{E} & \mathbf{200 m}\\
\hline \text { TX9 } & 40.78613 \mathrm{~N} & 29.45178 \mathrm{E} & \mathbf{225 m} \\
\hline \text { TX10 } & 40.78624 \mathrm{~N} & 29.45490 \mathrm{E} & \mathbf{250 m} \\
\hline \text { TX11 } & 40.78647 \mathrm{~N} & 29.45630 \mathrm{E} & \mathbf{275 m} \\
\hline \text { TX12 } & 40.78669 \mathrm{~N} & 29.45658 \mathrm{E} & \mathbf{300 m} \\
\hline \text { TX13 } & 40.78772 \mathrm{~N} & 29.45680 \mathrm{E} & \mathbf{325 m} \\
\hline \text { TX14 } & 40.78766 \mathrm{~N} & 29.45700 \mathrm{E} & \mathbf{350 m} \\
\hline \text { TX15 } & 40.78728 \mathrm{~N} & 29.45817 \mathrm{E} & \mathbf{375 m} \\
\hline \text { TX16 } & 40.78791 \mathrm{~N} & 29.45933 \mathrm{E} & \mathbf{400 m}\\
\hline
\end{array}
$$
\end{table}

\section{Measurement Results}

\hspace{0.95cm} Two types of results were analyzed for the A2G scenario. First, the anechoic chamber test was used to confirm that the system could detect a known delayed path. Second, the outdoor UAV measurements were processed to obtain PDP and path-loss results. The outdoor measurements are especially important because they show how the received signal changes as the UAV moves farther from the ground receiver.

\subsection{Power Delay Profiles of the A2G Measurements}

\hspace{0.95cm} As discussed in the measurement campaign section, the A2G measurements were collected at 16 horizontal transmitter distances. The outdoor environment produced several multipath components due to reflections and scattering. These components appear as peaks at different delay values in the PDP. Figures~5.6 and 5.7 show the PDPs for TX1--TX8 and TX9--TX16, respectively.

\begin{figure}[htbp]
\centering\includegraphics[width=0.92\textwidth,height=0.70\textheight,keepaspectratio]{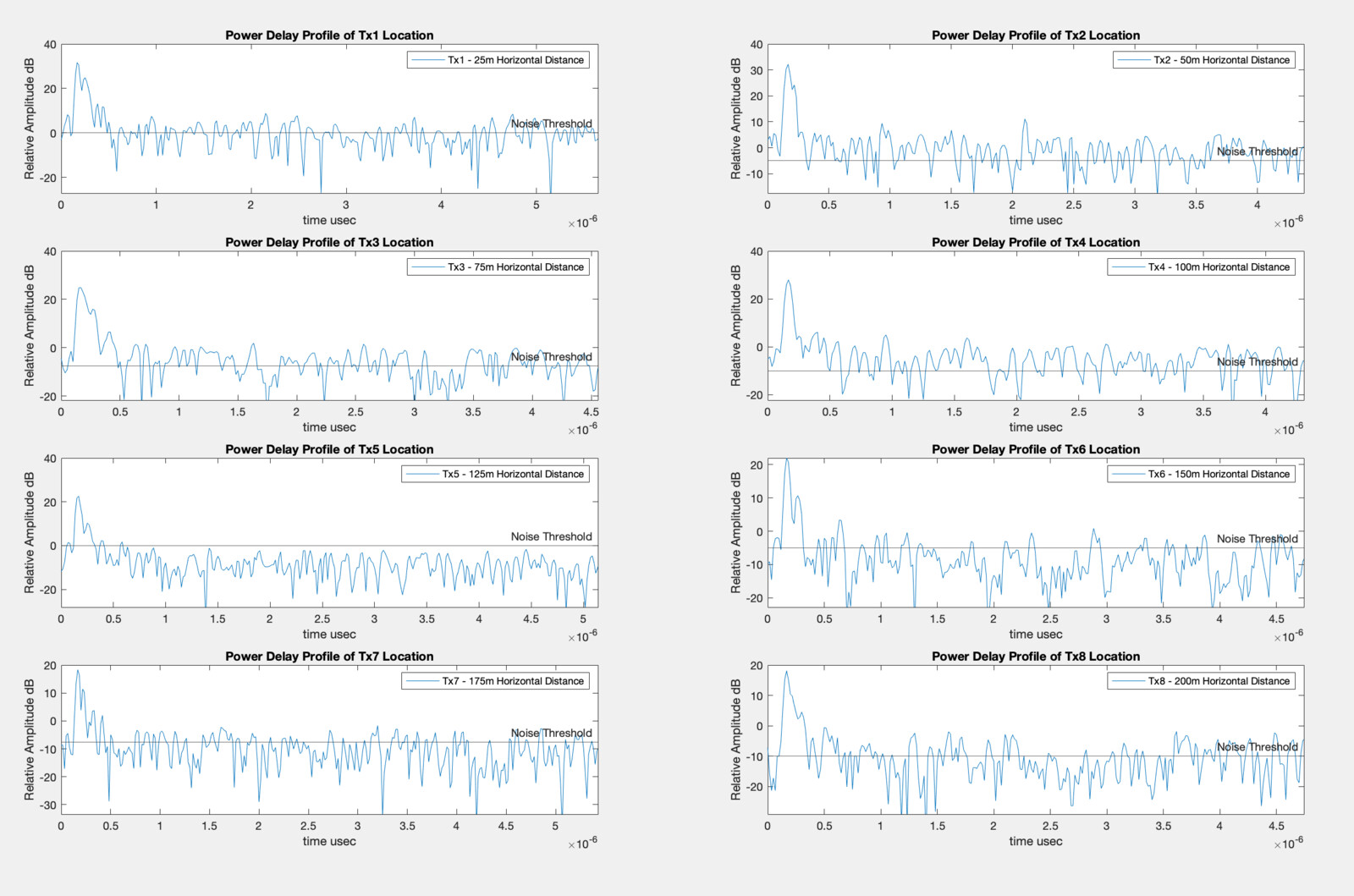}
\caption{Power delay profiles of the A2G measurements for TX1--TX8}
\end{figure}

\begin{figure}[htbp]
\centering\includegraphics[width=0.92\textwidth,height=0.70\textheight,keepaspectratio]{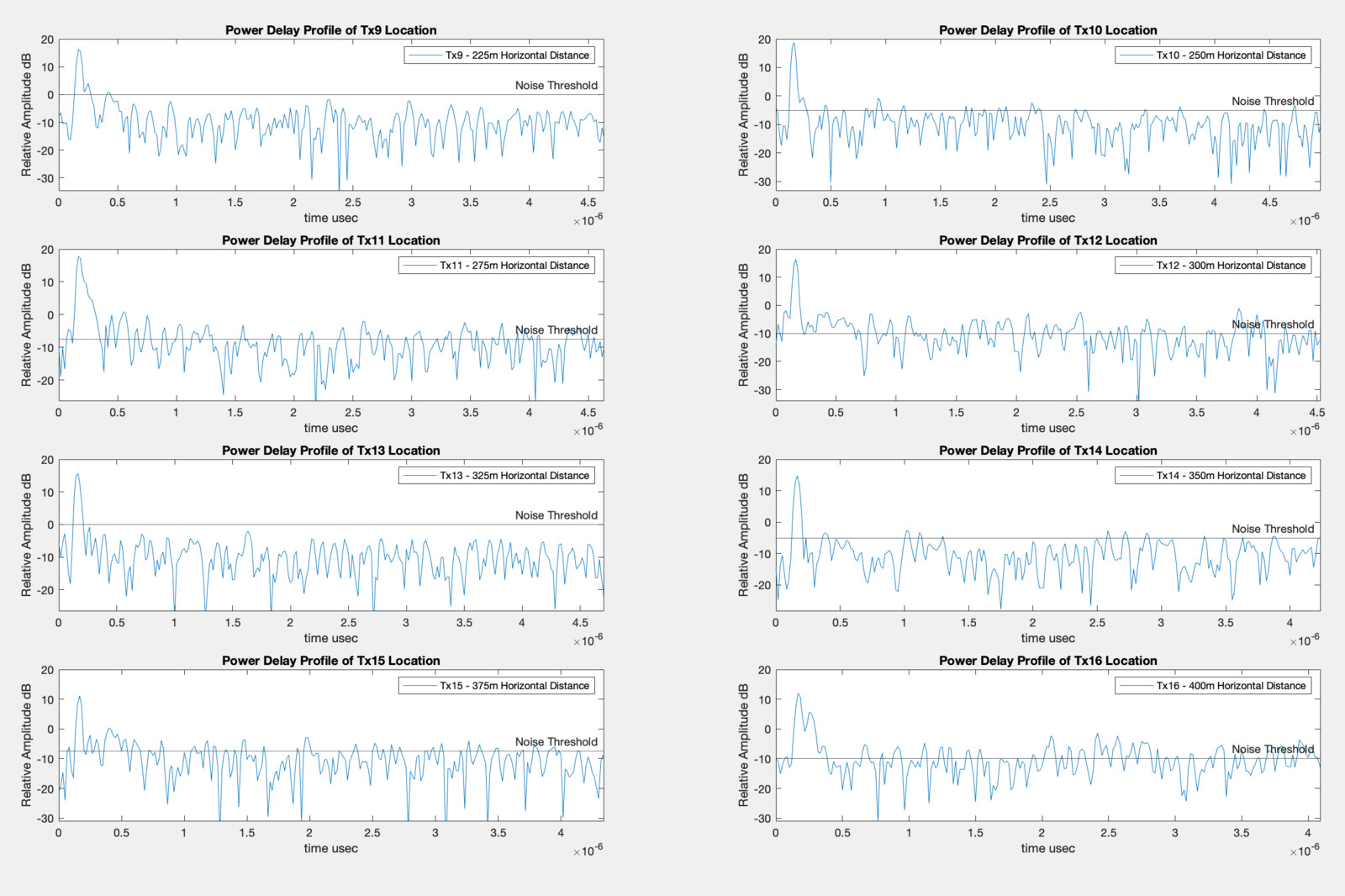}
\caption{Power delay profiles of the A2G measurements for TX9--TX16}
\end{figure}

There were 16 different A2G channel responses at 2.57 GHz. As shown in Figures~5.6 and 5.7, the relative gain generally decreases as the horizontal distance between the receiver and transmitter increases. TX1, located 25 m horizontally from the receiver, has the strongest peak among the measured points. In addition, several locations include delayed components above the noise threshold, showing that the A2G channel is not described only by the direct path.

\begin{figure}[htbp]
\centering\includegraphics[width=0.92\textwidth,height=0.70\textheight,keepaspectratio]{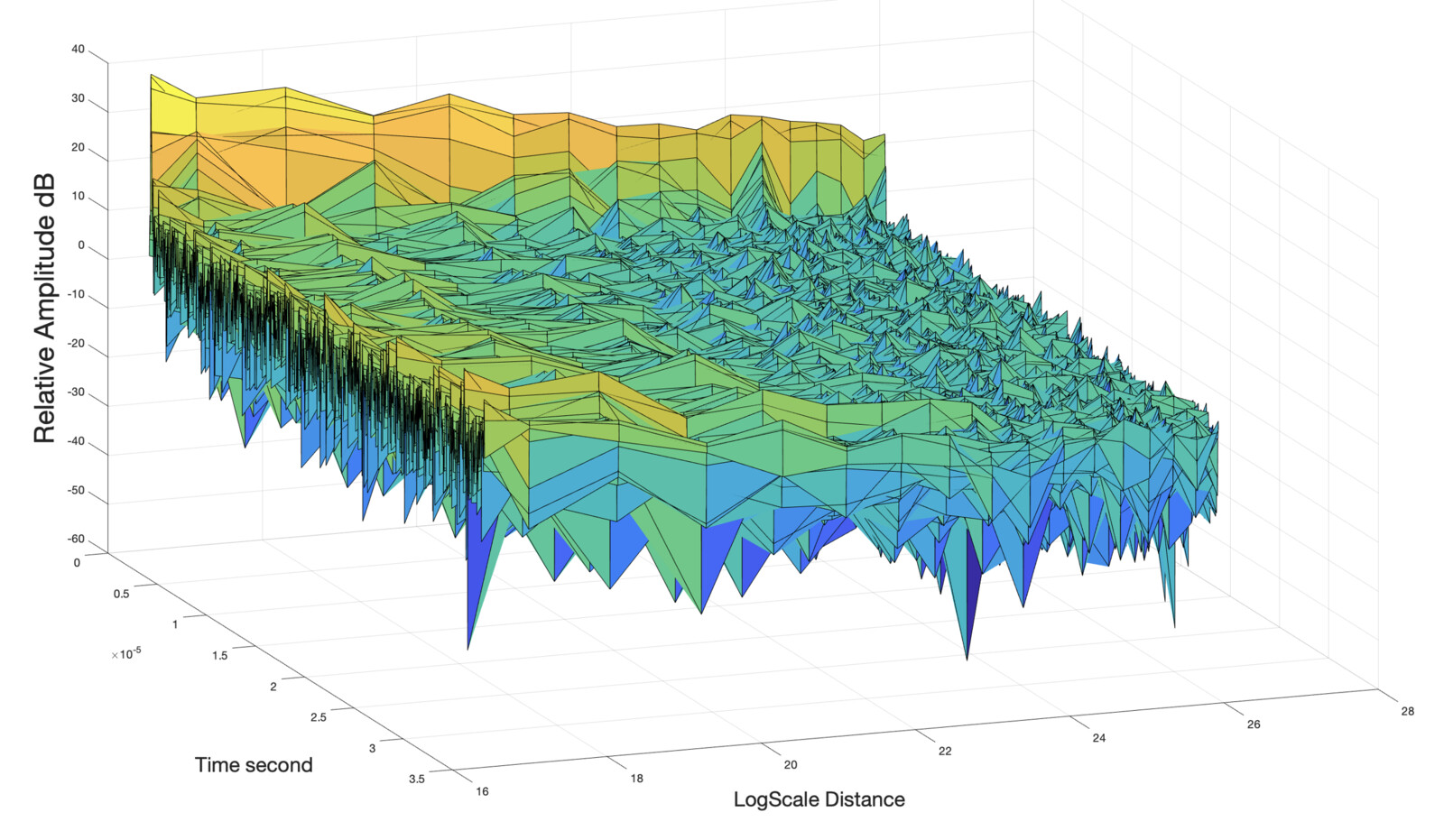}
\caption{Three-dimensional representation of the A2G power delay profiles}
\end{figure}
Figure~5.8 presents the A2G PDPs in a three-dimensional graph. The x-axis represents time in microseconds, the y-axis represents horizontal distance on a logarithmic scale, and the z-axis represents relative power in dB. This representation makes the distance-dependent attenuation and the delay-domain multipath structure visible in the same graph. The dominant peak generally weakens with distance, while weaker delayed components remain visible at several transmitter positions.

\subsection{Path Loss of the A2G Measurements}

\hspace{0.95cm} The main objective of the A2G measurement campaign was to characterize two important channel statistics: path loss and power delay profile. The path-loss exponent $\eta$ of the A2G wireless UAV channel was calculated using the least-squares method in MATLAB. Figure~5.9 shows the relative power values, defined with respect to a reference point, plotted against the receiver-transmitter distance on a logarithmic scale. Three IQ data sets were saved for each distance in order to reduce the effect of random measurement variations. The receiver gain was adjusted to 30 dB during the campaign. The estimated path-loss exponent is $\eta=-2.238$, and the fitted path-loss expression is:
$$
P_{L}(d B)=-2.238 \times d-18.99
$$

\begin{figure}[htbp]
\centering\includegraphics[width=0.75\textwidth,keepaspectratio]{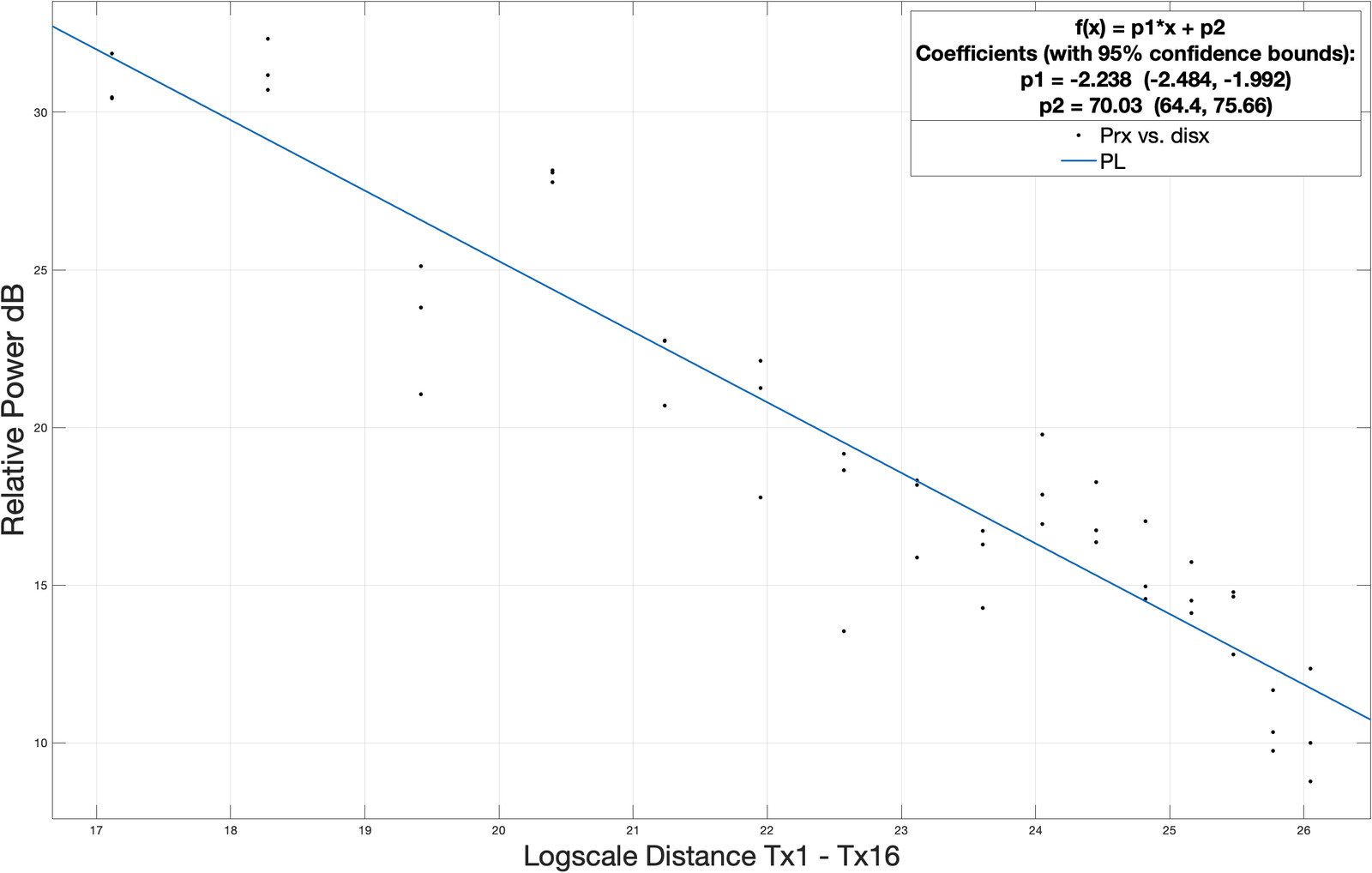}
\caption{A2G path-loss exponent}
\end{figure}

The A2G results show that the received power decreases with increasing UAV distance, as expected. However, the PDPs also show that delayed components remain present in the channel. Therefore, a reliable A2G channel model should include not only the large-scale path-loss trend but also the small-scale delay-domain behavior.

%%%%%%%%%%%%%%%%%%%%%%%%%%%%%%%%%%%%%%%%%%%%%%%%%%%%%%%%%%%%%%%%%
\chapter{Air to Air UAV Wireless Communication Channel Modeling}\label{Ch6}
%%%%%%%%%%%%%%%%%%%%%%%%%%%%%%%%%%%%%%%%%%%%%%%%%%%%%%%%%%%%%%%%%

\hspace{0.95cm} This chapter presents the air-to-air (A2A) UAV communication channel measurements. The A2A campaign was performed after the G2G and A2G campaigns, and it represents the most complex measurement scenario in this thesis. In the A2A case, both the transmitter and receiver are aerial platforms, so the measurement setup must be portable, synchronized, and remotely controllable. As in the previous campaigns, the maximum length sequence was used as the channel sounding signal because it provided a clear autocorrelation peak and reliable delay-domain behavior.

Compared with the A2G setup, the A2A setup required two measurement boxes instead of one. Additional components were also needed for remote control, position information, and synchronization. These changes increased the complexity of the hardware design, but they made it possible to conduct a practical A2A measurement campaign at 2.57 GHz. The following sections describe the measurement setup, the measurement campaign, and the obtained results.

\section{Measurement Setup}

\hspace{0.95cm} The A2A measurement setup had to be designed for two UAV-mounted boxes. Each box had to carry the SDR, antennas, power components, gateway equipment, GPS unit, and clock-related components required for synchronization. The system also had to remain mechanically stable during flight and provide reliable communication between the transmitter and receiver boxes. The subsections below describe the main components of the A2A measurement box.

\subsection{Measurement Box}

\hspace{0.95cm} In the A2A outdoor measurements, two measurement boxes were required. The A2G setup used one box on the UAV and one receiver on the ground. In contrast, the A2A setup required both the transmitter and receiver equipment to be mounted on aerial platforms. Therefore, the number of components increased, and the internal organization of the measurement box became more important.

Two antennas were used in each box. One antenna was used for the 2.57 GHz measurement link between the transmitter and receiver. The other antenna was used for the 915 MHz control and gateway communication between the boxes. BladeRF was selected as the software-defined radio instead of ADALM-Pluto because it provided better performance for this setup. A Microhard gateway was used to establish communication between the boxes, while a UBLOX GPS module and a clock were used to support positioning and synchronization. The general structure of the measurement box is shown in Figure~6.1.

\begin{figure}[htbp]
\centering\includegraphics[width=0.75\textwidth,height=0.6\textheight,keepaspectratio]{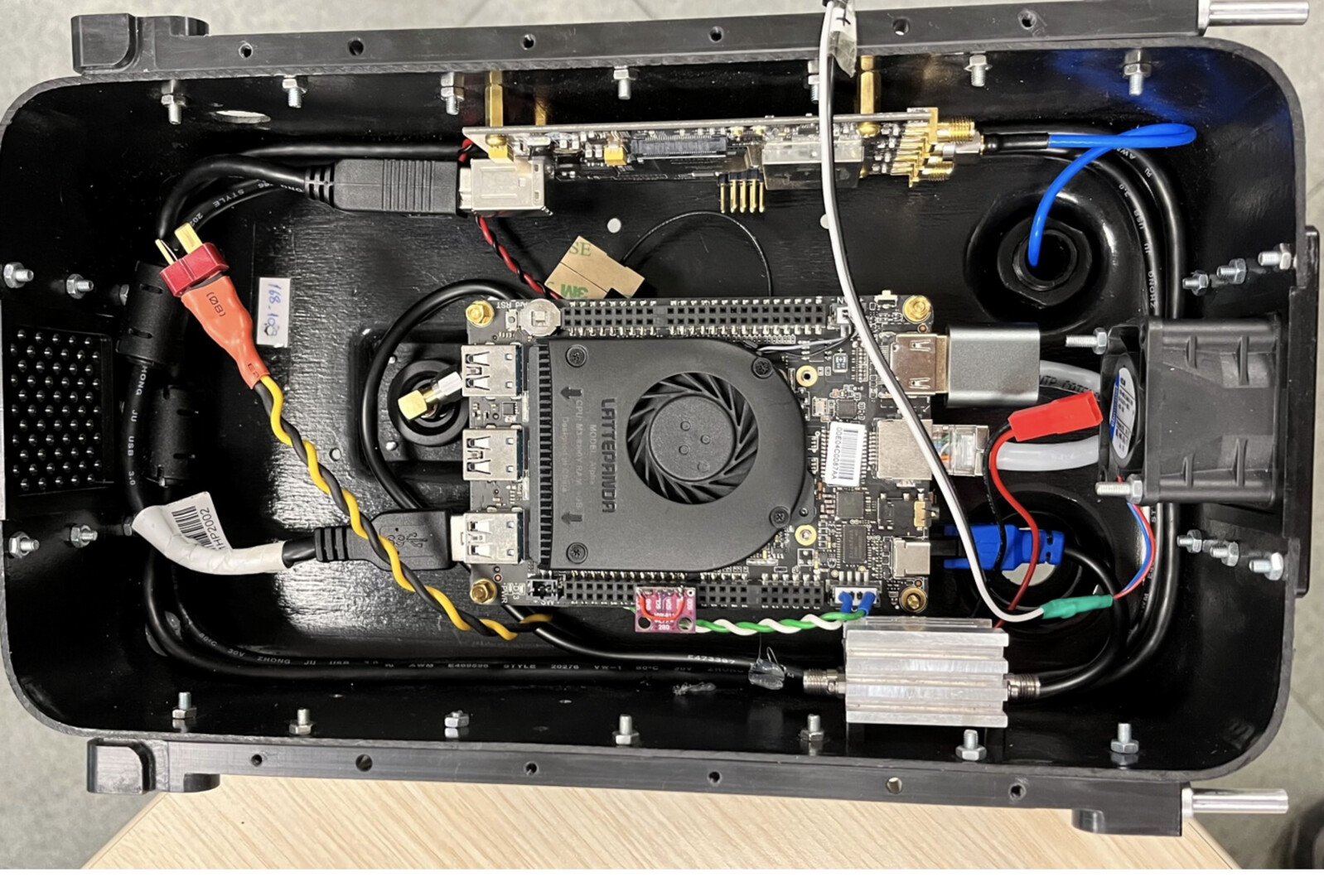}
\caption{Measurement box for A2A measurements}
\end{figure}

\subsubsection{Software-Defined Radio}

\hspace{0.95cm} BladeRF 2.0 Micro XA9 was selected as the SDR platform for both the transmitter and receiver. It was preferred over ADALM-Pluto because it provided higher resolution and a more accurate oscillator structure, which helped the system detect signals more reliably. In addition, the BladeRF platform was more suitable for integration with the gateway-based remote communication structure used in the A2A measurement boxes.

\begin{figure}[htbp]
\centering\includegraphics[width=0.92\textwidth,height=0.70\textheight,keepaspectratio]{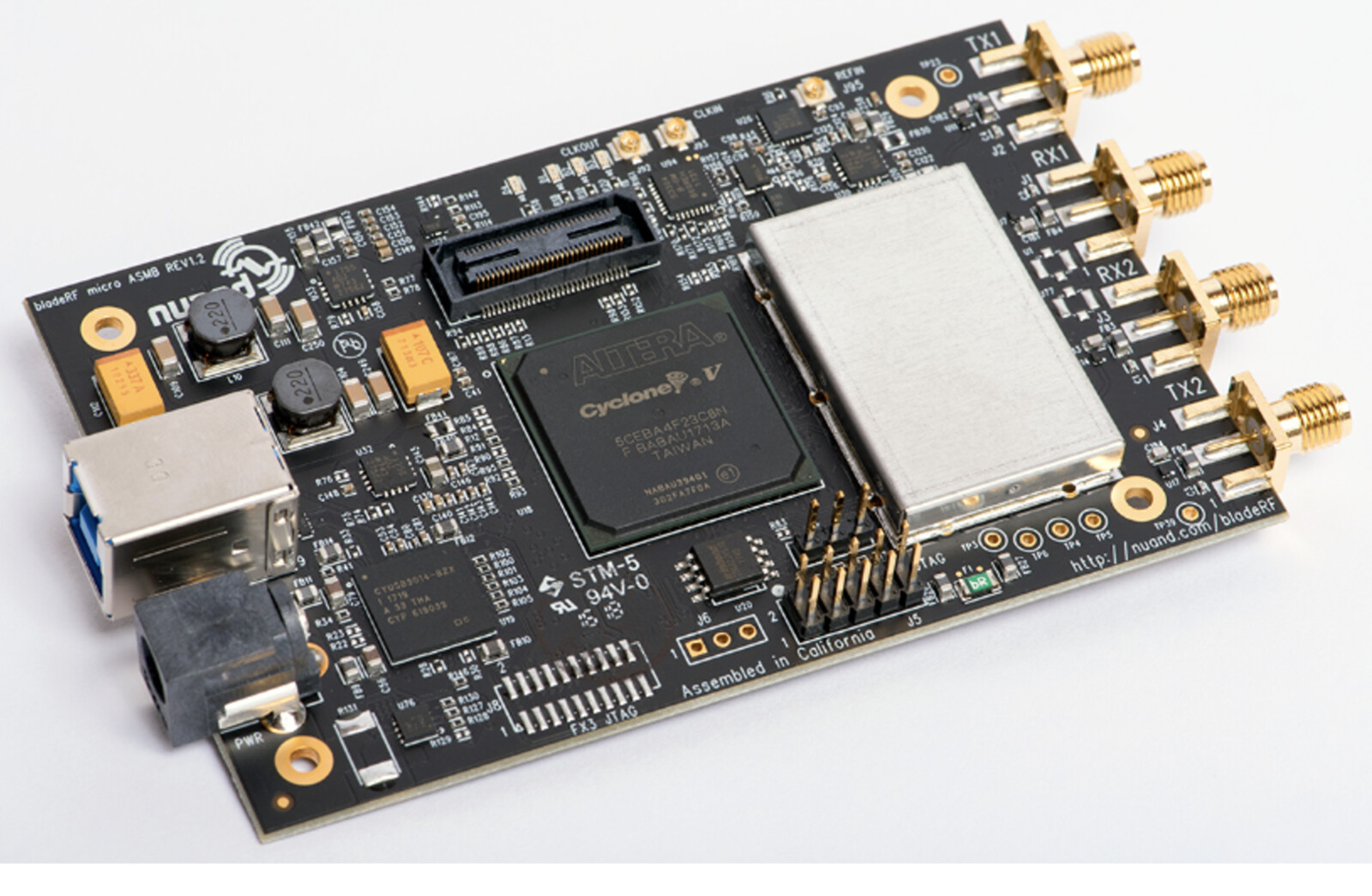}
\caption{BladeRF 2.0 Micro XA9}
\end{figure}

\subsubsection{Antennas and Power Amplifier}

\hspace{0.95cm} Antenna selection was important because the measurement link operated at 2.57 GHz. The ANT-LTE-WS-SMA antenna manufactured by Linx Technologies was used because it can operate in the 2.5 GHz--2.7 GHz range. The same antenna type was also used for the 915 MHz gateway communication link between the two boxes. Figure~6.3 shows the two antennas mounted in the measurement box.

\begin{figure}[htbp]
\centering\includegraphics[width=0.92\textwidth,height=0.70\textheight,keepaspectratio]{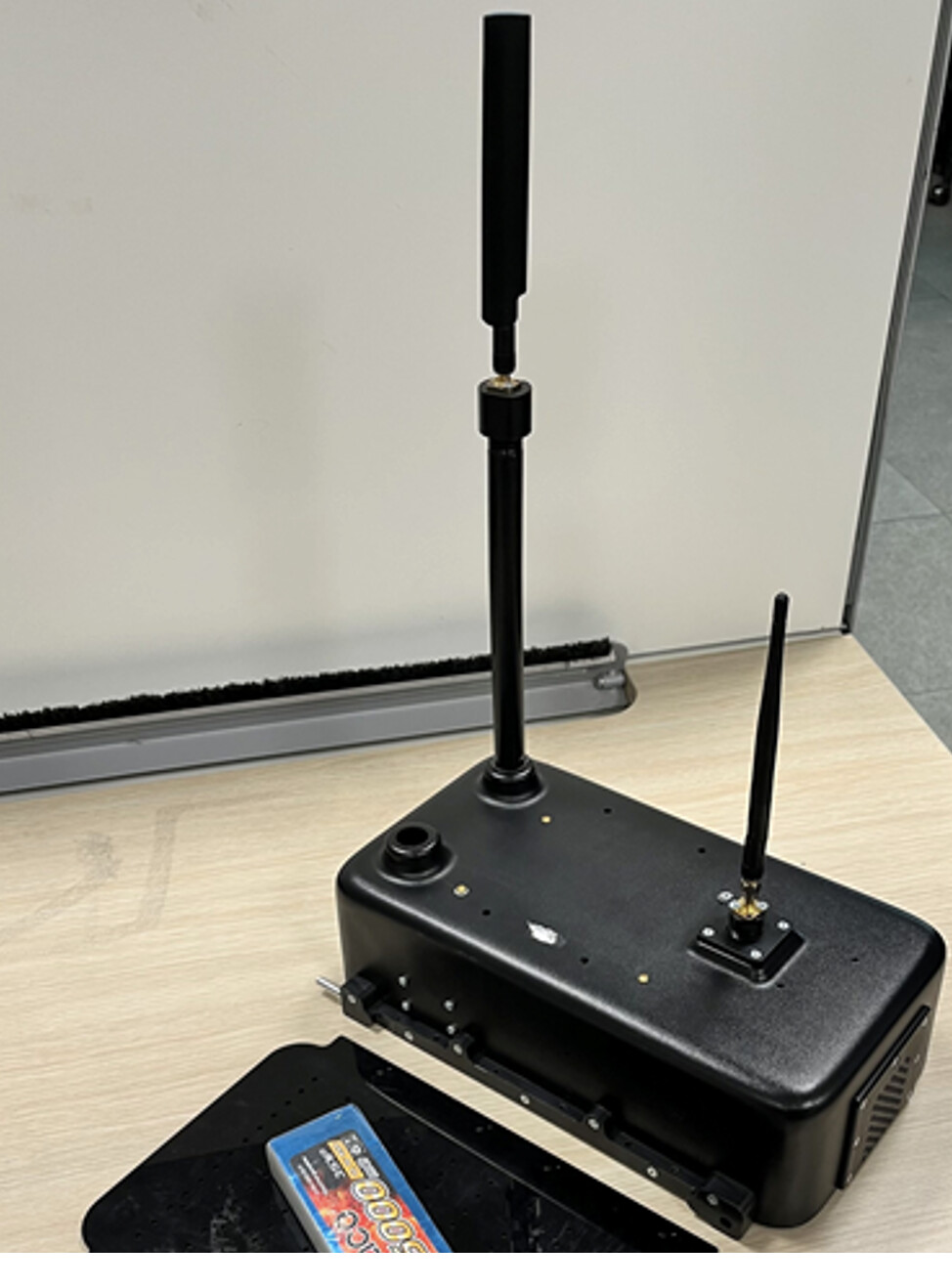}
\caption{Two antennas used in the measurement box}
\end{figure}

The link distance in the A2A campaign was larger than in the earlier campaigns, and the received signal became weaker as the transmitter moved away from the receiver. To improve the transmitted signal level, a ZX60-V63+ power amplifier manufactured by Mini-Circuits was added to the setup. The amplifier was connected between the antenna and the BladeRF SDR on the transmitter side and provided approximately 16 dB gain. Coaxial cables were used to connect the external antennas to the SDR inside the box. Because of cable-length limitations, the power amplifier was mounted on the side of the box, as shown in Figure~6.1.

\subsubsection{Microhard Gateway}

\hspace{0.95cm} The Microhard BulletPlus-NA2 was used as a gateway in the A2A measurement setup. It provides cellular connectivity, Ethernet interfaces, serial communication, programmable input-output pins, and GPS support. In this thesis, its main role was to create a communication bridge between the transmitter and receiver boxes. Through this gateway, the BladeRF SDRs and measurement boxes could be managed remotely from a computer. The gateway communication used the 915 MHz antennas mounted on the boxes. This control link was separate from the 2.57 GHz channel-sounding link. Separating the measurement link from the control link improved the practicality of the A2A system because the boxes could be monitored and controlled without interfering with the main channel measurement.

\subsubsection{UBLOX GPS and Clock}

\hspace{0.95cm} A UBLOX GPS module was included in the measurement setup to obtain accurate position information. Position data were important because the receiver-transmitter distance had to be known for the path-loss analysis. A clock was also used to improve synchronization between the boxes. Synchronization is especially important in A2A measurements because both platforms may move, and timing errors can directly affect the interpretation of delay-domain results.

\begin{figure}[htbp]
\centering\includegraphics[width=0.75\textwidth,keepaspectratio]{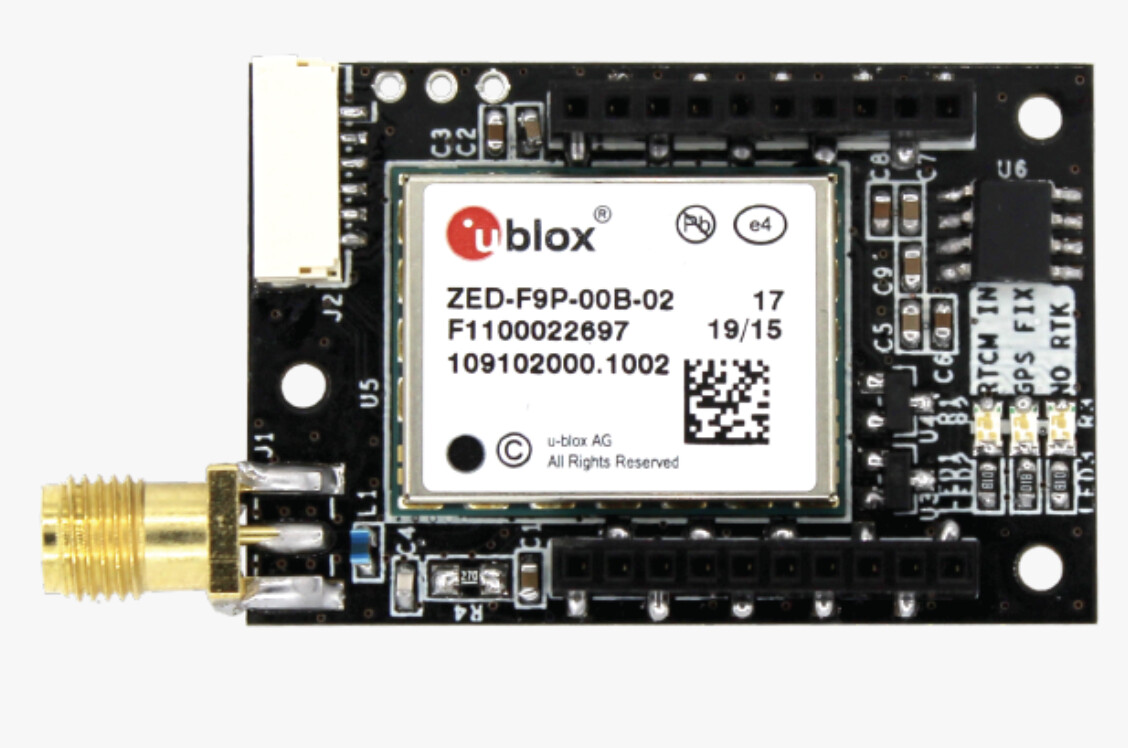}
\caption{ZED-F9P GPS module}
\end{figure}

\subsubsection{Computer, Flash Disk, and Power Bank}

\hspace{0.95cm} A computer was used on the receiver side to collect the received data. A flash disk was used on the transmitter side to store the sounding signal transmitted through the channel. A power bank supplied the power amplifier with 5 V. These components allowed the measurement boxes to operate as portable and self-contained units during the A2A campaign.

\subsection{Measurement Campaign}

\hspace{0.95cm} The A2A measurement campaign was conducted at 2.57 GHz for eight transmitter locations. The transmitter and receiver UAVs were flown at an altitude of approximately 50 m. The transmitter UAV moved horizontally in 100 m steps, and measurements were collected until the horizontal separation from the receiver reached 800 m. Figure~6.5 shows the measurement locations, and Figure~6.6 shows the UAV measurement environment.

\begin{figure}[htbp]
\centering\includegraphics[width=0.92\textwidth,height=0.70\textheight,keepaspectratio]{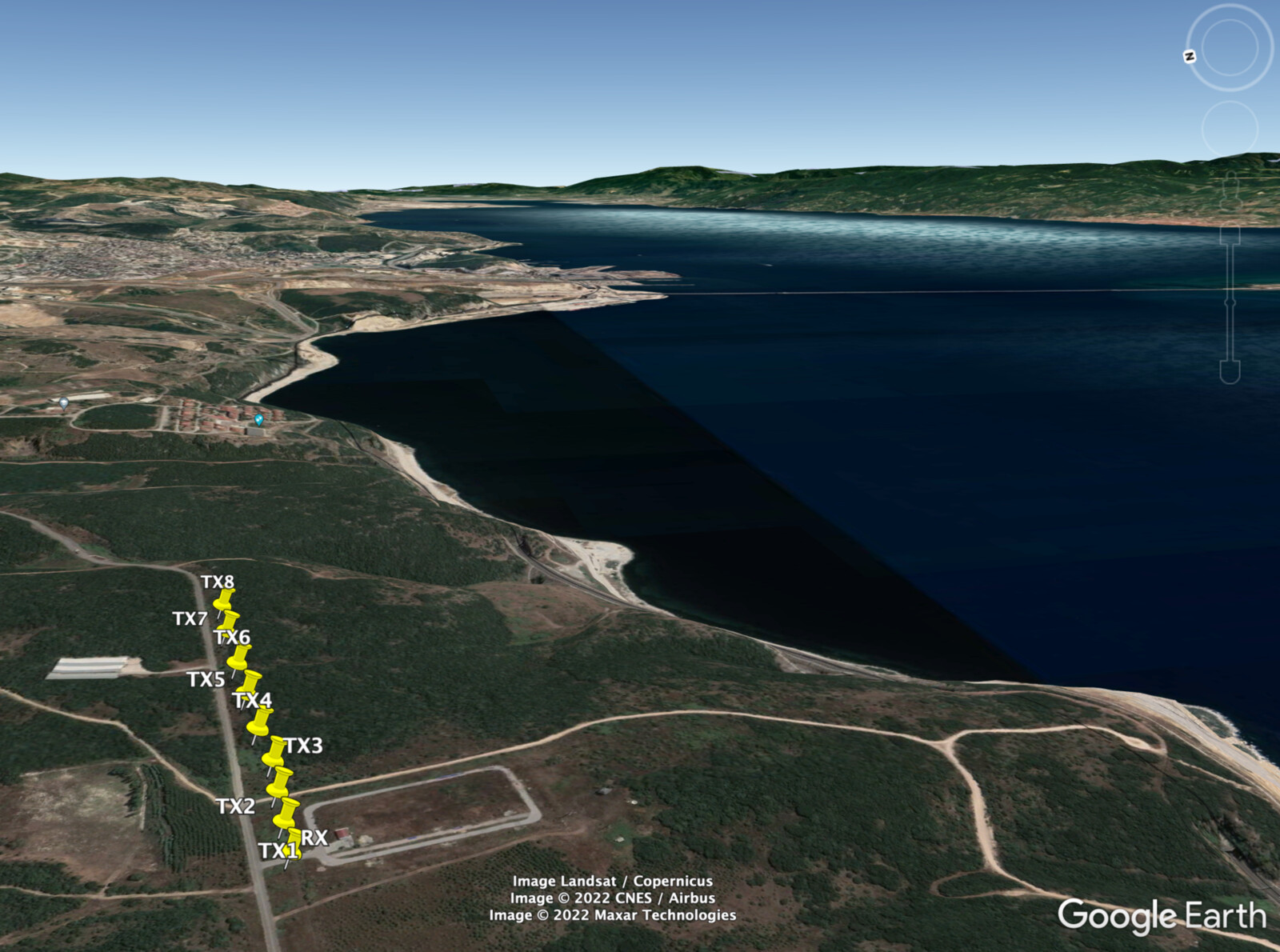}
\caption{A2A measurement locations}
\end{figure}

\begin{figure}[htbp]
\centering\includegraphics[width=0.92\textwidth,height=0.70\textheight,keepaspectratio]{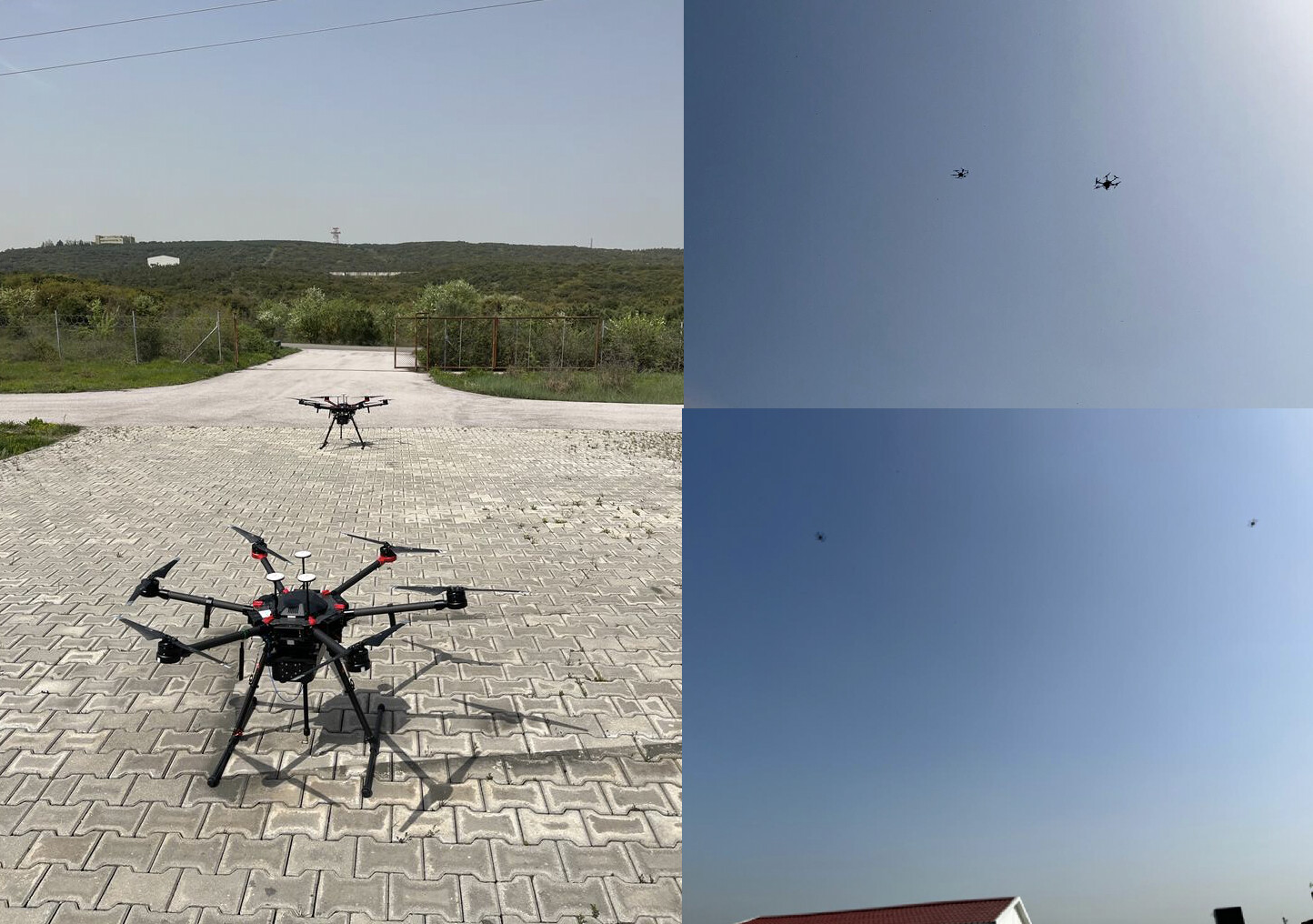}
\caption{UAVs and A2A measurement environment}
\end{figure}

\begin{table}[htbp]
\caption{A2A Measurement Locations}
$$
\begin{array}{|c|c|c|c|}
\hline \text { Devices } & \text { Latitude } & \text { Longitude } & \text { Receiver-Transmitter Distance } \\
\hline \text { Receiver } & 40.464335 \mathrm{~N}  & 29.275847 \mathrm{E} & - \\
\hline \text { TX1 } & 40.464381 \mathrm{~N} & 29.275989 \mathrm{E} & \mathbf{100 m} \\
\hline \text { TX2 } & 40.464433 \mathrm{~N} & 29.28160 \mathrm{E} & \mathbf{200 m} \\
\hline \text { TX3 } & 40.464236 \mathrm{~N} & 29.28347 \mathrm{E} & \mathbf{300 m} \\
\hline \text { TX4 } & 40.464543 \mathrm{~N} & 29.28584 \mathrm{E} & \mathbf{400 m} \\
\hline \text { TX5 } & 40.464582 \mathrm{~N} & 29.28810 \mathrm{E} & \mathbf{500 m} \\
\hline \text { TX6 } & 40.464621 \mathrm{~N} & 29.281054 \mathrm{E} & \mathbf{600 m} \\
\hline \text { TX7 } & 40.464670 \mathrm{~N} & 29.281345 \mathrm{E} & \mathbf{700 m} \\
\hline \text { TX8 } & 40.464698 \mathrm{~N} & 29.281588 \mathrm{E} & \mathbf{800 m}\\
\hline
\end{array}
$$
\end{table}

\section{Measurement Results}

\hspace{0.95cm} The A2A results were evaluated using the same general processing method applied in the earlier campaigns. The received IQ data were correlated with the transmitted M-sequence to obtain the delay response of the channel. The PDP was then calculated from the delay response, and the relative received power was used to estimate the path-loss exponent. Because both platforms were aerial, the channel was expected to include a strong LOS component, but reflections and scattering from the surrounding environment could still produce delayed components.

\subsection{Power Delay Profiles of the A2A Measurements}

\hspace{0.95cm} The A2A measurements were collected at eight receiver-transmitter distances. The outdoor environment produced several multipath components, which appear as peaks at different delay values in the PDP. Figure~6.7 shows the PDPs for the transmitter locations listed in Table~6.1.

\begin{figure}[htbp]
\centering\includegraphics[width=0.92\textwidth,height=0.70\textheight,keepaspectratio]{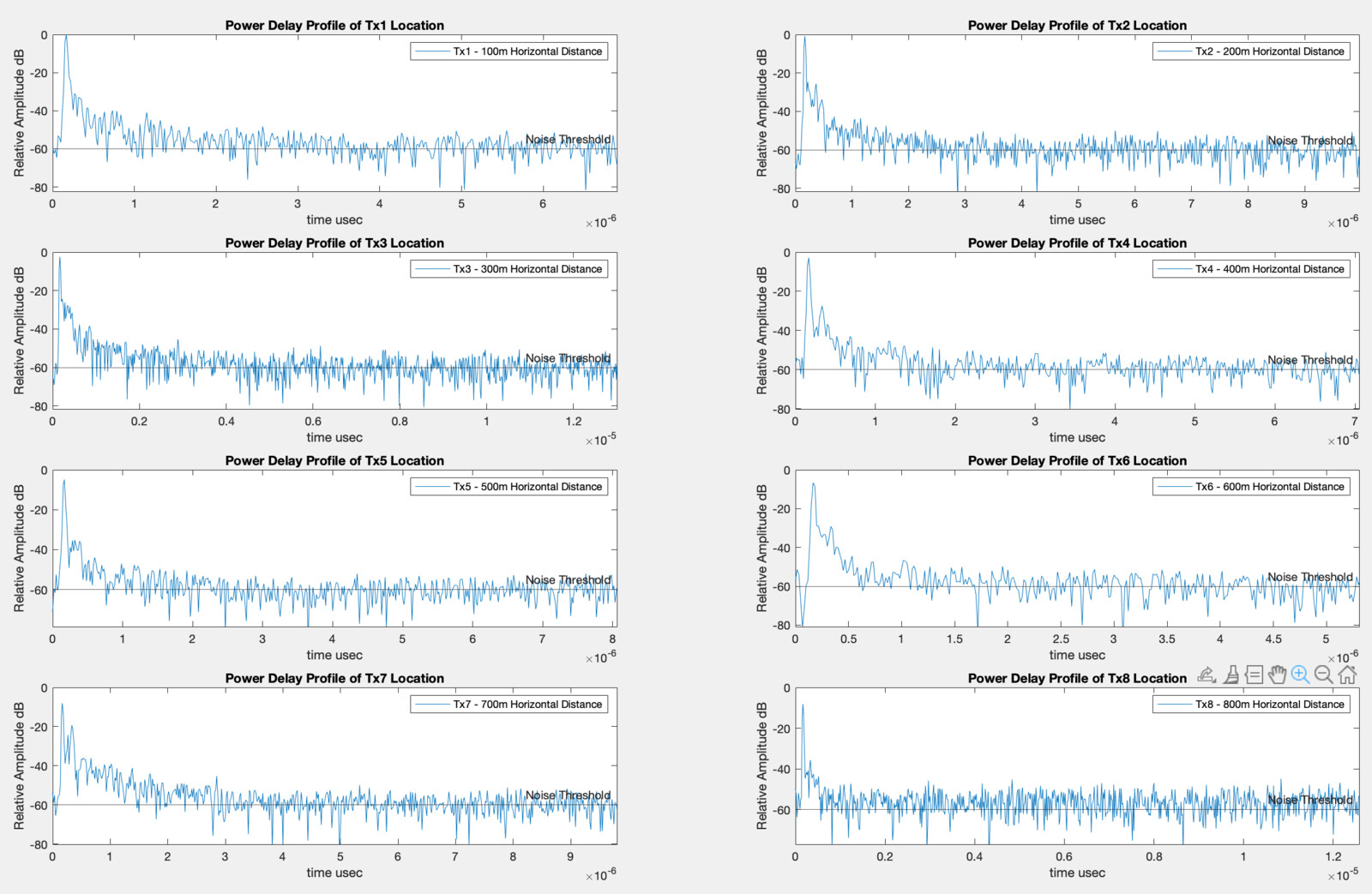}
\caption{Power delay profiles of the A2A measurements}
\end{figure}

There are eight different A2A channel responses at 2.57 GHz. Figure~6.8 shows these PDPs in terms of time, logarithmic distance, and relative power. As the distance between the transmitter and receiver increases, the relative gain generally decreases. TX1, which is 100 m from the receiver, has the strongest peak among the measured locations. Delayed components are also visible at several locations, showing that the A2A channel is not purely a single-path LOS channel.

\begin{figure}[htbp]
\centering\includegraphics[width=0.92\textwidth,height=0.70\textheight,keepaspectratio]{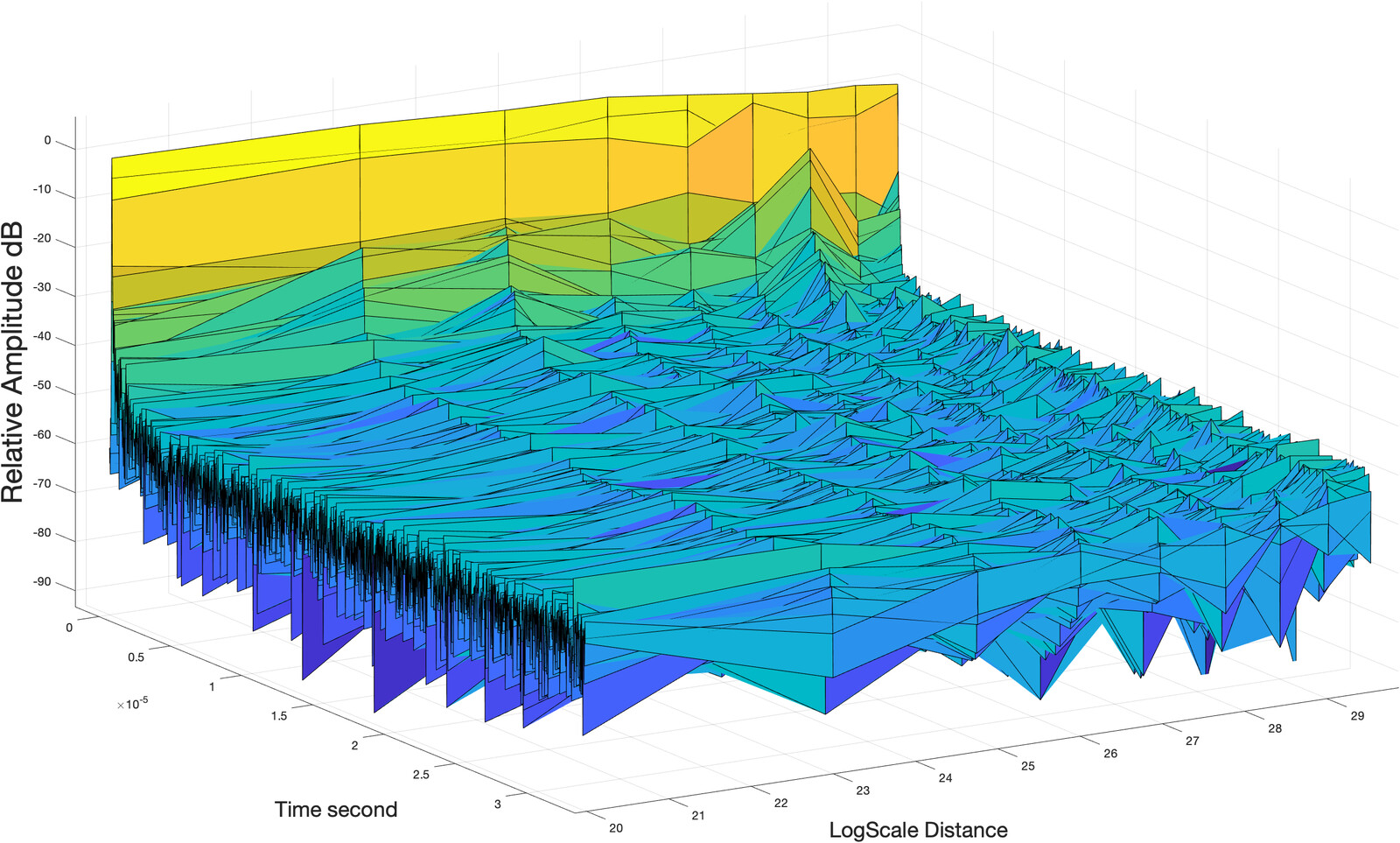}
\caption{Three-dimensional representation of the A2A power delay profiles}
\end{figure}

The three-dimensional PDP representation helps compare the eight locations simultaneously. The dominant component becomes weaker with distance, while the delayed components provide information about reflections and scattering in the measurement area. These observations confirm that PDP analysis is necessary even when the measurement scenario appears to be LOS-dominant.

\subsection{Path Loss of the A2A Measurements}

\hspace{0.95cm} The main objective of the A2A measurement campaign was to obtain the path-loss and PDP characteristics of the aerial channel. The path-loss exponent $\eta$ was calculated using the least-squares method in MATLAB. Figure~6.9 shows the relative power values, defined with respect to the reference point, and the receiver-transmitter distance on a logarithmic scale. IQ data were saved for each distance to reduce possible errors in the calculation. The receiver gain was adjusted to 30 dB during the measurements. The estimated path-loss exponent is $\eta=-2.079$, and the fitted path-loss expression is:

$$
P_{L}(d B)=-2.079 \times d-16.53
$$

\begin{figure}[htbp]
\centering\includegraphics[width=0.92\textwidth,height=0.70\textheight,keepaspectratio]{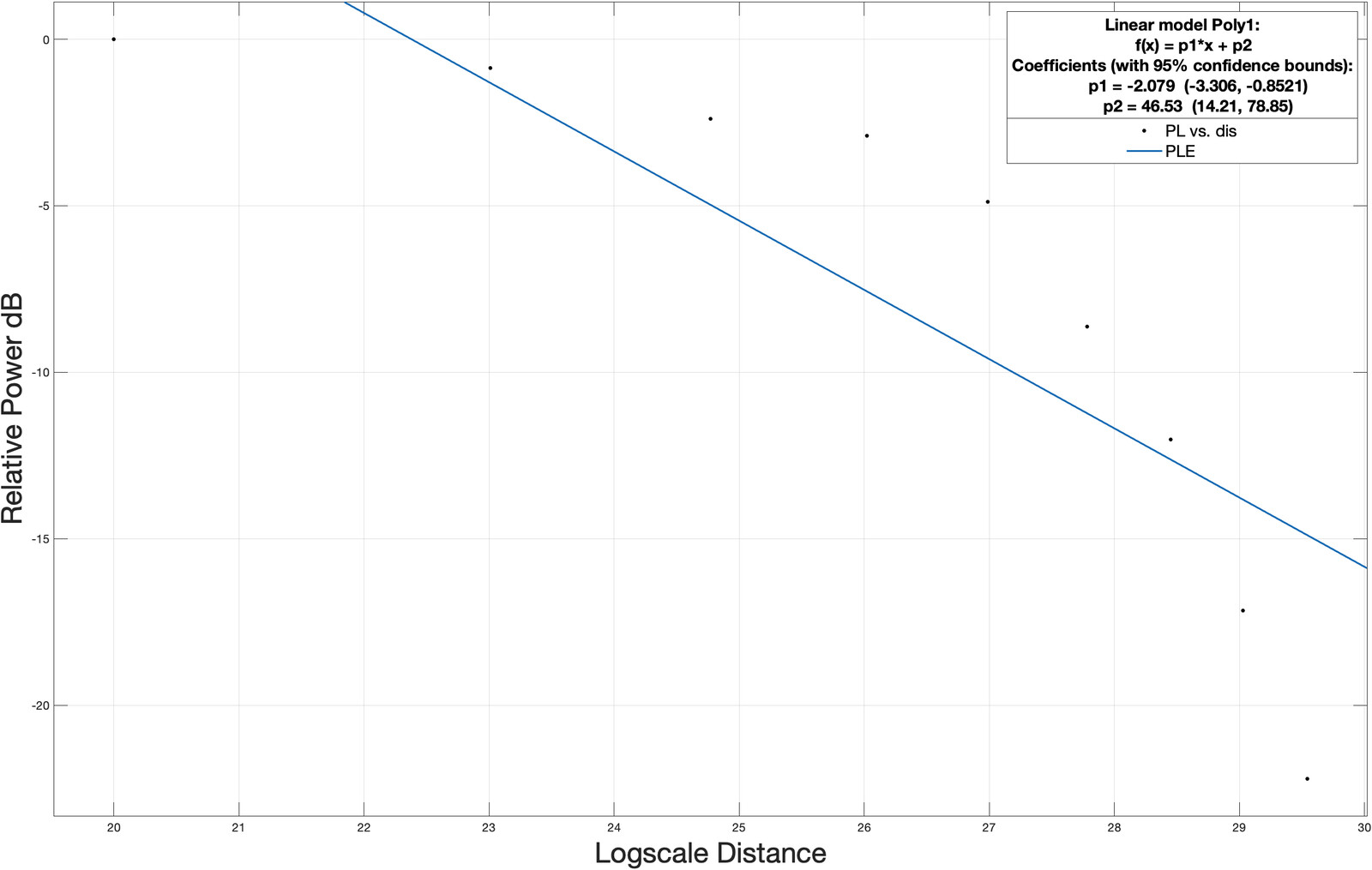}
\caption{A2A path-loss exponent}
\end{figure}

The A2A path-loss result indicates a clear distance-dependent attenuation trend. At the same time, the PDPs show that delayed components remain present in the received signal. Therefore, the A2A channel should be evaluated with both path-loss and delay-domain statistics, especially when the measurement environment includes possible ground reflections, nearby objects, or changes in antenna orientation.

%%%%%%%%%%%%%%%%%%%%%%%%%%%%%%%%%%%%%%%%%%%%%%%%%%%%%%%%%%%%%%%%%
\chapter{Conclusions and Improvements}\label{Ch7}
%%%%%%%%%%%%%%%%%%%%%%%%%%%%%%%%%%%%%%%%%%%%%%%%%%%%%%%%%%%%%%%%%

\section{Conclusions}

\hspace{0.95cm} The objective of this thesis was to characterize UAV communication channels using both large-scale and small-scale statistics. Many previous studies in the literature mainly focus on large-scale path-loss behavior. In this work, path loss was evaluated together with channel impulse response and power delay profile in order to obtain a more complete description of the propagation channel. Three measurement scenarios were investigated: ground-to-ground, air-to-ground, and air-to-air.

A suitable measurement setup was designed separately for each scenario. The G2G campaign was used as the first outdoor validation stage. It allowed the transmitter, receiver, antenna, power amplifier, SDR configuration, and processing method to be tested under practical conditions. The reflector experiment in the G2G scenario was especially important because it demonstrated that the system could detect an intentionally created multipath component and estimate its delay consistently with the physical path-length difference.

The A2G campaign extended the system from a fixed terrestrial transmitter to a UAV-mounted transmitter. For this purpose, a compact measurement box was designed and mounted on the drone. Before the outdoor A2G measurements, an anechoic chamber test was performed using a known cable delay. The measured delay and theoretical delay were very close, which verified the delay resolution and reliability of the channel sounder. The outdoor A2G measurements then showed how the PDP and received power changed as the UAV moved horizontally away from the ground receiver.

The A2A campaign was the most complex measurement scenario. In this case, two aerial platforms were used, and the measurement setup required two boxes, gateway communication, GPS support, synchronization, and a more advanced SDR platform. The A2A measurements showed a clear distance-dependent reduction in received power, but the PDPs also revealed delayed components. This result confirms that even aerial links, which are often assumed to be LOS-dominant, may still include multipath effects that should be considered in channel modeling.

\

\

One of the most important improvements during the project was increasing the sampling rate in order to improve the delay resolution. With a sampling rate of 60 MS/s, the distance resolution was approximately 4.9 m. This resolution made it possible to observe resolvable multipath components in the measurement campaigns. The reflector experiment and the anechoic chamber test were also critical steps because they verified the measurement setup before the UAV-based outdoor campaigns were performed.

After completing the outdoor measurements, the received data were processed to obtain the main channel statistics. Path loss, channel impulse response, and power delay profile were calculated using the methods described in Chapter~2. The G2G, A2G, and A2A results showed that path loss provides useful information about average attenuation, while the PDP provides additional insight into delay-domain multipath behavior. Therefore, a more realistic UAV channel model should not depend only on path-loss measurements.

\section{Recommendations for Future Work}

\hspace{0.95cm} Several improvements can be made in future studies to obtain a more detailed UAV channel model. First, the delay resolution can be improved by using a larger sounding bandwidth or a higher sampling rate. A finer resolution would allow the receiver to separate multipath components that are close to each other in time. This would be especially useful in urban or near-urban environments, where reflections from buildings and other structures may produce dense multipath components.

Second, future measurement campaigns should include a wider variety of environments. The measurements in this thesis were mainly performed under conditions where the LOS component was expected to be strong. Additional campaigns in urban, suburban, rural, over-water, and obstructed-line-of-sight environments would make it possible to compare how terrain and surrounding objects affect the UAV channel. In particular, A2A measurements in different terrains and at different altitudes would provide valuable information about the generality of the measured path-loss and PDP behavior.

Third, Doppler behavior should be investigated in more detail. UAV movement can introduce time variation in the channel, especially when the platform speed or antenna orientation changes during the measurement. Future studies may include controlled flight trajectories, different UAV speeds, and Doppler-spectrum estimation in addition to PDP and path-loss analysis. This would provide a more complete time-frequency characterization of UAV communication channels.

\

\

Fourth, antenna orientation and polarization should be examined systematically. In practical UAV systems, the antenna direction may change due to UAV motion, vibration, and wind. These changes can affect the received power and the multipath structure. Repeating the measurements with different antenna orientations and polarizations would help quantify these effects and improve the robustness of UAV channel models.

Finally, the measurement system can be improved by increasing automation and synchronization accuracy. Automatic data logging, GPS-based time stamping, better clock synchronization, and real-time monitoring would reduce measurement errors and make the campaign easier to repeat. With these improvements, future work can build more reliable and general UAV channel models that include path loss, CIR, PDP, delay spread, coherence bandwidth, and Doppler characteristics together.

\cleardoublepage
\phantomsection

\appendix
\frontchapter{APPENDICES}
The appendices provide supplementary material that supports the measurement and processing methodology used in the thesis. They are intended to keep the main chapters focused while still documenting the most important auxiliary definitions and practical steps.

\vspace{0.8cm}
\begin{description}[leftmargin=3.3cm,style=nextline,itemsep=0.8em]
  \item[\textbf{APPENDIX A:}] Supplementary channel-statistics definitions used in the processing of the measured data.
  \item[\textbf{APPENDIX B:}] Practical notes on measurement data organization and post-processing workflow.
\end{description}

\newpage

\chapter{Supplementary Statistics Definitions}
\label{App:ChannelStatistics}

\hspace{0.95cm} This appendix summarizes several expressions that are useful when processing measurement data for UAV communication channels. The expressions are provided as supporting material for the methodology discussed in the main chapters. They are not intended to replace the detailed derivations in the literature; rather, they provide a compact reference for the parameters evaluated in this thesis.

\section{Power Delay Profile}

The power delay profile describes the distribution of received power over propagation delay. If the estimated channel impulse response is denoted by $h(t,\tau)$, the instantaneous PDP can be written as
\begin{equation}
P(t,\tau)=\left|h(t,\tau)\right|^2 .
\label{EqAppA:PDP}
\end{equation}
For measurement records that contain several snapshots, averaging can be applied in order to reduce random variations:
\begin{equation}
\overline{P}(\tau)=\frac{1}{N}\sum_{n=1}^{N} P_n(\tau),
\label{EqAppA:AveragePDP}
\end{equation}
where $N$ is the number of available measurement snapshots.

\section{Mean Excess Delay and RMS Delay Spread}

Delay-spread parameters provide a compact description of the time dispersion of a multipath channel. For a discrete PDP with delay samples $\tau_k$ and power values $P_k$, the mean excess delay can be calculated as
\begin{equation}
\overline{\tau}=\frac{\sum_k P_k \tau_k}{\sum_k P_k}.
\label{EqAppA:MeanDelay}
\end{equation}
The RMS delay spread is then obtained from
\begin{equation}
\sigma_\tau=\sqrt{\frac{\sum_k P_k \tau_k^2}{\sum_k P_k}-\overline{\tau}^{\,2}}.
\label{EqAppA:RMSDelay}
\end{equation}
These parameters are important because they are related to frequency selectivity and inter-symbol interference in wideband systems.

\section{Log-Distance Path-Loss Fitting}

The average path-loss behavior can be represented by a log-distance model:
\begin{equation}
PL(d)=PL(d_0)+10\eta\log_{10}\left(\frac{d}{d_0}\right)+X_\sigma,
\label{EqAppA:LogDistance}
\end{equation}
where $d_0$ is the reference distance, $\eta$ is the path-loss exponent, and $X_\sigma$ represents shadowing or measurement variation. In practical processing, $\eta$ can be estimated by fitting measured path-loss values to the logarithmic distance term using a least-squares approach.

\chapter{Measurement Data Organization and Processing Notes}
\label{App:ProcessingNotes}

\hspace{0.95cm} A consistent data organization structure is important for repeatable UAV channel measurements. The measurement scenarios in this thesis involve different transmitter and receiver configurations, but the same general processing logic is followed for each scenario.

\section{Recommended Data Structure}

For future measurement campaigns, the following folder structure is recommended:
\begin{itemize}[itemsep=0.25em]
  \item \texttt{raw\_iq/}: unprocessed IQ files recorded during the campaign;
  \item \texttt{metadata/}: distance, altitude, GPS, date, time, antenna orientation, and measurement notes;
  \item \texttt{calibration/}: cable-delay tests, chamber measurements, and reference measurements;
  \item \texttt{processed/}: estimated CIR, PDP, path-loss values, and delay-spread parameters;
  \item \texttt{figures/}: plots generated from the processed results.
\end{itemize}
This structure separates raw data from processed outputs and reduces the risk of overwriting original measurement files.

\section{Processing Workflow}

The post-processing workflow can be summarized as follows:
\begin{enumerate}[itemsep=0.25em]
  \item import the recorded IQ data and the corresponding measurement metadata;
  \item apply synchronization and identify the useful part of the received sounding sequence;
  \item estimate the channel impulse response through correlation or matched filtering;
  \item compute the PDP from the magnitude-squared impulse response;
  \item normalize or calibrate the received power according to the selected reference case;
  \item fit the path-loss model and extract delay-domain parameters;
  \item compare G2G, A2G, and A2A scenarios using consistent plotting and reporting rules.
\end{enumerate}

\section{Documentation Notes}

\hspace{0.95cm} Each measurement record should include enough information to reproduce the processing result. At minimum, the documentation should include carrier frequency, sampling rate, antenna type, transmit gain, receiver gain, distance, altitude, flight condition, environment description, and file naming convention. These details are especially important for UAV measurements because small changes in antenna placement, platform motion, and surrounding objects may noticeably affect the measured channel response.

\frontchapter{RESUME}
\section*{Necati Kağan Erkek}
Necati Kağan Erkek completed his undergraduate studies at Istanbul Technical University, Faculty of Electrical and Electronics Engineering. His bachelor thesis work focused on measurement-based UAV communication channel modeling, including SDR-based channel sounding, path-loss estimation, and delay-domain channel analysis.

\vspace{0.5cm}
\section*{Emre Balcı}
Emre Balcı completed his undergraduate studies at Istanbul Technical University, Faculty of Electrical and Electronics Engineering. Within this thesis, he contributed to the experimental measurement setup, UAV-based campaign preparation, and post-processing of the collected channel measurement data.

\vspace{0.5cm}
\section*{Berkin Halay}
Berkin Halay completed his undergraduate studies at Istanbul Technical University, Faculty of Electrical and Electronics Engineering. His contribution to the thesis included system integration, measurement verification, and interpretation of UAV channel statistics for G2G, A2G, and A2A scenarios.

\begin{figure}[t]
\centering\includegraphics[width=0.92\textwidth,height=0.70\textheight,keepaspectratio]{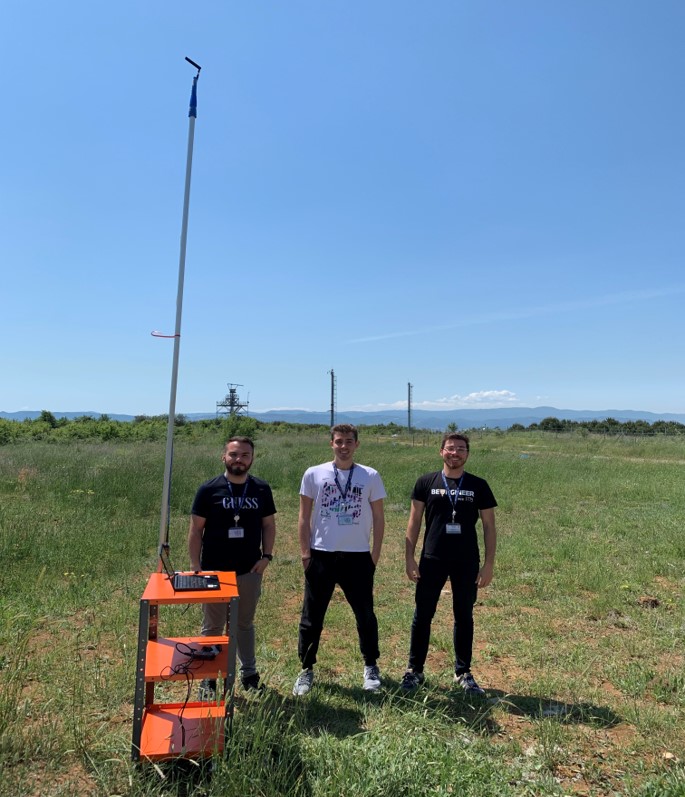}
\end{figure}

\end{document}